\documentclass[10pt]{article}
\usepackage{fullpage}
\usepackage{amsmath}
\usepackage{amssymb}
\usepackage[dvips]{epsfig}
\usepackage{color}
\usepackage{subfigure}

\def\##1{{\bf #1}}
\def\=#1{\underline{\underline #1}}
\def\.{\mbox{ \tiny{$^\bullet$} }}

\def\epso{\epsilon_{\scriptscriptstyle 0}}
\def\alphao{\alpha_{\scriptscriptstyle 0}}
\def\alphal{\alpha_\ell}
\def\lambdao{\lambda_{\scriptscriptstyle 0}}

\def\muo{\mu_{\scriptscriptstyle 0}}
\def\etao{\eta_{\scriptscriptstyle 0}}
\def\etal{\eta_\ell}

\def\ko{k_{\scriptscriptstyle 0}}
\def\kl{k_\ell}
\def\nl{n_\ell}

\def\eps{\epsilon}
\def\epsa{\epsilon_a}
\def\epsb{\epsilon_b}
\def\epsc{\epsilon_c}
\def\epsa2{\epsilon_{a2}}
\def\epsb2{\epsilon_{b2}}
\def\epsc2{\epsilon_{c2}}

\def\ux{\#u_x}
\def\uy{\#u_y}
\def\uz{\#u_z}

\def\obs{^{obs}}

\def\un{{\#u}_n}
\def\ut{{\#u}_\tau}
\def\ub{{\#u}_b}

\def\le{\left(}
\def\ri{\right)}
\def\les{\left[}
\def\ris{\right]}
\def\lec{\left\{}
\def\ric{\right\}}

\def\c#1{\cite{#1}}
\def\l#1{\label{#1}}
\def\r#1{(\ref{#1})}

\begin{document}

\begin{center}

{\bf {\LARGE  On  chemiluminescent emission from an \vspace{8pt}
\\ infiltrated chiral sculptured thin film}}

\vspace{5mm}

{\bf {\large Siti S. Jamaian${}^{a}$ and Tom G. Mackay${}^{a,b,}$\footnote{Email: T.Mackay@ed.ac.uk}}}\\

\vspace{5mm}

${}^{a}$School of Mathematics and
   Maxwell Institute for Mathematical Sciences,\\
University of Edinburgh, Edinburgh EH9 3JZ, UK \vspace{4pt}  \\
${}^{b}$NanoMM---Nanoengineered Metamaterials Group, Department of Engineering Science and Mechanics,\\
Pennsylvania State University, University Park, PA 16802-6812, USA

\end{center}
\vspace{5mm}

\begin{abstract}

The theory describing the far--field emission from a dipole source
embedded inside a chiral sculptured thin film (CSTF), based on a
spectral Green function formalism,
 was further
developed to allow for infiltration of the void regions of the CSTF
by a fluid. In  doing so, the extended Bruggeman homogenization
formalism~---~which accommodates constituent particles that are
small compared to wavelength but not vanishingly small~---~was used
to estimate the relative permittivity parameters of the infiltrated
CSTF. For a numerical example, we found  that left circularly
polarized (LCP) light was preferentially emitted through one face of
the CSTF while right circularly polarized (RCP) light was
preferentially emitted through the opposite face, at wavelengths
within the Bragg regime. The centre wavelength for the preferential
emission of LCP/RCP light was red shifted as the refractive index of
the infiltrating fluid increased from unity, and this red shift was
accentuated when the size of the constituent particles in our
homogenization model was increased. Also, the bandwidth of the
preferential LCP/RCP emission regime decreased as the refractive
index of the infiltrating fluid increased from unity.

\end{abstract}

\vspace{15mm}

\noindent {\bf Keywords:} Inverse Bruggeman homogenization
formalism; spectral Green function; dipole radiation

\vspace{15mm}

\section{Introduction} \l{intro}

Chiral sculptured thin films (CSTFs) constitute a remarkable class
of engineered materials. A CSTF  comprises an array of parallel
helical nanowires which can be grown on a substrate using vapour
deposition techniques \c{STF_Book,HW}. By careful control of the
fabrication process, both the optical properties and the
multi--scale porosity of the CSTF can be tailored to order
\c{Messier}. Accordingly, CSTFs are promising candidates as
platforms for optical sensing, as well as a host of other
applications
 \c{Polo,LDHX,Wong}.

 Three quite different methods of utilizing  CSTFs for
  optical sensing applications have been proposed. The first relies
on the circular Bragg phenomenon, which CSTFs exhibit just as
cholesteric liquid crystals  do \c{Gennes}. That is, within a
wavelength regime known as the Bragg regime, a structurally
right--handed CSTF almost completely reflects normally--incident
right circularly polarized (RCP) plane waves whereas
normally--incident left circularly polarized (LCP) plane waves are
reflected very little. And vice versa for a structurally
left--handed CSTF. The spectral shift in the circular Bragg
phenomenon  induced by infiltration of the void regions between the
CSTF's nanowires may be exploited for sensing applications
\c{ML_STF_PJ}. The second sensing methodology is based on the
excitation of surface--plasmon--polariton waves at the interface of
an infiltrated CSTF and a suitable metal \c{ML_STF_SJ}. The third
methodology for CSTF--based optical sensing~---~which provides the
main motivation for the present communication~---~involves the
emission of radiation from a chemiluminescent source embedded within
a CSTF \c{L01}.

Chemiluminescent radiation may be generated within a CSTF~---~and
harnessed for optical  biosensing~---~as follows. Suppose that
certain biorecognition sites are immobilized on the surface of a
CSTF's nanowires, and the CSTF becomes infiltrated with a solution
containing complementary analyte biomolecules.  The binding of the
analytes to the biorecognition sites, perhaps in the presence of a
transition metal complex, results in the generation of
chemiluminescent photons \c{Fan,Campana}. For example, the
biorecognition sites and the analytes could be fragments of
complementary single--stranded DNA, which combine to produce
chemiluminescence in the presence of a ruthenium complex
\c{Beck,Ru,Chang}. Inspired by the prospects of this
chemiluminescent sensing scenario, the theory of emission from a
dipole source embedded inside a CSTF was recently established
\c{ML_OE}. In the following, we extend the theory  in order to
investigate the effects on infiltration upon the far--field dipole
radiation.

As regards notation, vectors and matrixes are represented in
boldface, with the Cartesian unit vectors given by $\#u_{x,y,z}$;
dyadics are double--underlined; and both 4--vectors and 4$\times$4
matrixes are enclosed within square brackets. The inverse of matrix
$\les \#Z \ris$ is written as $\les \#Z \ris^{-1}$, while the
inverse, adjoint and trace of the  dyadic $\=Z$ are written as
$\=Z^{-1}$, $\=Z^{adj}$ and $\mbox{tr} \le \=Z \ri$, respectively.
The permittivity and the permeability of free space are denoted by
$\epso$ and $\muo$, respectively; $\ko=\omega\sqrt{\epso\muo}$ is
the free--space wavenumber; $\lambdao=2\pi/\ko$ is the free--space
wavelength; and $\etao=\sqrt{\muo/\epso}$ is the intrinsic impedance
of free space.
 An
$\exp(-i\omega t)$ time--dependence is implicit, with $\omega$ being
the angular frequency.

\section{Theory}

In this section we present an overview of the theory which describes
the emission of electromagnetic radiation from a dipole source
embedded inside a CSTF that is infiltrated by a fluid of refractive
index $n_\ell$. The theory rests upon two recently--established
pillars: First, in \S\ref{inv_homog} we describe how the
constitutive parameters of an infiltrated CSTF can be estimated
using an inverse homogenization technique. Second, in
\S\ref{radiation_from_dipole} the far--field radiative emission from
a dipole source embedded in an infiltrated CSTF is estimated  via a
spectral Green function approach. Comprehensive details of these two
pillars are available elsewhere \c{ML_STF_PJ,ML_OE,ML_STF_JNP}. Let
us note that the theory presented in \S\ref{inv_homog} represents an
extension of that presented in \c{ML_STF_PJ} insofar as here the
\emph{extended} Bruggeman homogenization formalism is implemented.
Also, the theory presented in \S\ref{radiation_from_dipole} differs
from that developed in \c{ML_OE} insofar as here we consider an
\emph{infiltrated} CSTF whose upper surface is immersed in a fluid
  of refractive
index $n_\ell$. A schematic diagram of the scenario under
consideration is provided in Fig.~\ref{schematic}.

\subsection{Constitutive parameters of an infiltrated CSTF} \l{inv_homog}

A CSTF consists of an array of parallel nanowires. The `chirality'
of the CSTF stems from the fact that the nanowires are helical
 \cite{STF_Book}.
Such structures  can be grown on a
  planar substrate~---~parallel to the plane $z=0$, say~---~
 by the deposition of an evaporated bulk material.
The helical shape of the nanowires is achieved by means of uniform
rotation of
 the
 substrate  about
the $z$ axis during the deposition process. We suppose that
 the deposited material is
 an isotropic dielectric material of refractive index $n_s$.
Notice that the value of $n_s$ needs to be determined as it may
differ somewhat from the refractive index of the bulk material that
was evaporated, depending upon the precise nature of the deposition
conditions \c{MTR1976,BMYVM,WRL03}.

It is convenient here to regard the individual helixes of a CSTF as
strings of highly elongated ellipsoidal inclusions, wound
end--to--end around the $z$ axis \cite{Sherwin,Lakh_Opt}.
 The position
vector
\begin{equation} \l{r_U}
\#r (\vartheta, \varphi) =   \frac{ \eta }{ \sqrt{\gamma_\tau
\gamma_b}} \, \=U \cdot \les \, \sin \vartheta \cos \varphi \, \#u_n
+ \cos \vartheta \, \#u_\tau + \sin \vartheta \sin \varphi \,\#u_b\,
\ris, \qquad \vartheta \in \les 0, \pi \ris, \:\: \varphi \in \les
0, 2 \pi \ris
\end{equation}
 prescribes the surface of a particular ellipsoid relative to
its centroid. Herein $\eta$ is a linear measure of  size while  the
shape dyadic
\begin{equation}
\=U = \un \, \un + \gamma_\tau \, \ut \, \ut + \gamma_b \, \ub \,
\ub
\end{equation}
is expressed in terms of  the normal, tangential, and binormal basis
vectors per
\begin{equation}
\left. \begin{array}{l}
 \un = - \ux \, \sin \chi + \uz \, \cos \chi\,\quad \vspace{0pt} \\
 \ut =  \ux \, \cos \chi + \uz \, \sin \chi\,\quad \vspace{0pt} \\
\ub = - \uy
\end{array}
\right\},
\end{equation}
with $\chi$ denoting the inclination angle relative to the $xy$
plane. An elongated ellipsoidal shape  is achieved
 by selecting the shape parameters $\gamma_{b} \gtrsim 1$ and
$\gamma_\tau \gg 1$. Since increasing $\gamma_\tau$ beyond 10 does
not result in  significant effects for slim inclusions
\cite{Lakh_Opt}, the value
  $\gamma_\tau = 15$ is taken for the numerical results presented in \S\ref{Numerica}.

 The proportion of a CSTF's total volume occupied by helical
nanowires is represented by $f \in \le 0, 1 \ri $. That is to say,
 the  volume fraction  $1 -
f$ of a CSTF is not occupied by nanowires.

Let us suppose now that the CSTF under consideration occupies the
region $-L \leq z \leq L$, and is  unbounded in extent in directions
perpendicular to the $z$ axis. At length scales much greater than
the nanoscale, the CSTF is characterized by the relative
permittivity dyadic
\begin{eqnarray}
&& \=\eps_{\,cstf} = {\=S}_{\,z} \les h  \frac{\pi (z+L)}{\Omega}
\ris \cdot {\=S}_{\,y} \le \chi \ri \cdot \=\eps^{(\nu)}_{\,cstf}
\cdot {\=S}^T_{\,y} \le \chi \ri  \cdot {\=S}^T_{\,z} \les h
\frac{\pi (z+L)}{\Omega} \ris,\qquad  z \in \les -L,  L \ris,
\l{eps1_dyadic}
\end{eqnarray}
with the rotation dyadics
\begin{equation}
\left.
\begin{array}{r}
{\=S}_{\,y} \le \chi \ri = \#u_y\, \#u_y + \le \#u_x\, \#u_x +
\#u_z\, \#u_z \ri \cos \chi  + \le \#u_z\, \#u_x - \#u_x\, \#u_z \ri
\sin \chi \vspace{4pt} \\
{\=S}_{\,z} \le \sigma \ri =
 \#u_z\, \#u_z +
\le \#u_x\, \#u_x + \#u_y\, \#u_y \ri \cos  \sigma  + \le \#u_y\,
\#u_x - \#u_x\, \#u_y \ri \sin  \sigma
\end{array}
\right\}.
\end{equation}
 The structural period is   $2 \Omega$, and the handedness parameter $h = + 1 (-1)$ for a structurally
right (left)--handed CSTF.
 The reference
relative permittivity dyadic $\=\eps^{(\nu)}_{\,cstf}$ characterizes
the local orthorhombic symmetry; i.e.,
\begin{equation}
\=\eps^{(\nu)}_{\,cstf} = \eps_{a \nu}   \,\un\,\un +\eps_{b
\nu}\,\ut\,\ut \, +\,\eps_{c \nu}\,\ub\,\ub\, , \l{eps1_ref_dyadic}
\end{equation}
with $\nu =1$ denoting  an uninfiltrated CSTF (in which case
 the void regions between nanowires are assumed to be vacuous) and $\nu =2$ denoting a CSTF in which the
void regions are filled with a fluid of refractive index $n_\ell$.

We are required to estimate the relative permittivity parameters
$\lec \eps_{a2}, \eps_{b2}, \eps_{c2} \ric$ for an infiltrated CSTF
from a knowledge of the corresponding parameters  $\lec \eps_{a1},
\eps_{b1}, \eps_{c1} \ric$ for an uninfiltrated CSTF. A two--step
strategy is employed. The first step is the estimation of the
nanoscale parameters $\lec n_s, f, \gamma_b \ric$~---~which are not
readily determined by experimental means~---~from a knowledge of
$\lec \eps_{a1}, \eps_{b1}, \eps_{c1} \ric$. As described in detail
elsewhere \c{ML_STF_JNP}, this can be achieved by applying the
inverse Bruggeman
 homogenization formalism. Once $\lec n_s,
f, \gamma_b \ric$ have been estimated, the second step can be taken
wherein  these  parameters characterizing the uninfiltrated CSTF are
combined with  $\lec n_\ell,\gamma_\tau\ric$ in order to determine
the relative permittivity parameters $\lec \eps_{a2}, \eps_{b2},
\eps_{c2} \ric$ for the infiltrated CSTF, by applying the Bruggeman
homogenization formalism  in its usual forward sense
\c{ML_STF_PJ,Lakh_Opt}. For this second step we implement the
extended version of the Bruggeman formalism which takes into account
the nonzero size of the ellipsoidal inclusion particles \c{M_depol}.
That is, the size parameter $\eta$ is taken to be small relative to
wavelength(s)  but nonzero. Details of the extended Bruggeman
formalism, applicable to the locally anisotropic dielectric
materials under consideration here, is provided in the Appendix.

\subsection{Radiation from dipole source inside an infiltrated CSTF}
\l{radiation_from_dipole}

\subsubsection{Spectral Green function formulation}

Let us now introduce an electric dipole of moment $p/2$,
 oriented in the direction of the unit vector $\#u_{J}$, which  is embedded
inside the CSTF at ${\bf r} = d\,\#u_z$, $d \in \le -L,  L \ri$. The
corresponding source current density phasor is given by
\begin{equation}
{\bf J}(\#r,\omega) = - \frac{i\omega p}{2} \, \#u_{J}^{}
\,\delta(z-d)\,\delta(x)\,\delta(y)\,, \l{so}
\end{equation}
where $\delta(\cdot) $ is the Dirac delta function. With a view to
implementing  a spectral--Green--function formalism \c{LW97b},  the
spatial Fourier transform representation
\begin{equation}
\#J_{} (\#r, \omega) =\frac{1}{4\pi^2}\int_0^\infty\,d\kappa
\,\kappa\int_0^{2\pi} \,d\psi\; \#j_{} (z, \kappa, \psi, \omega) \,
\exp \les i \kappa \le x \, \cos \psi + y \, \sin \psi \ri \ris
\end{equation}
is adopted,  with $ \#j_{}(z, \kappa, \psi, \omega) = - (i \omega
p/2)\, \#u_{J} \, \delta \le z - d \ri $ in accordance with eq.
\r{so}.

Our attention is focused on
 a single spatial--Fourier  component of $\#J_{} (\#r, \omega)$, i.e.,
\begin{equation}
\#J_{} (\#r, \kappa, \psi, \omega) = \#j_{} (z, \kappa, \psi,
\omega) \, \exp \les i \kappa \le x \, \cos \psi + y \, \sin \psi
\ri \ris, \qquad z \in \le -L, L \ri. \l{j}
\end{equation}
Similarly, we write  the electromagnetic field phasors inside the
CSTF as
\begin{equation}
\left. \begin{array}{l}
 \#E (\#r, \kappa, \psi,  \omega) = \#e (z,
\kappa, \psi, \omega) \, \exp \les i
\kappa \le x \, \cos \psi + y \, \sin \psi \ri \ris\,, \vspace{6pt}\\
\#H (\#r, \kappa, \psi,  \omega)  = \#h (z, \kappa, \psi, \omega) \,
\exp \les i \kappa \le x \, \cos \psi + y \, \sin \psi \ri \ris
\end{array}
\right\}, \qquad z \in \le -L, L \ri . \l{hp}
\end{equation}
The procedure whereby
   the Fourier components \r{j} and \r{hp} are combined with the constitutive relations
   for the CSTF and
 the frequency--domain Maxwell curl postulates, and then solved to find expressions
  for the phasors inside the CSTF,  is comprehensively described
  elsewhere \c{ML_OE}. Therefore, here we simply state the particular solution
\begin{eqnarray}
 \les \#f (L, \kappa, \psi, \omega) \ris &=& \les \#M (L,  \kappa,
\psi, \omega) \ris \les \#f (-L, \kappa, \psi, \omega) \ris
\nonumber
\\ && - i \omega \frac{p}{2} \les \#M (L, \kappa, \psi, \omega) \ris \les \#M (d, \kappa,
\psi, \omega) \ris^{-1} \les \tilde{\#g} ( d, \kappa, \psi, \omega)
\ris, \l{fL}
\end{eqnarray}
wherein the column vectors
 \begin{equation} \les \#f (z, \kappa, \psi, \omega) \ris = \les
\begin{array}{c}
\#e(z, \kappa, \psi, \omega) \cdot \#u_x  \\
\#e(z, \kappa, \psi, \omega) \cdot \#u_y  \\
\#h(z, \kappa, \psi, \omega) \cdot \#u_x  \\
\#h(z, \kappa, \psi, \omega) \cdot \#u_y
\end{array} \ris
\end{equation}
and
\begin{equation} \les \tilde{\#g} (z, \kappa, \psi, \omega) \ris =
\frac{\#u_{J}\cdot\#u_z}{\eps_{a \nu} \, \cos^2 \chi + \eps_{b \nu}
\, \sin^2 \chi}
 \les
\begin{array}{c}
\frac{\kappa \cos \psi}{\omega \epso}
 \\
\frac{\kappa \sin \psi}{\omega \epso}
 \\
 \le  \eps_{a \nu} - \eps_{b \nu} \ri \, \sin \chi \, \cos \chi \, \sin \les h
\frac{ \pi (z+L)}{\Omega}\ris \vspace{2pt}
 \\
 \le  \eps_{b \nu} - \eps_{a \nu} \ri \, \sin \chi \, \cos \chi \, \cos
 \les h \frac{ \pi (z+L)}{\Omega}\ris
 \end{array} \ris +
  \les
\begin{array}{c}
0 \\
0 \\
\#u_{J}\cdot\#u_y \\
-\#u_{J}\cdot\#u_x \end{array} \ris\,.
\end{equation}
The expressions for the 4$\times$4
 matrizants $\les \#M (L,  \kappa, \psi, \omega) \ris$ and $\les
\#M (d, \kappa, \psi, \omega) \ris$~---~which are straightforwardly
derived, but too cumbersome to reproduce here~---~are available in
standard works \cite[Chap. 9]{STF_Book}. We note that the piecewise
uniform approximation technique provides a convenient method for
their evaluation \cite[Chap. 9]{STF_Book}.

\subsubsection{Boundary value problem}

Next we turn to the   two half--spaces $z < -L$ and $z > L$. The
half--space $z < -L$ is vacuous while the half--space $z > L$ is
filled by a fluid of refractive index $\nl $. In consonance with
eqs. \r{hp}, the electromagnetic field phasors for these two
half--spaces may be expressed as
\begin{equation}
\left.
\begin{array}{l}
 \#E (\#r, \kappa, \psi,  \omega) = \displaystyle{ \frac{1}{\sqrt{2}}} \les
-b_L \le \kappa, \psi ,\omega\ri \le i \#s - \#p_- \ri + b_R \le
\kappa, \psi,\omega \ri \le i \#s + \#p_- \ri \ris \,  \\
\hspace{70pt}  \times \exp \lec i \les \kappa \le x \,
\cos \psi + y \, \sin \psi \ri  - \alphao \le z + L \ri  \ris \ric\, \vspace{6pt} \\
\#H (\#r, \kappa, \psi,  \omega) =  \displaystyle{ \frac{i}{\etao
\sqrt{2}}} \les b_L \le \kappa, \psi,\omega \ri \le i \#s - \#p_-
\ri
+ b_R \le \kappa, \psi,\omega \ri \le i \#s + \#p_- \ri \ris \, \\
\hspace{70pt} \times \exp \lec i \les \kappa \le x \, \cos \psi + y
\, \sin \psi \ri   - \alphao \le z + L \ri \ris \ric\,\end{array}
\right\}, \qquad z < -L , \l{hhp}
\end{equation}
and
\begin{equation}
\left. \begin{array}{l} \#E (\#r, \kappa, \psi,  \omega) =
 \displaystyle{ \frac{1}{\sqrt{2}}} \les c_L \le \kappa, \psi, \omega\ri \le i \#s -
\#p_+ \ri - c_R \le
\kappa, \psi, \omega \ri \le i \#s + \#p_+ \ri \ris \,  \\
\hspace{70pt} \times \exp \lec i \les \kappa \le x \, \cos \psi + y
\, \sin \psi \ri   + \alphal \le z - L \ri \ris \ric
\, \vspace{6pt} \\
\#H (\#r, \kappa, \psi,  \omega) = - \displaystyle{ \frac{i}{\etal
\sqrt{2}}} \les c_L \le \kappa, \psi, \omega \ri  \le i \#s - \#p_+
\ri + c_R \le
\kappa, \psi, \omega \ri \le i \#s + \#p_+ \ri \ris \,  \\
\hspace{70pt} \times \exp \lec i \les \kappa \le x \, \cos \psi + y
\, \sin \psi \ri   + \alphal \le z - L \ri \ris \ric \end{array}
\right\} , \qquad z > L, \l{hpp}
\end{equation}
where $\alphao=+\sqrt{\ko^2-\kappa^2}$,
$\alphal=+\sqrt{\kl^2-\kappa^2}$, $\kl = \ko \nl$ and $\etal = \etao
/ \nl$. The complex--valued amplitudes $b_L \le \kappa, \psi
,\omega\ri$ and $c_L \le \kappa, \psi ,\omega\ri$ represent the LCP
components, while  $b_R \le \kappa, \psi ,\omega\ri$ and $c_R \le
\kappa, \psi ,\omega\ri$ likewise represent the RCP components. The
unit vectors
\begin{equation}
\#s = - \#u_x \sin \psi + \#u_y \cos \psi
\end{equation}
and
\begin{equation}
\#p_\pm = \left\{ \begin{array}{l} \mp (\alphao/\ko)\le \#u_x \cos
\psi + \#u_y \sin \psi \ri  + (\kappa/\ko)\, \#u_z , \quad   z < -L
\vspace{6pt}\\
\mp (\alphal/\kl)\le \#u_x \cos \psi + \#u_y \sin \psi \ri  +
(\kappa/\kl)\, \#u_z ,\quad  z > L
\end{array}
\right.
\end{equation}
relate to the perpendicular-- and parallel--polarization states of
the plane wave, respectively.

The tangential components of $\#E (\#r, \kappa, \psi, \omega)$ and
$\#H (\#r, \kappa, \psi,  \omega)$ are required to be continuous
across the pupils at $z=-L$ and $z=L$ of the CSTF. Thus we have that
\begin{equation} \les \#f (-L, \kappa, \psi, \omega) \ris =  \frac{1}{\sqrt{2}} \les \, \#K (\kappa, \psi ,\omega, \alphao, \ko, \etao) \, \ris \les
\begin{array}{c}
0 \\
0 \\
-i \les b_L \le \kappa, \psi,\omega \ri  - b_R \le \kappa, \psi ,\omega\ri  \ris \\
b_L \le \kappa, \psi ,\omega\ri  + b_R \le \kappa, \psi ,\omega\ri
   \end{array} \ris
\l{f0_bc}
\end{equation}
and
\begin{equation}
 \les \#f (L, \kappa,
\psi, \omega) \ris = \frac{1}{\sqrt{2}} \les \, \#K (\kappa, \psi
,\omega, \alphal, \kl, \etal) \, \ris \les
\begin{array}{c}
i \les c_L \le \kappa, \psi,\omega \ri   - c_R \le \kappa, \psi ,\omega\ri   \ris \\
- \les c_L \le \kappa, \psi ,\omega\ri  + c_R  \le \kappa, \psi ,\omega\ri  \ris \\
0 \\ 0  \end{array} \ris \l{fL_bc}
\end{equation}
where
\begin{equation}
\les\#K(\kappa,\psi,\omega, \alpha, k, \eta)\ris =
\les\begin{array}{cccc}
-\sin\psi & -\frac{\alpha \cos\psi}{k} & -\sin\psi & \frac{\alpha \cos\psi}{k} \hspace{6pt} \\
\cos\psi & - \frac{\alpha \sin\psi}{k} & \cos\psi & \frac{\alpha \sin\psi}{k} \hspace{6pt} \\
- \frac{\alpha
\cos\psi}{ k \eta } & \frac{\sin\psi}{\eta } & \frac{\alpha \cos\psi}{ k \eta } & \frac{\sin\psi}{\eta }\hspace{6pt}\\
- \frac{\alpha \sin\psi}{ k \eta } &- \frac{\cos\psi}{  \eta } &
\frac{\alpha \sin\psi}{ k \eta } & - \frac{\cos\psi}{  \eta }
\end{array}\ris .
\label{defK}
\end{equation}
The unknown four amplitudes $b_{L,R} \le \kappa, \psi ,\omega\ri  $
and $c_{L,R} \le \kappa, \psi,\omega \ri $ can now be determined
using standard algebraic manipulations, by combining eqs. \r{f0_bc}
and \r{fL_bc} with the particular solution  \r{fL}.

\subsubsection{Emitted far--field phasors}

The emitted electromagnetic phasors in the half--spaces $z < -L$ and
$z>L$ are found by summing the corresponding spatial Fourier
components  \r{hhp} and \r{hpp} per
\begin{equation}
\left. \begin{array}{l} \displaystyle{ \#E (\#r, \omega) =
\frac{1}{4 \pi^2} \int_{0}^\infty\,d \kappa \int_{0}^{2\pi} \,d
\psi\; \,\kappa \,\#E (\#r,\kappa,\psi, \omega)} \vspace{8pt}\\
\displaystyle{ \#H (\#r, \omega) = \frac{1}{4 \pi^2}
\int_{0}^\infty\,d \kappa \int_{0}^{2\pi} \,d \psi\; \,\kappa \,\#H
(\#r,\kappa,\psi, \omega) } \end{array}\right\}, \qquad
z\notin[-L,L].
\end{equation}
Asymptotic approximations to these integrals, representing the
emitted field phasors in the far zone, are provided by \c{BW,Wang}
\begin{equation}
\left. \begin{array}{l} \displaystyle{ \#E (\#r\obs, \omega) \approx
\frac{i \cos \theta\obs}{ 2 \sqrt{2} \, \pi} \les -b\obs_L \le i
\#s\obs - \#p\obs_- \ri + b\obs_R \le i \#s\obs + \#p\obs_- \ri \ris
\, \frac{ \exp \le i
\ko \tilde{r}_-\obs \ri}{\ko  \tilde{r}_-\obs }} \vspace{8pt} \\
\displaystyle{\#H (\#r\obs,  \omega) \approx - \frac{\cos
\theta\obs}{\etao 2 \sqrt{2} \, \pi} \les b\obs_L \le i \#s\obs -
\#p\obs_- \ri + b\obs_R \le i \#s\obs + \#p\obs_- \ri \ris \, \frac{
\exp \le i \ko \tilde{r}_-\obs \ri}{\ko  \tilde{r}_-\obs } }
\end{array} \right\}, \qquad  z\obs< -L \l{ahhp}
\end{equation}
 and
\begin{equation}
\left. \begin{array}{l}
 \displaystyle{\#E (\#r\obs,  \omega) \approx \frac{i \cos
\theta\obs}{ 2 \sqrt{2} \, \pi} \les c\obs_L  \le i \#s\obs -
\#p\obs_+ \ri - c\obs_R \le i \#s\obs + \#p\obs_+ \ri \ris \,\frac{
\exp \le i \kl \tilde{r}_+\obs \ri}{\kl \tilde{r}_+\obs}
} \vspace{8pt}\\
\displaystyle{\#H (\#r\obs,   \omega) \approx  \frac{\cos
\theta\obs}{\etal 2 \sqrt{2} \, \pi} \les c\obs_L
 \le i \#s\obs - \#p\obs_+ \ri + c\obs_R  \le i
\#s\obs + \#p\obs_+ \ri \ris \,\frac{ \exp \le i \kl \tilde{r}_+\obs
\ri}{\kl \tilde{r}_+\obs}} \end{array} \right\}, \qquad z\obs > L
\l{ahpp},
\end{equation}
wherein $\tilde{r}_\pm\obs = \vert \#r\obs \mp L \#u_z \vert$ and
the superscript $obs$ denotes evaluation at the distant
 observation  point $\#r\obs\equiv\le
r\obs,\theta\obs,\psi\obs\ri$ with $z\obs = \#r\obs \cdot \#u_z$.
Note that the
 approximations \r{ahhp} and \r{ahpp} are appropriate at distances far
from the CSTF pupils but not in the vicinity of  $\theta\obs =
\pi/2$ \c{BW,Wang}.

For practical purposes,  the radiation field in the far zone is
conveniently characterized in terms of the  time--averaged Poynting
vector, which we write as the sum of LCP and RCP contributions per
\begin{equation}
\label{defP} \#P(\#r\obs, \omega) =  \#P_{LCP} (\#r\obs, \omega) +
\#P_{RCP} (\#r\obs, \omega),
\end{equation}
  wherein
\begin{equation}
\label{defPL} \#P_{LCP} (\#r\obs, \omega) \approx \left\{
\begin{array}{lr} \displaystyle{ \frac{1}{2 \etao} \le | b\obs_L
  |^2  \ri \le
\frac{\cos \theta\obs }{2 \pi \ko \tilde{r}_-\obs }\ri^2\,
\hat{\#r}\obs }, & z\obs <
-L \vspace{6pt} \\
 \displaystyle{\frac{1}{2 \etal} \le | c\obs_L
   |^2  \ri \le
\frac{\cos \theta\obs }{2 \pi \kl  \tilde{r}_+\obs } \ri^2}
 \, \hat{\#r}\obs, & z\obs > L
\end{array}
\right.
\end{equation}
and
\begin{equation}
\label{defPR} \#P_{RCP} (\#r\obs, \omega) \approx \left\{
\begin{array}{lr}\displaystyle{ \frac{1}{2 \etao} \le | b\obs_R
  |^2  \ri \le
\frac{\cos \theta\obs }{2 \pi \ko \tilde{r}_-\obs }\ri^2\,
\hat{\#r}\obs} , & z\obs <
-L \vspace{6pt} \\
\displaystyle{ \frac{1}{2 \etal} \le | c\obs_R
   |^2  \ri \le
\frac{\cos \theta\obs }{2 \pi \kl  \tilde{r}_+\obs } \ri^2
 \, \hat{\#r}\obs}, & z\obs > L
\end{array}
\right.
\end{equation}
at the  observation point $\#r\obs=r\obs\,\hat{\#r}\obs$.

\section{Numerical results} \l{Numerica}

For our numerical investigations we selected a structurally
right--handed CSTF (i.e., $h=+1$) with structural half--period
$\Omega = 200$ nm. Experimentally--determined  values for the
relative permittivity parameters $\lec \eps_{a1}, \eps_{b1},
\eps_{c1} \ric$, which characterize the uninfiltrated scenario, were
used. In light of the absence of appropriate  data for CSTFs, we
chose the relative permittivity parameters
\begin{equation}
\left.
\begin{array}{l}
\eps_{a1} = \displaystyle{\les 1.0443 + 2.7394 \le \frac{2
\chi_v}{\pi} \ri - 1.3697
\le \frac{2 \chi_v}{\pi} \ri^2 \ris^2} \vspace{6pt} \\
\eps_{b1} = \displaystyle{ \les 1.6765 + 1.5649 \le \frac{2
\chi_v}{\pi} \ri - 0.7825 \le \frac{2 \chi_v}{\pi} \ri^2 \ris^2}
 \vspace{6pt} \\
\eps_{c1} = \displaystyle{ \les 1.3586 + 2.1109 \le \frac{2
\chi_v}{\pi} \ri - 1.0554 \le \frac{2 \chi_v}{\pi} \ri^2 \ris^2}
\end{array}
\right\} \l{tio1}
\end{equation}
with
\begin{equation}
\tan \chi = 2.8818 \, \tan \chi_v, \l{tio2}
\end{equation}
which were determined by  measurements on a columnar thin film
 made from patinal${}^{\mbox{\textregistered}}$
titanium oxide \c{HWH_AO,Chiadini1}. The angle $\chi_v$ (radians) in
eqs. \r{tio1} represents  the average direction of the vapour flux
relative to the substrate during the deposition process. As
described in a previous study \c{ML_STF_JNP}, the inverse Bruggeman
homogenization formalism yields
 the corresponding nanoscale model parameter values: $n_s = 3.0517$,
 $f= 0.5039$ and $\gamma_b = 1.8381$ for the vapour flux
angle $\chi_v = 30^\circ$.

The relative permittivity parameters $\lec \eps_{a2}, \eps_{b2},
\eps_{c2} \ric$ for the  infiltrated CSTF, as computed using the
extended Bruggeman homogenization formalism, are graphed as
functions of the refractive index of the infiltrating fluid  $\nl
\in \le 1.0, 1.5 \ri$ and the relative size parameter $\ko \eta \in
\le 0, 0.2 \ri $ in Fig.~\ref{constitutive_params}. The real parts
of $\lec \eps_{a2}, \eps_{b2}, \eps_{c2} \ric$ increase uniformly as
$\nl$ increases; they also increase uniformly as $\eta$ increases
but more slowly. The imaginary parts of  $\lec \eps_{a2}, \eps_{b2},
\eps_{c2} \ric$ are fairly insensitive to $\nl$, but these
quantities increase exponentially as the size parameter increases
from zero. The manifestation of constitutive parameters with nonzero
imaginary parts for homogenized composite materials,  when the
component materials are themselves nondissipative, is a
well--recognized phenomenon. This phenomenon~---~which arises in
higher--order homogenization theories, such as the
strong--property--fluctuation theory \c{TK81,3rd_SPFT} and extended
variants of the Bruggeman and Maxwell Garnett formalisms
\c{Prinkey,SL93a,S96}~---~may be attributed to radiative scattering
loss from the macroscopic coherent field \c{Kranendonk}. In
particular, we note that $\mbox{Im} \, \eps_{a2,b2,c2} \to 0$ in the
limit as $\eta \to 0$.

Now let us turn to the radiation emitted from a dipole source
embedded within the CSTF. We chose the source to be located
relatively close to the upper surface of the CSTF at $d = L - 40$
nm; and the source orientation was given by $\#u_J = \=S_{\,z} (d,
h) \cdot \#u_n$\footnote{We also considered the source orientations
$\#u_J = \underline{\underline{S}}_{\,z} (d, h) \cdot \#u_\tau$ and
$\#u_J = \underline{\underline{S}}_{\,z} (d, h) \cdot \#u_b$. The
results for these cases are not presented here as they were
qualitatively similar to the case $\#u_J =
\underline{\underline{S}}_{\,z} (d, h) \cdot \#u_n$.}. From the
point of view of sensor applications, we are particularly interested
in the ability of the CSTF to discriminate between LCP and RCP
light. Accordingly, the wavelength ranges considered here were
selected to include the circular Bragg regime. The location and
extent of the the circular Bragg regime were conveniently  estimated
by the centre wavelength \c{STF_Book}
\begin{equation} \l{Bragg}
\lambdao^{Br} (   \eps_{a\nu}, \eps_{b\nu}, \eps_{c\nu}, \theta\obs
) \simeq \Omega \le \sqrt{\left| \eps_{c\nu} \right|} + \sqrt{
\left| \frac{\eps_{a\nu} \eps_{b\nu}}{\eps_{a\nu} \cos^2 \chi +
\eps_{a\nu} \sin^2 \chi} \right| }\ri \sqrt{\cos \theta\obs},
\end{equation}
and  the full--width--at--half--maximum bandwidth  \c{STF_Book}
\begin{equation} \l{Bragg_width}
\le \Delta \lambdao \ri^{Br}  (   \eps_{a\nu}, \eps_{b\nu},
\eps_{c\nu}, \theta\obs )\simeq  2 \Omega \le \sqrt{\left|
\eps_{c\nu} \right|} - \sqrt{ \left| \frac{\eps_{a\nu}
\eps_{b\nu}}{\eps_{a\nu} \cos^2 \chi + \eps_{a\nu} \sin^2 \chi}
\right| }\ri \sqrt{\cos \theta\obs}.
\end{equation}
 The thickness ratio $L / \Omega = 30$ was chosen to ensure that the circular Bragg phenomenon is fully developed.

\begin{figure}[!ht]
\centering
\subfigure[]{\includegraphics[width=2in]{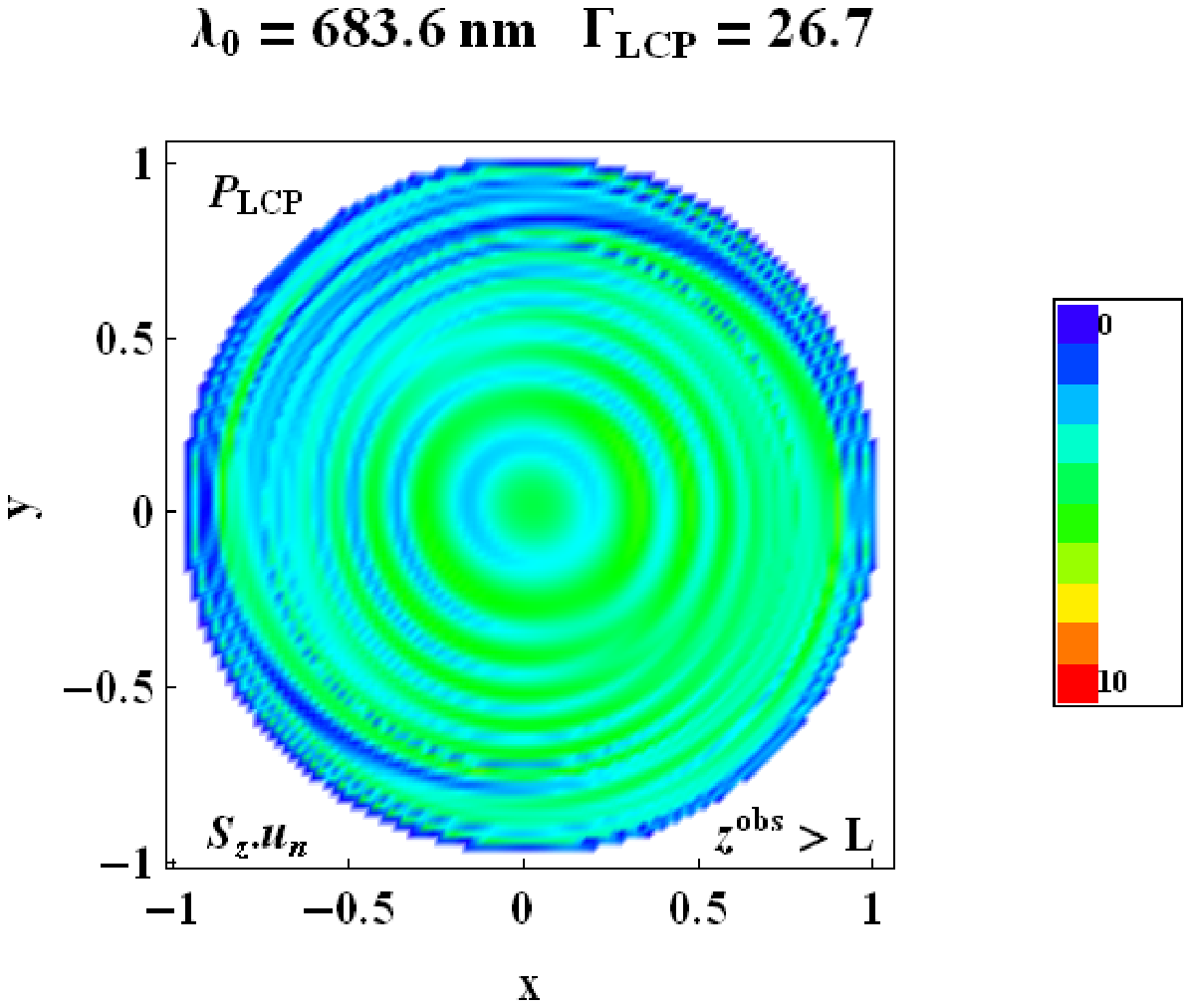}}
\subfigure[]{\includegraphics[width=2in]{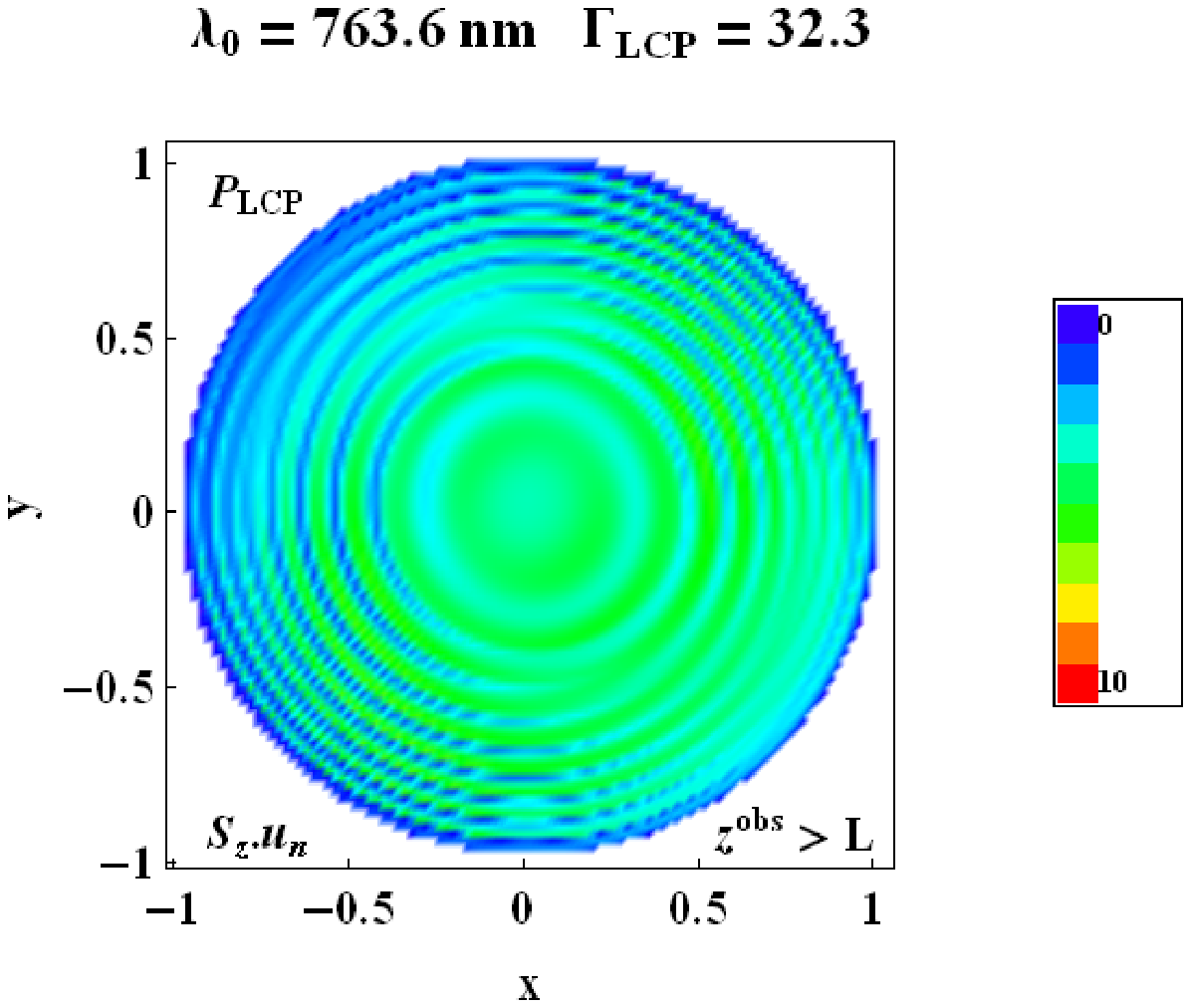}}
\subfigure[]{\includegraphics[width=2in]{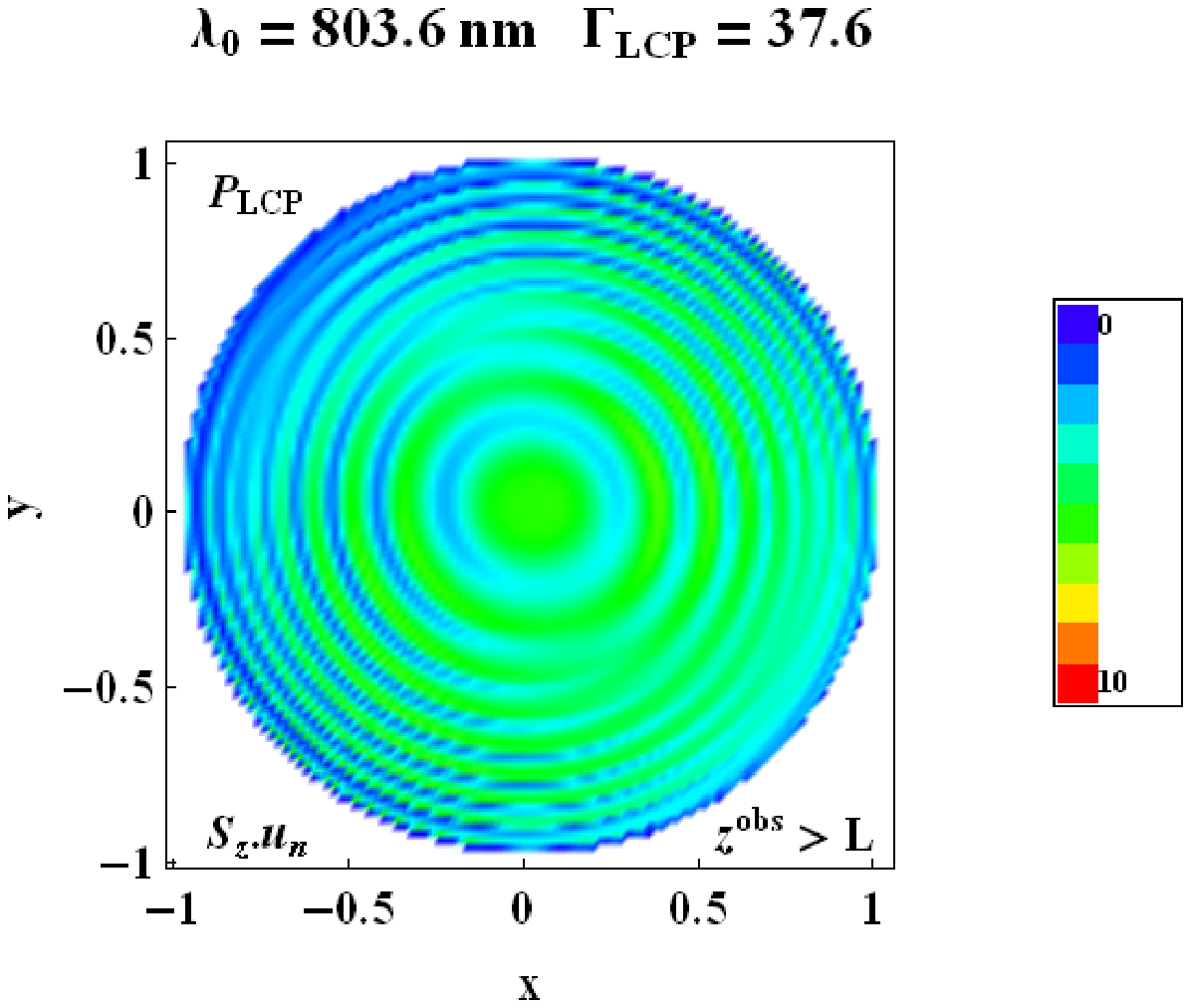}}\\
\subfigure[]{\includegraphics[width=2in]{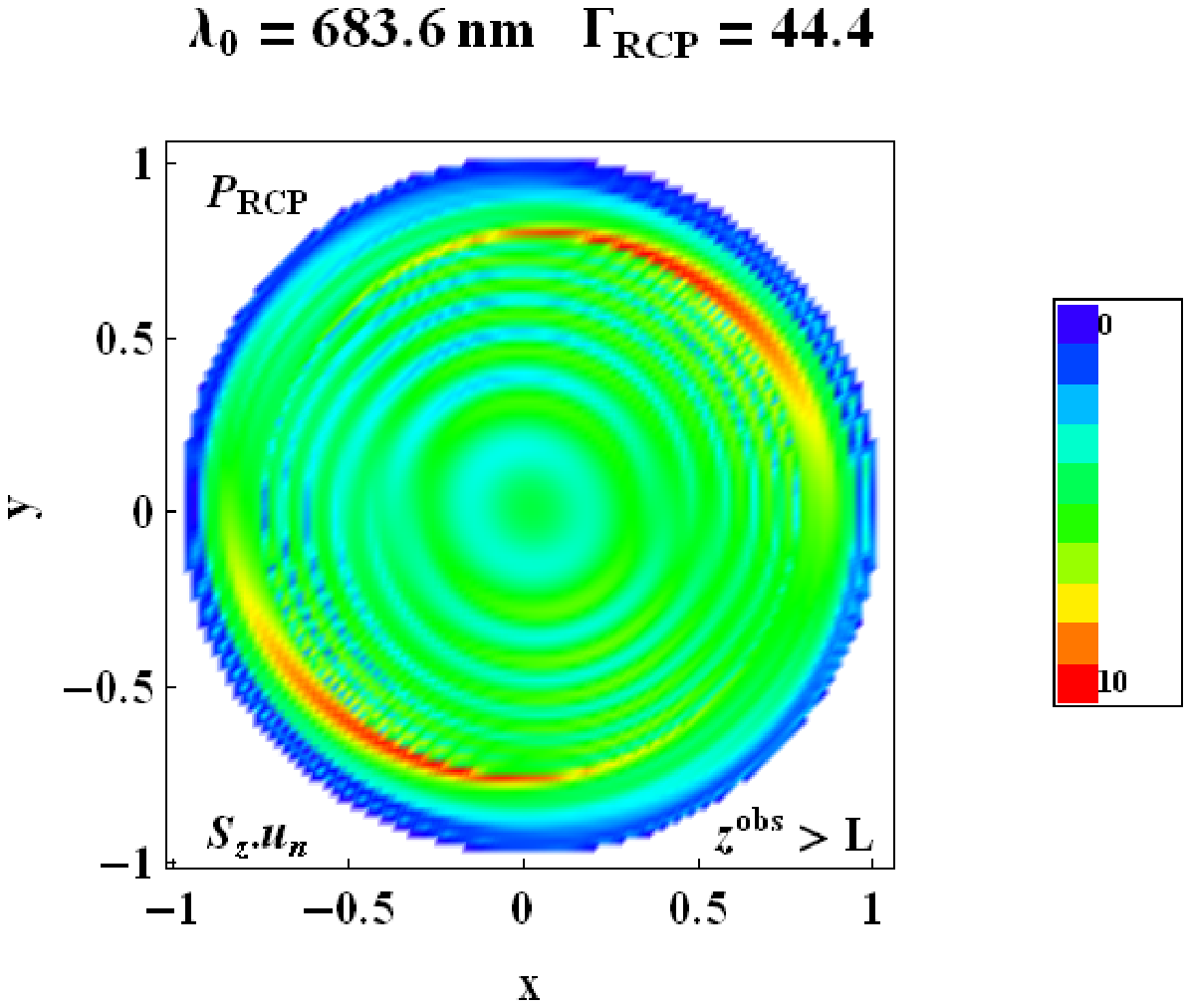}}
\subfigure[]{\includegraphics[width=2in]{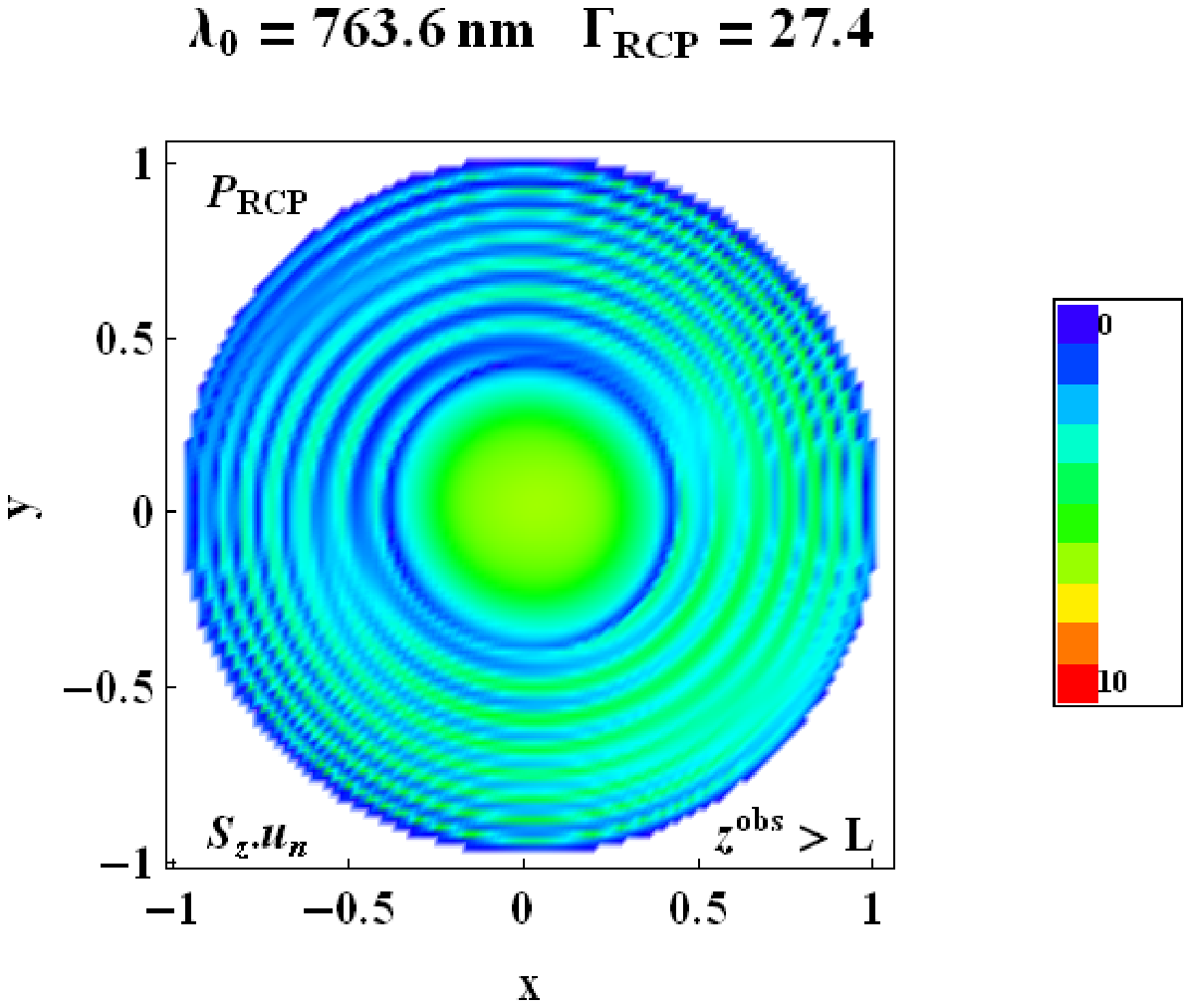}}
\subfigure[]{\includegraphics[width=2in]{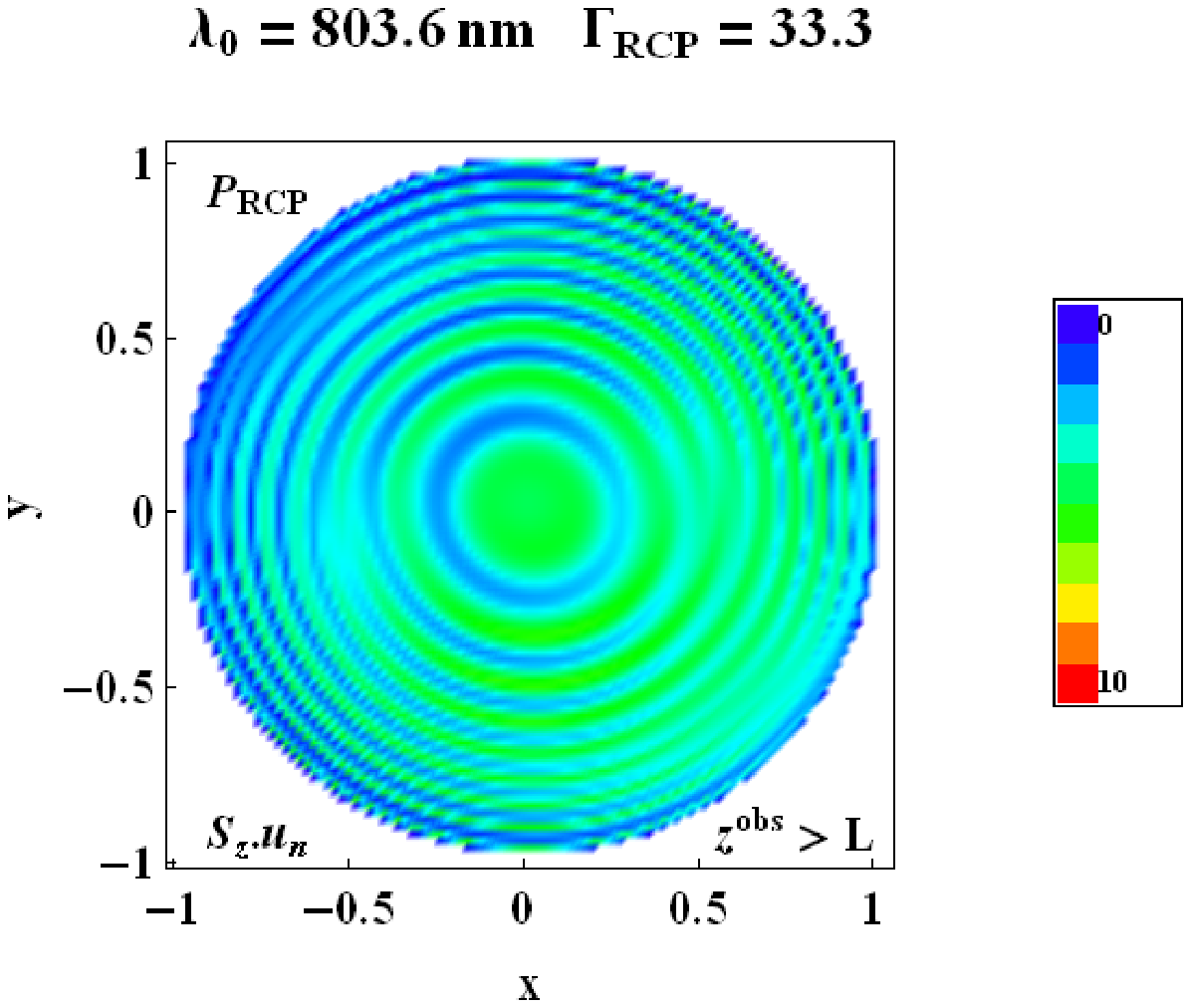}}\\
\subfigure[]{\includegraphics[width=2in]{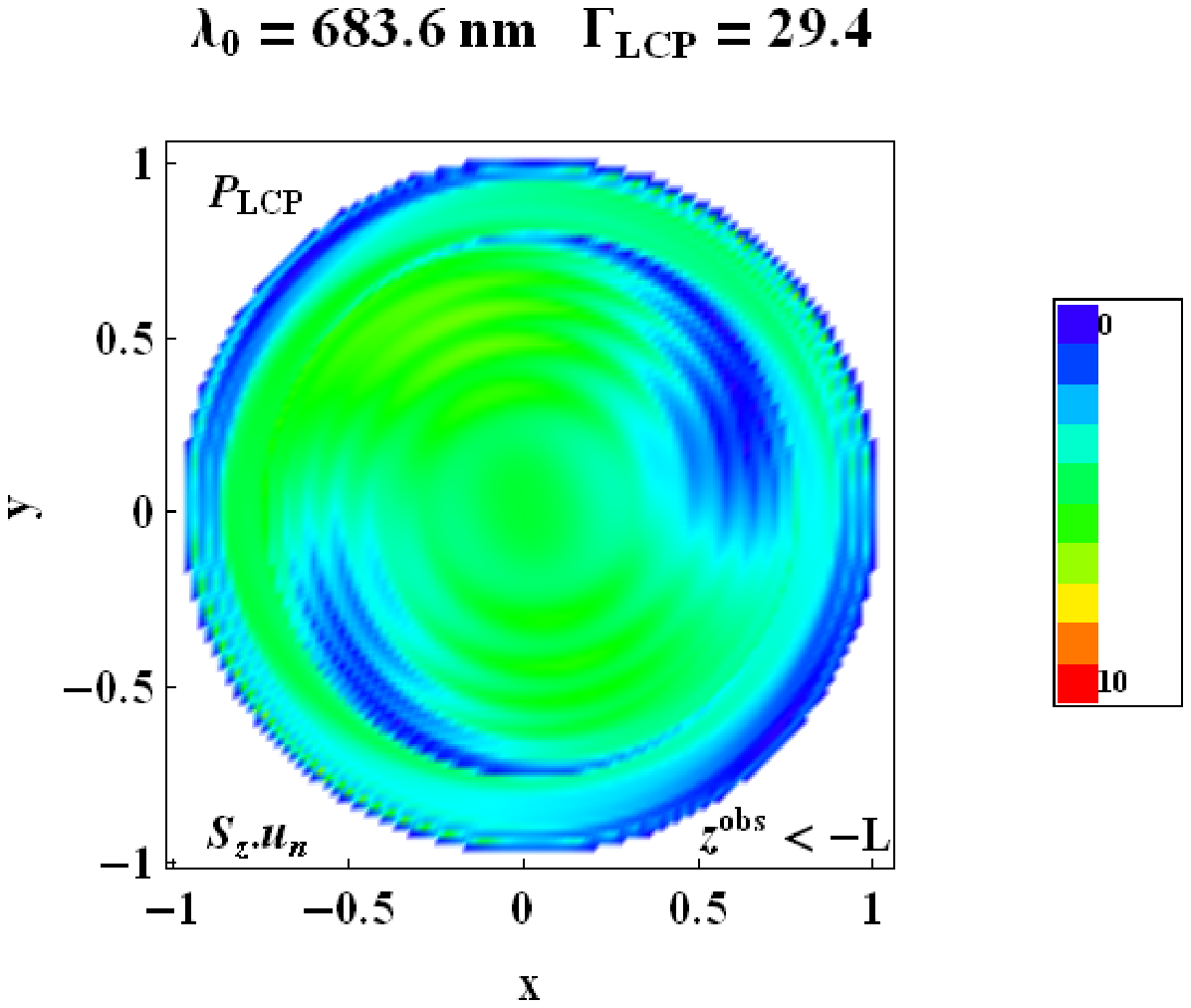}}
\subfigure[]{\includegraphics[width=2in]{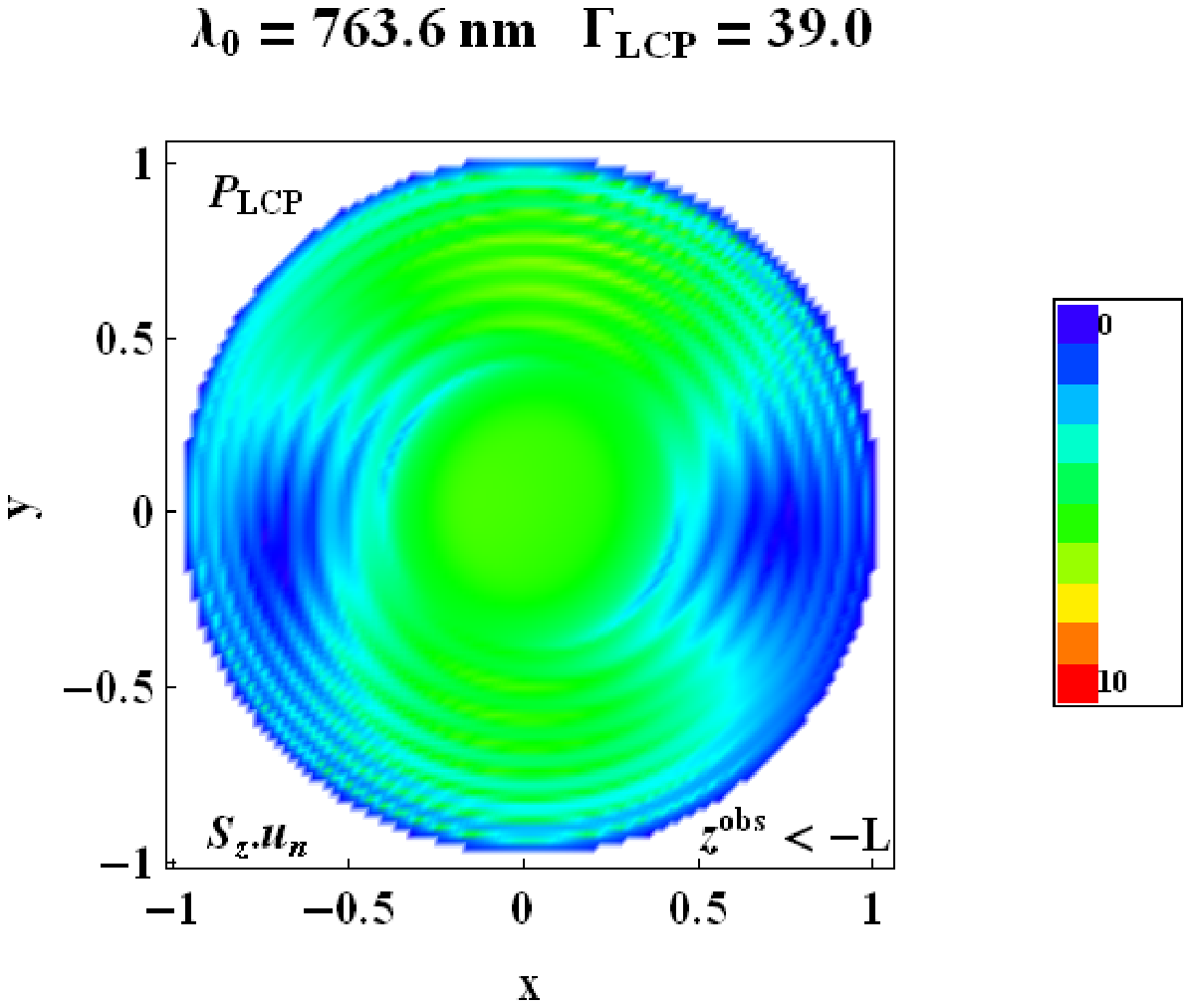}}
\subfigure[]{\includegraphics[width=2in]{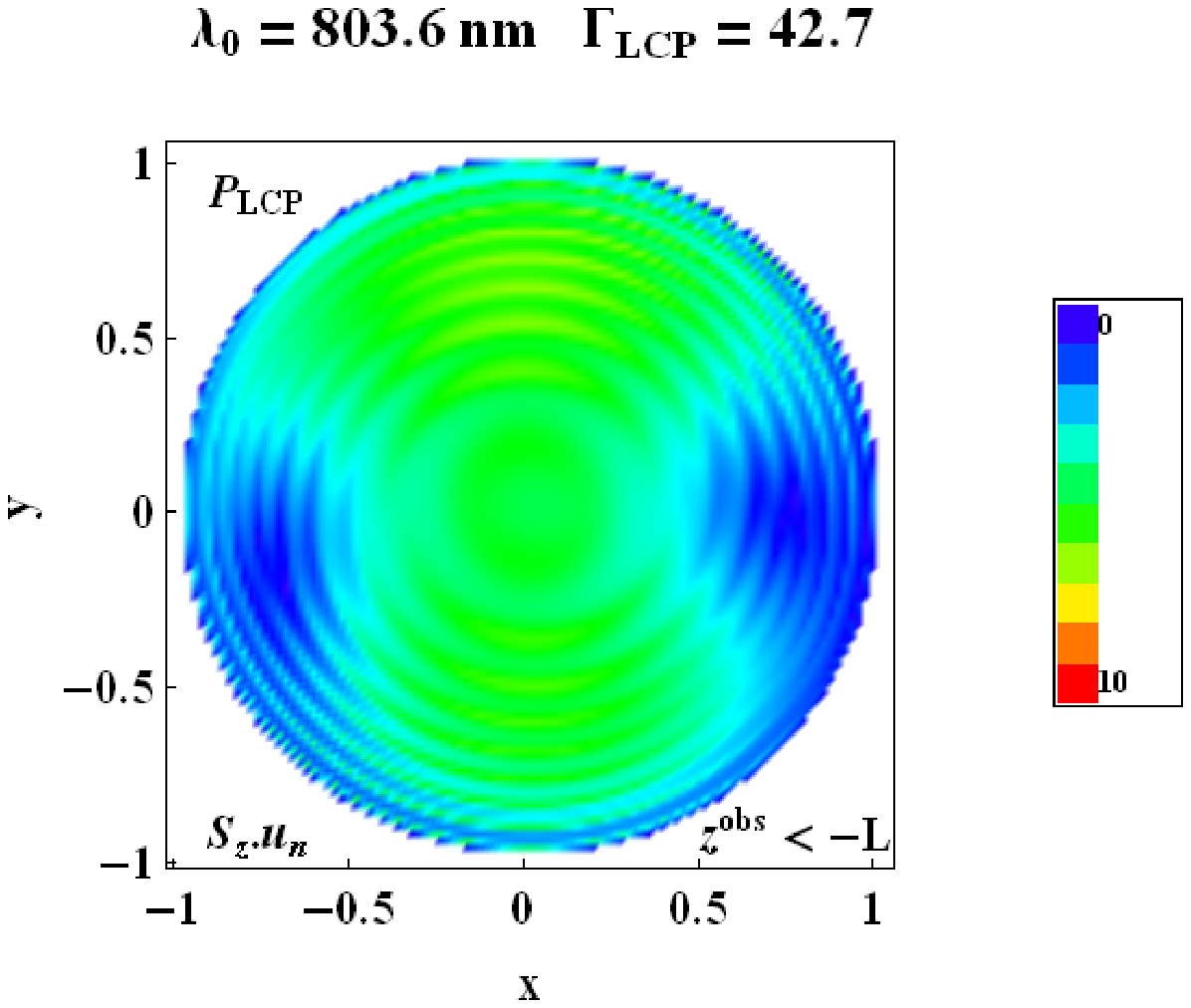}}\\
\subfigure[]{\includegraphics[width=2in]{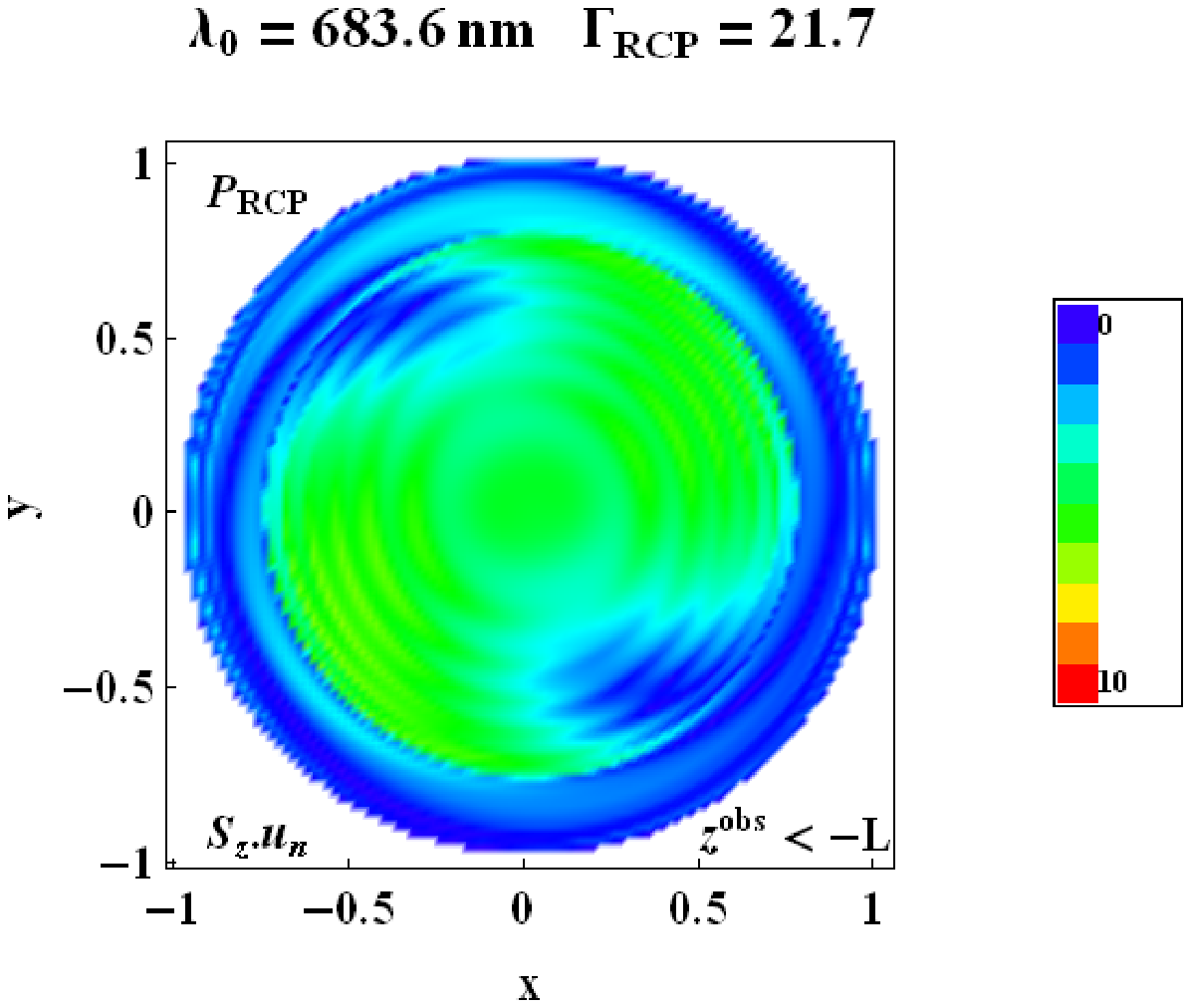}}
\subfigure[]{\includegraphics[width=2in]{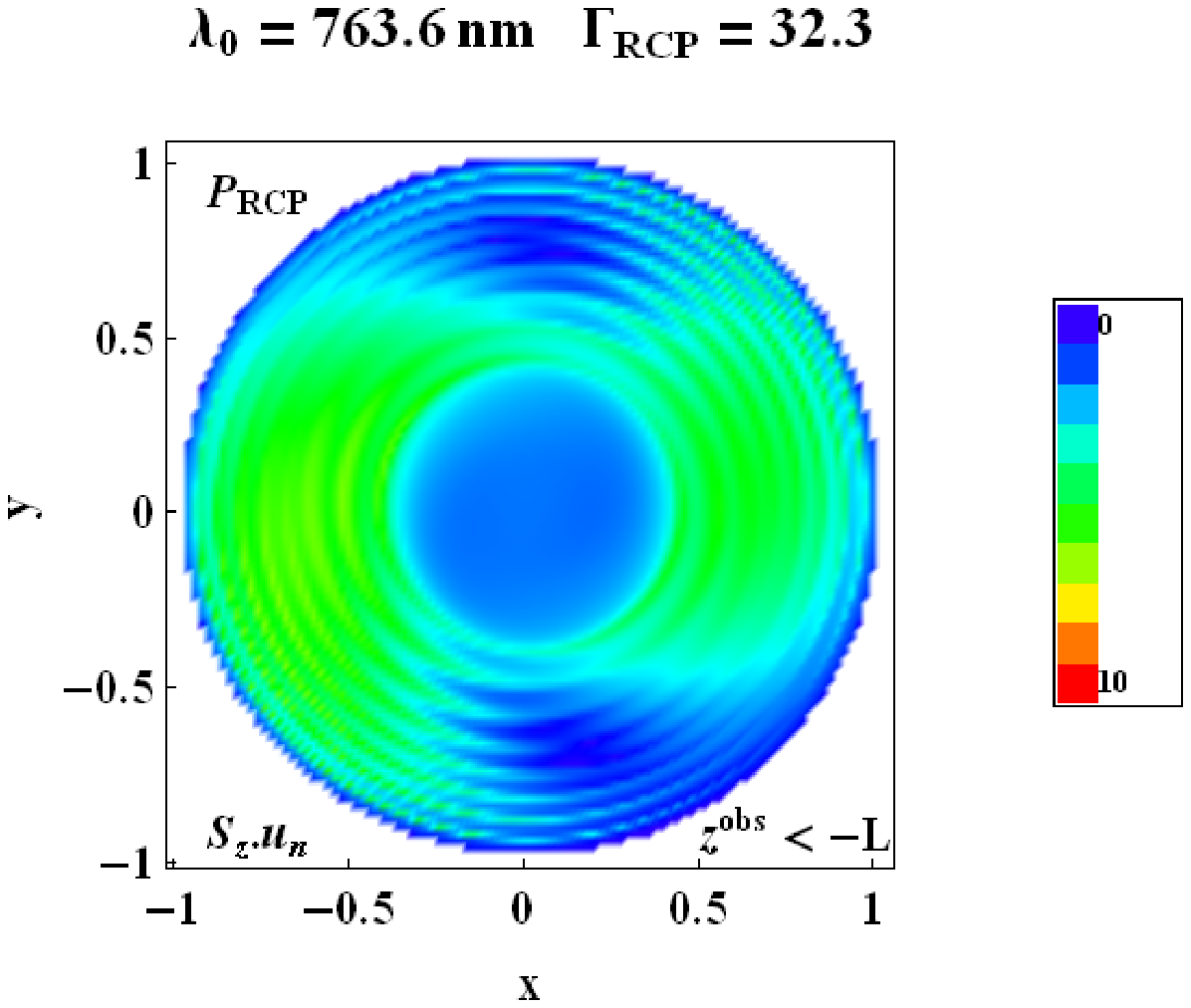}}
\subfigure[]{\includegraphics[width=2in]{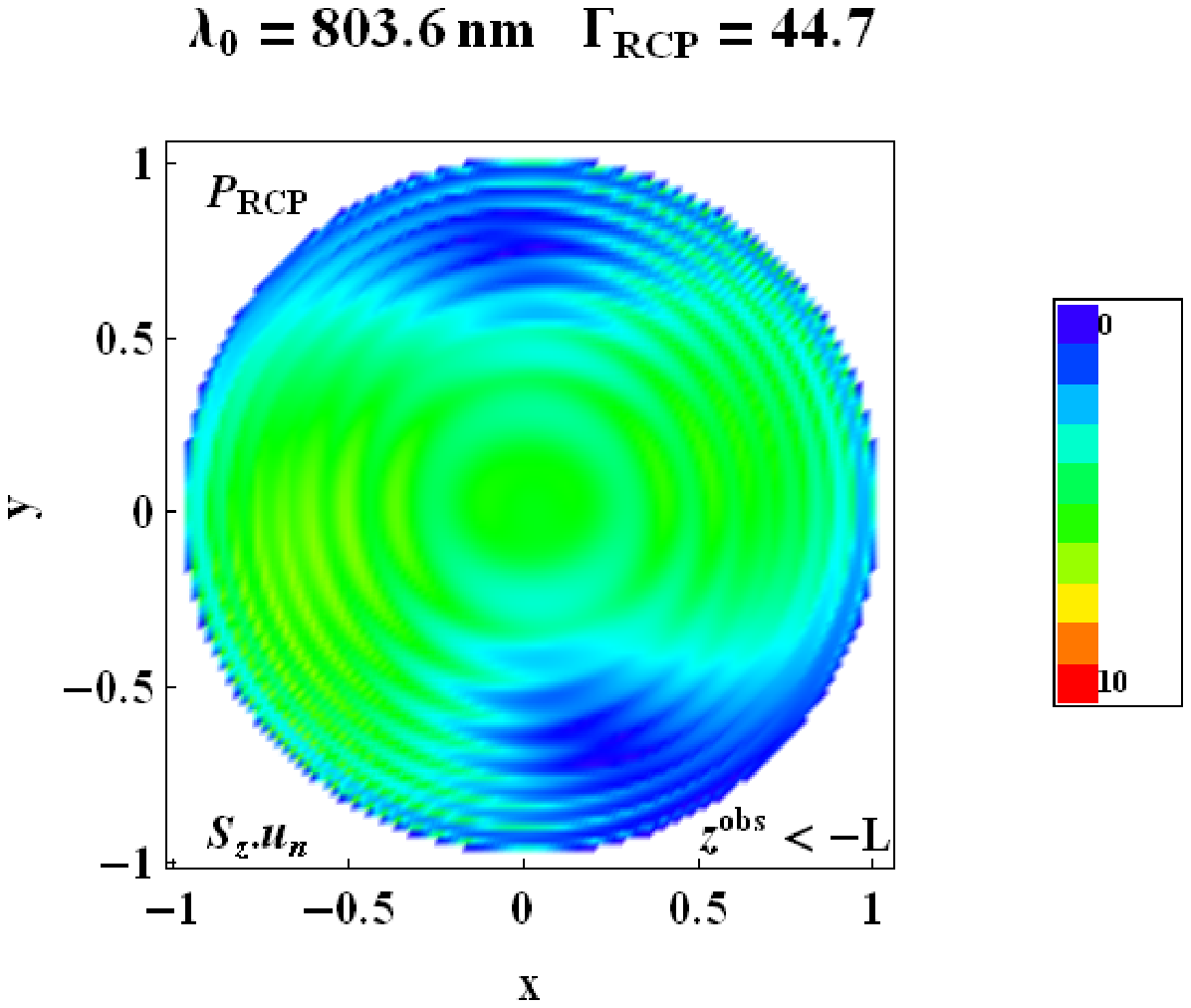}}
\caption{Projections of $\left| \#P_{LCP} \right|$ and $\left|
\#P_{RCP} \right|$ (scaled by $10^{13} \omega^{-2} \left| p
\right|^{-2} $) onto the $z=0$ plane for $z\obs
> L$ and $z\obs < -L$. Here
$\lambdao \in \lec 683.6 \mbox{nm}, 763.6 \mbox{nm}, 803.6
\mbox{nm}\ric$ and $\nl = 1.0$.} \label{fn1.0}
\end{figure}

In order to appreciate the effects of infiltration, we must first
consider the uninfiltrated scenario. In Fig.~\ref{fn1.0}, the
projections of $\left| \#P_{LCP} ( \#r\obs, \omega) \right|$ and
$\left| \#P_{RCP} (\#r\obs, \omega ) \right|$ onto the $z=0$ plane
are mapped for $\nl = 1$ (i.e., $\nu = 1$). The contributions to the
time--averaged Poynting vector were scaled by a factor of $10^{13}
\omega^{-2} \left| p \right|^{-2} $.
 The radiation emitted through the CSTF's pupils at $z = L$ and $z
= -L$ are both represented. Results are presented for $\lambdao =
\lambdao^{Br} (   \eps_{a1}, \eps_{b1}, \eps_{c1}, 0^\circ ) = 763.6
$ nm, as well as for $\lambdao =  \lambdao^{Br} ( \eps_{a1},
\eps_{b1}, \eps_{c1}, 0^\circ ) - 80 $ nm and $\lambdao =
\lambdao^{Br} ( \eps_{a1}, \eps_{b1}, \eps_{c1}, 0^\circ ) + 40 $
nm. Also presented in Fig.~\ref{fn1.0} are evaluations of the
real--valued parameter
\begin{equation} \Gamma_j = \frac{10^{16}}{\omega^2 \left| p \right|^2} \, \int^{\rho_2}_{\theta\obs = \rho_1}
d \theta\obs \int^{2 \pi}_{\psi\obs = 0} d \psi\obs \; \le r\obs
\ri^2 \sin \theta\obs \, \left|  \#P_{j}(\#r\obs, \omega ) \right|
\,, \quad j\in\lec LCP, RCP\ric\,, \l{gam_def}
\end{equation}
wherein $\rho_1 = 0$, $\rho_2 = 0.95 \pi/2$ for $z\obs > L$, and
$\rho_1 = \pi - \le 0.95 \pi/2 \ri $, $\rho_2 =  \pi$ for $z\obs <
-L$. The quantity $\Gamma_j$ delivers a measure of the total rate of
energy flow into the half--spaces $z > L$ and $z < -L$.

There are several notable features in Fig.~\ref{fn1.0}, especially
concerning  differences between the LCP and RCP emission
characteristics,
 which relate to the
 circular Bragg phenomenon: (i) For emission which is approximately
   normal to the two pupils of the CSTF, RCP radiation is
   preferentially emitted through the pupil at $z = L$ for  $\lambdao = 763.6 $ nm
   whereas at the same wavelength
LCP radiation is
   preferentially emitted through the pupil at $z = -L$.
(ii) At $\lambdao = 683.6 $ nm, the differences between LCP and RCP
emission are very small
 for $\vert \cos\theta\obs\vert \simeq 1$, but
we can see
 that
 RCP radiation is
   preferentially emitted through the upper pupil
   whereas
LCP radiation is
   preferentially emitted through the lower pupil
for $0.4 \lesssim  \vert\cos\theta^{obs}\vert \lesssim 0.8$. This
observation is  in accordance with the blue shift of the circular
Bragg phenomenon for obliquely incident plane waves \c{STF_Book}.
(iii)  The distinction between the LCP and RCP patterns is barely
noticeable when $\vert\cos\theta_{obs}\vert \lesssim 0.4$, for all
three wavelengths considered. This is indicative of the severe
diminishment of the circular Bragg phenomenon for highly oblique
planewave incidence \c{PLP}. (iv) There is very little evidence of
the CSTF discriminating between LCP and RCP radiation at $\lambdao =
803.6$ nm;  this wavelength represents an upper bound on the
circular Bragg phenomenon for all angles of incidence.

Let us remark too upon the distinctive pattern of concentric rings
that appears in Fig.~\ref{fn1.0}. These are Fabry--Perot
interference rings, arising due to the finite thickness of the CSTF.
Indeed, even if the CSTF were replaced by a homogeneous  isotropic
dielectric material the ring pattern would still be observed
\c{ML_OE}.

\begin{figure}[!ht]
\centering
\subfigure[]{\includegraphics[width=2in]{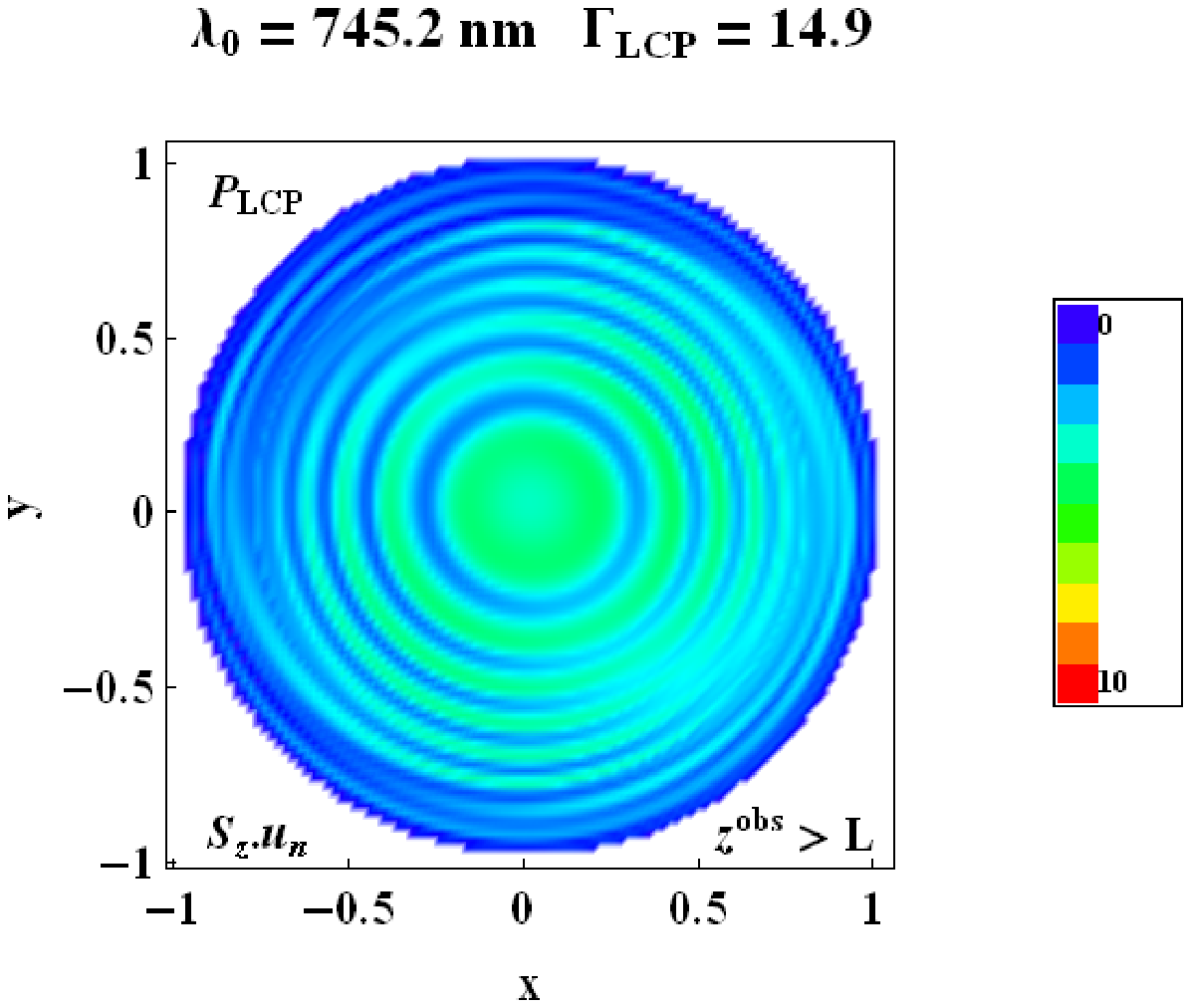}}
\subfigure[]{\includegraphics[width=2in]{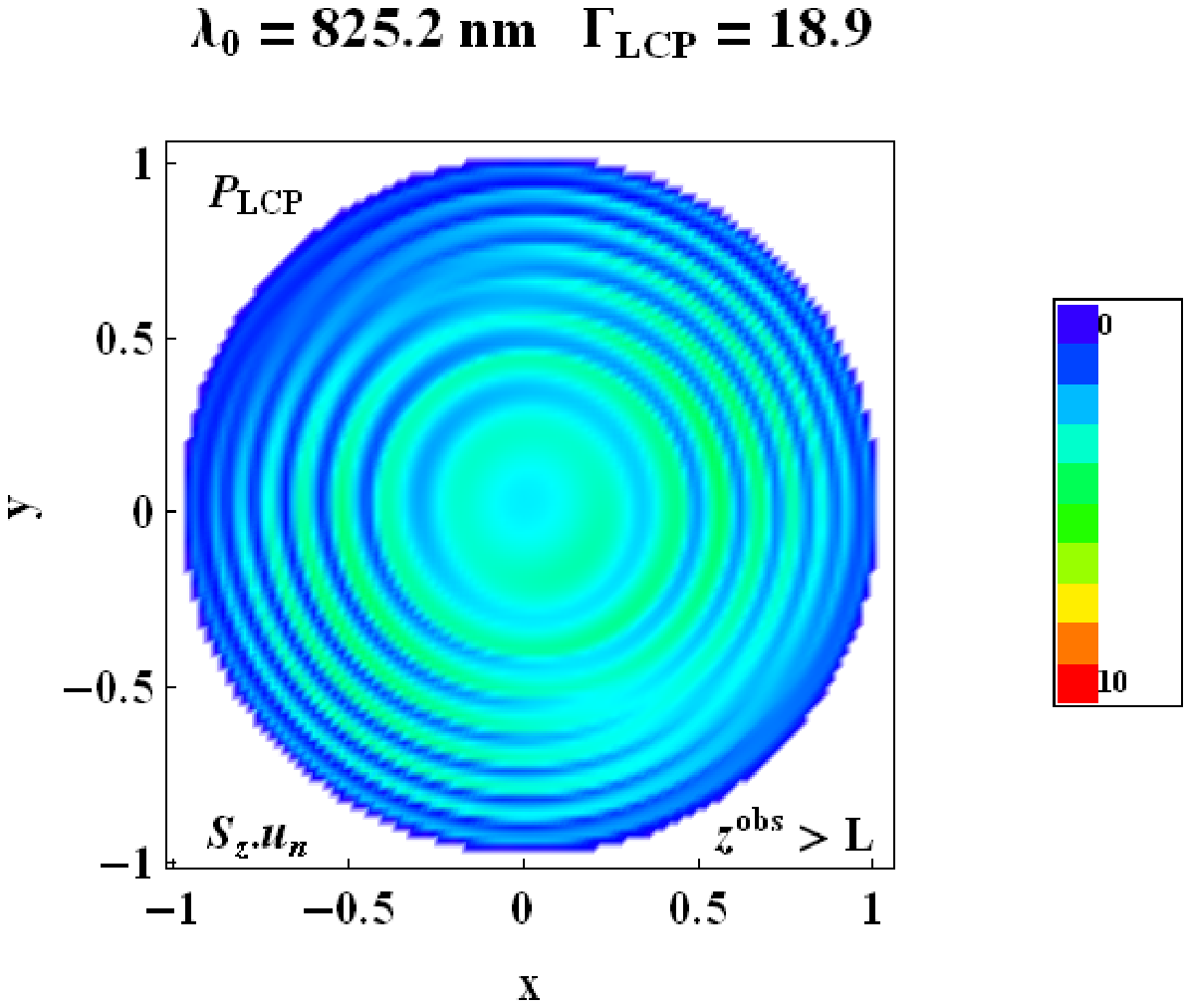}}
\subfigure[]{\includegraphics[width=2in]{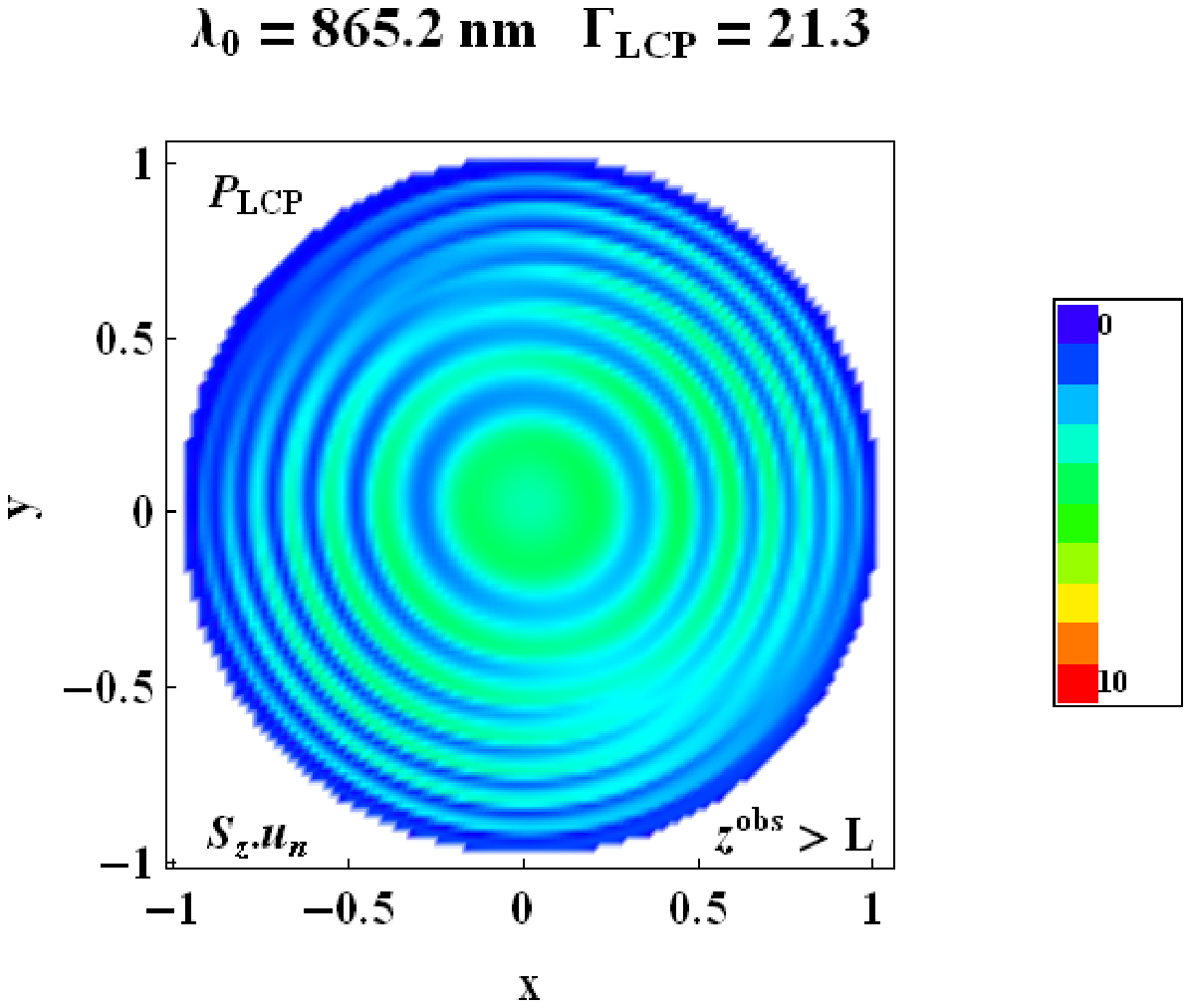}}\\
\subfigure[]{\includegraphics[width=2in]{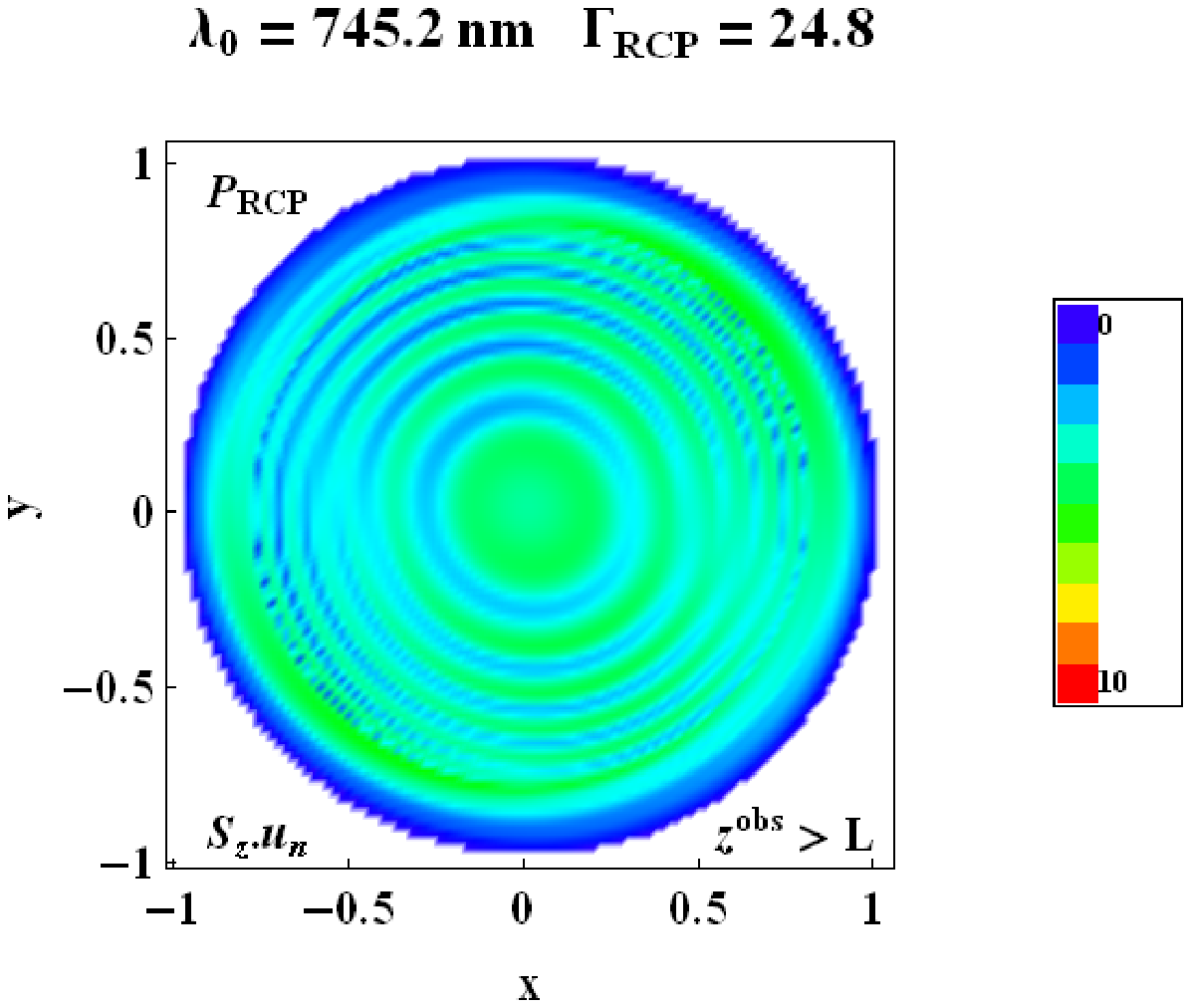}}
\subfigure[]{\includegraphics[width=2in]{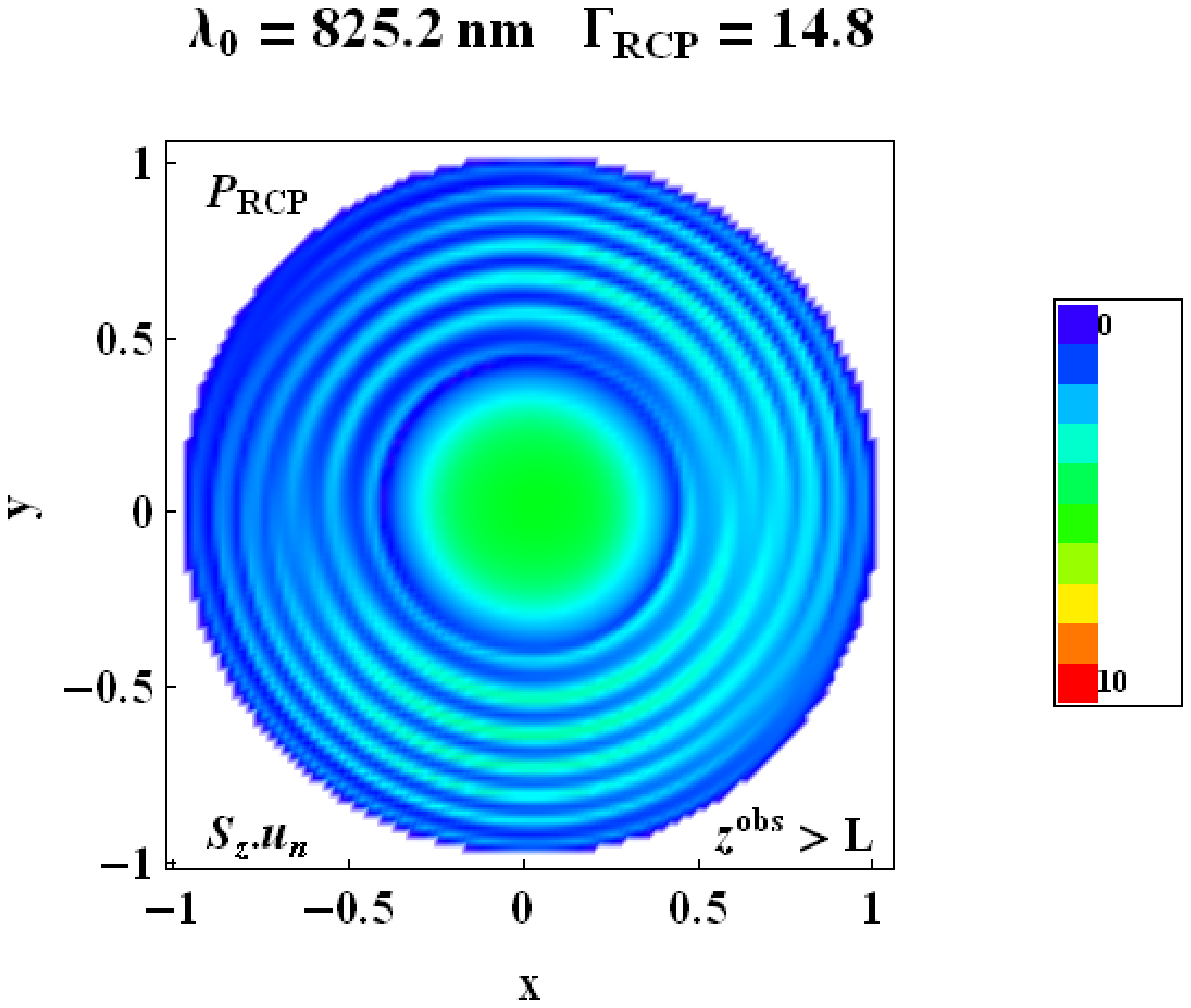}}
\subfigure[]{\includegraphics[width=2in]{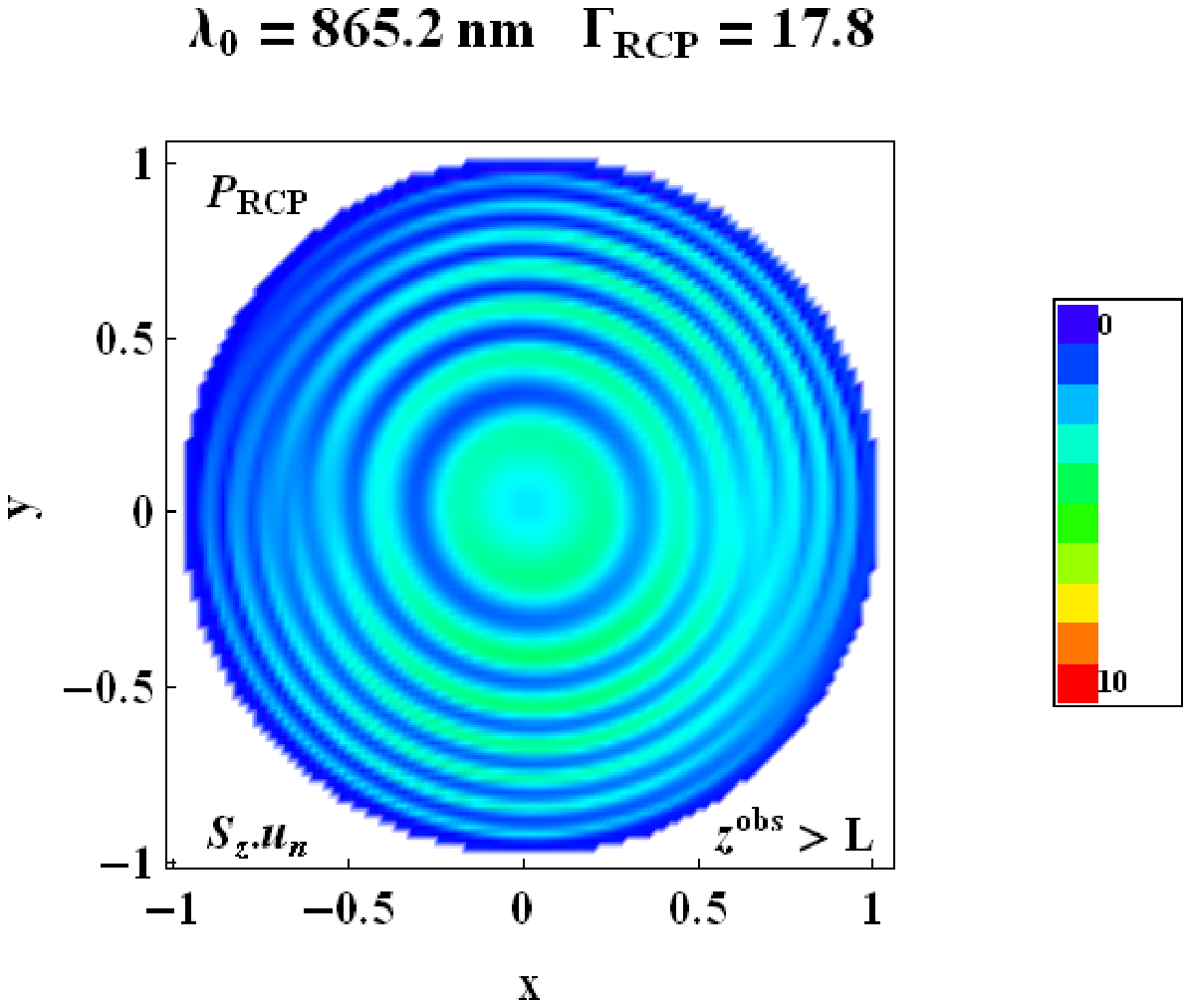}}\\
\subfigure[]{\includegraphics[width=2in]{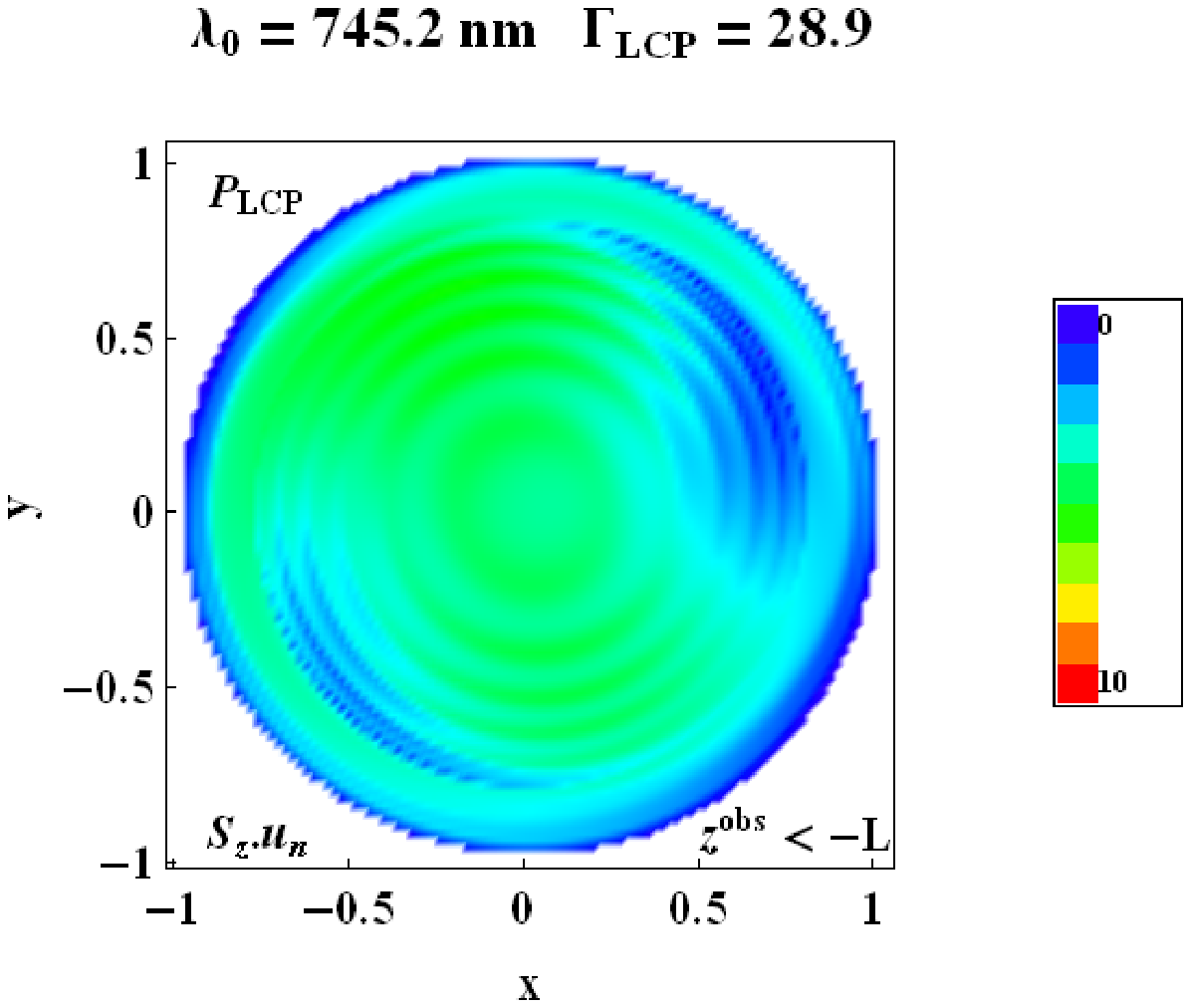}}
\subfigure[]{\includegraphics[width=2in]{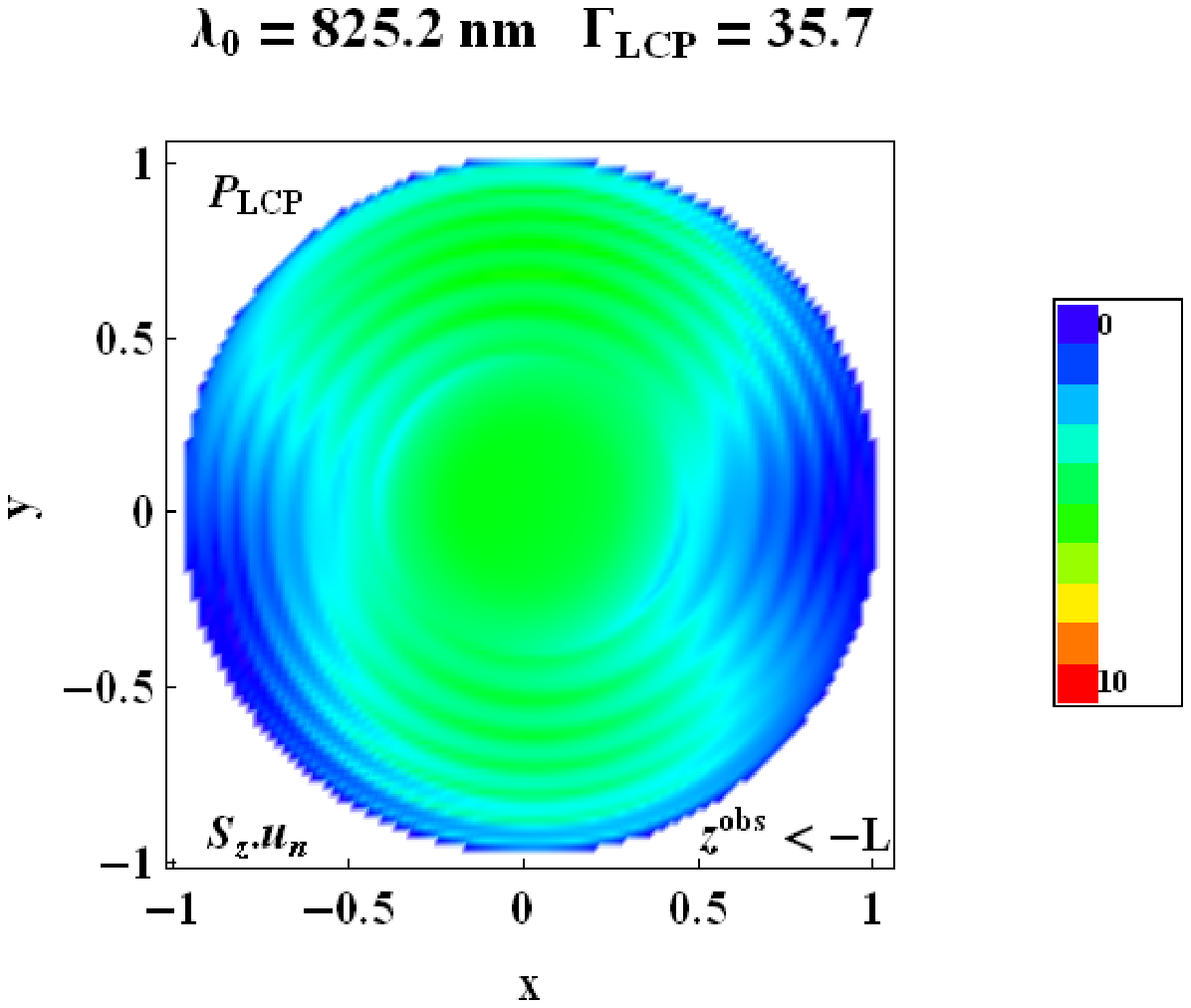}}
\subfigure[]{\includegraphics[width=2in]{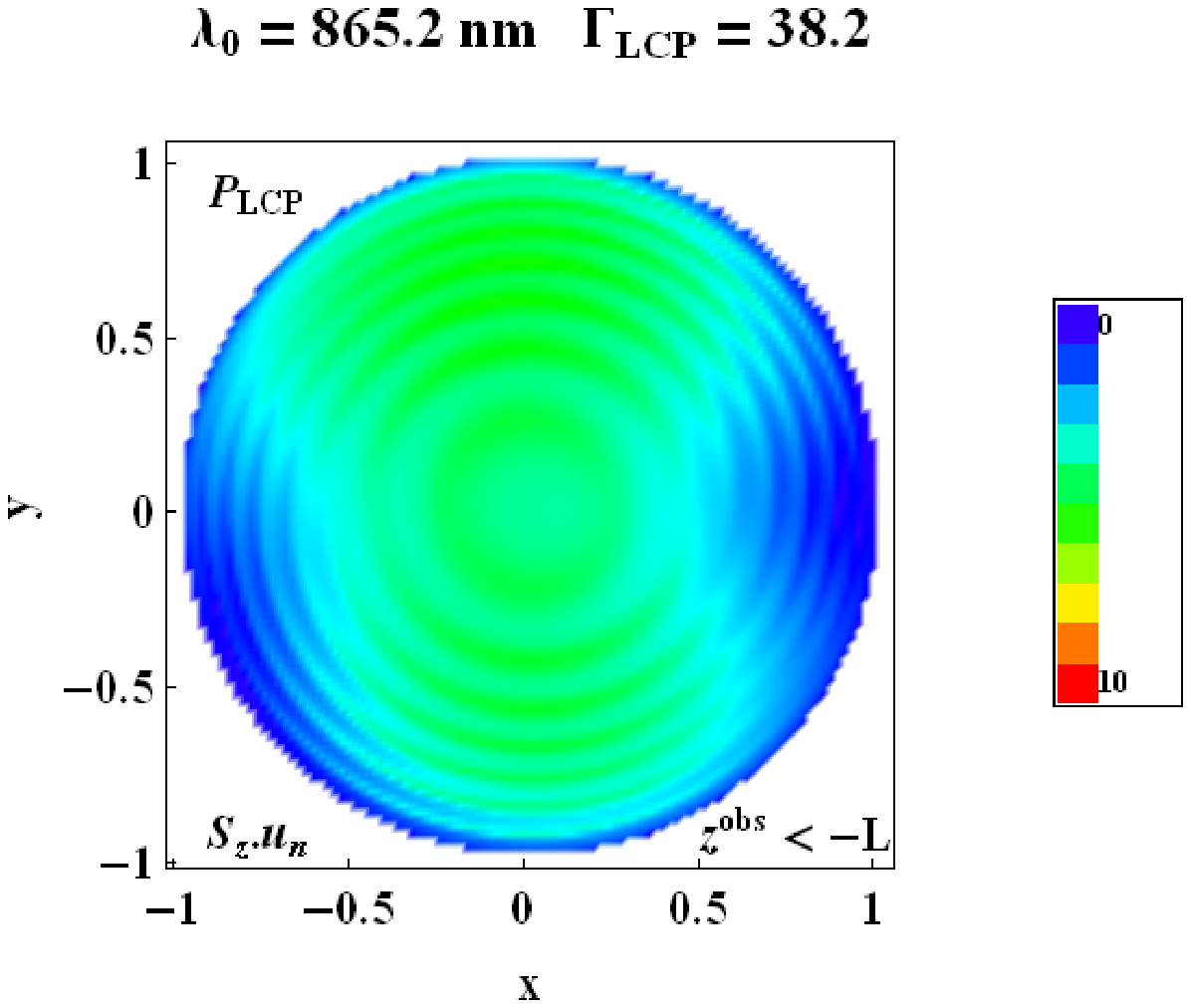}}\\
\subfigure[]{\includegraphics[width=2in]{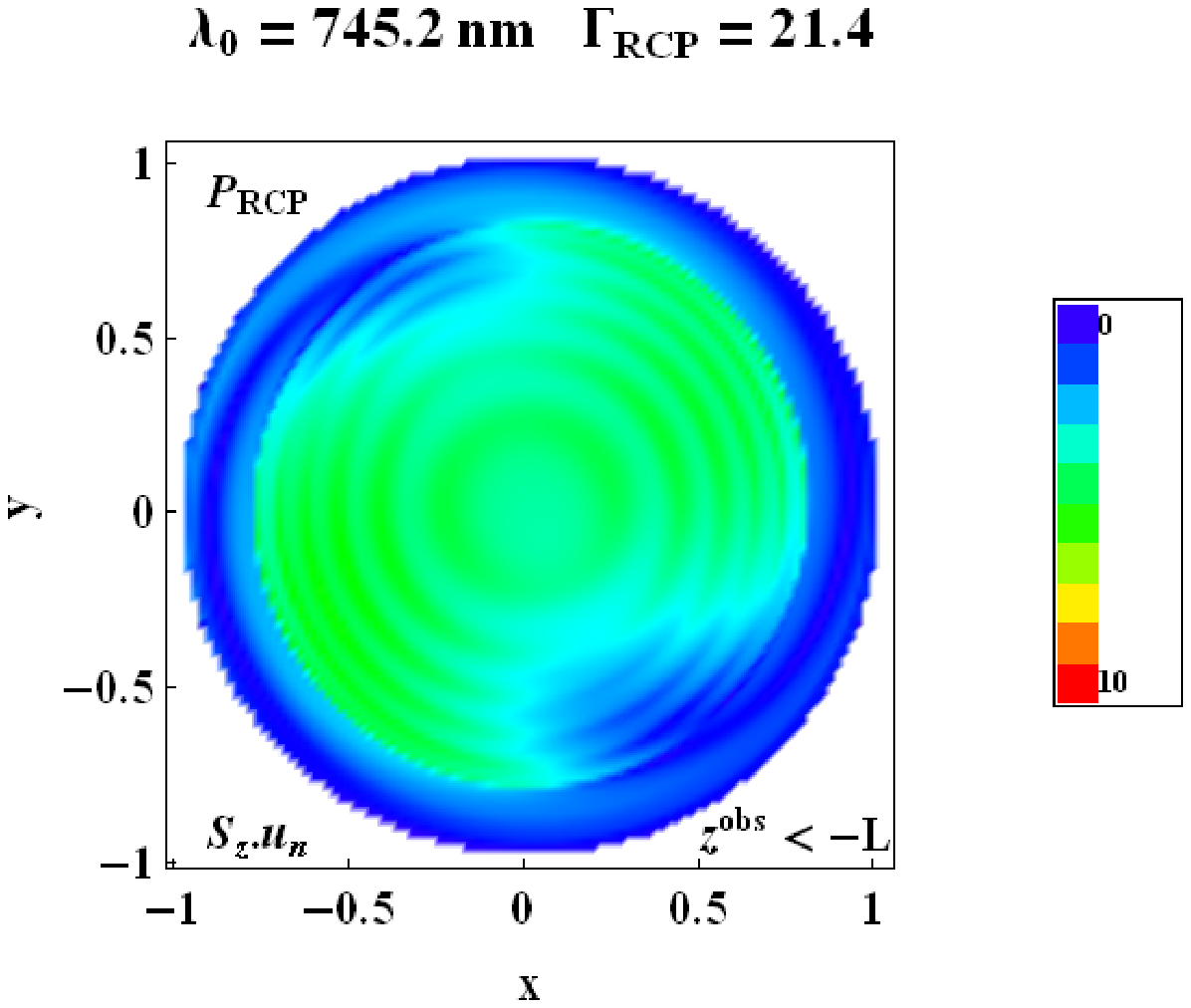}}
\subfigure[]{\includegraphics[width=2in]{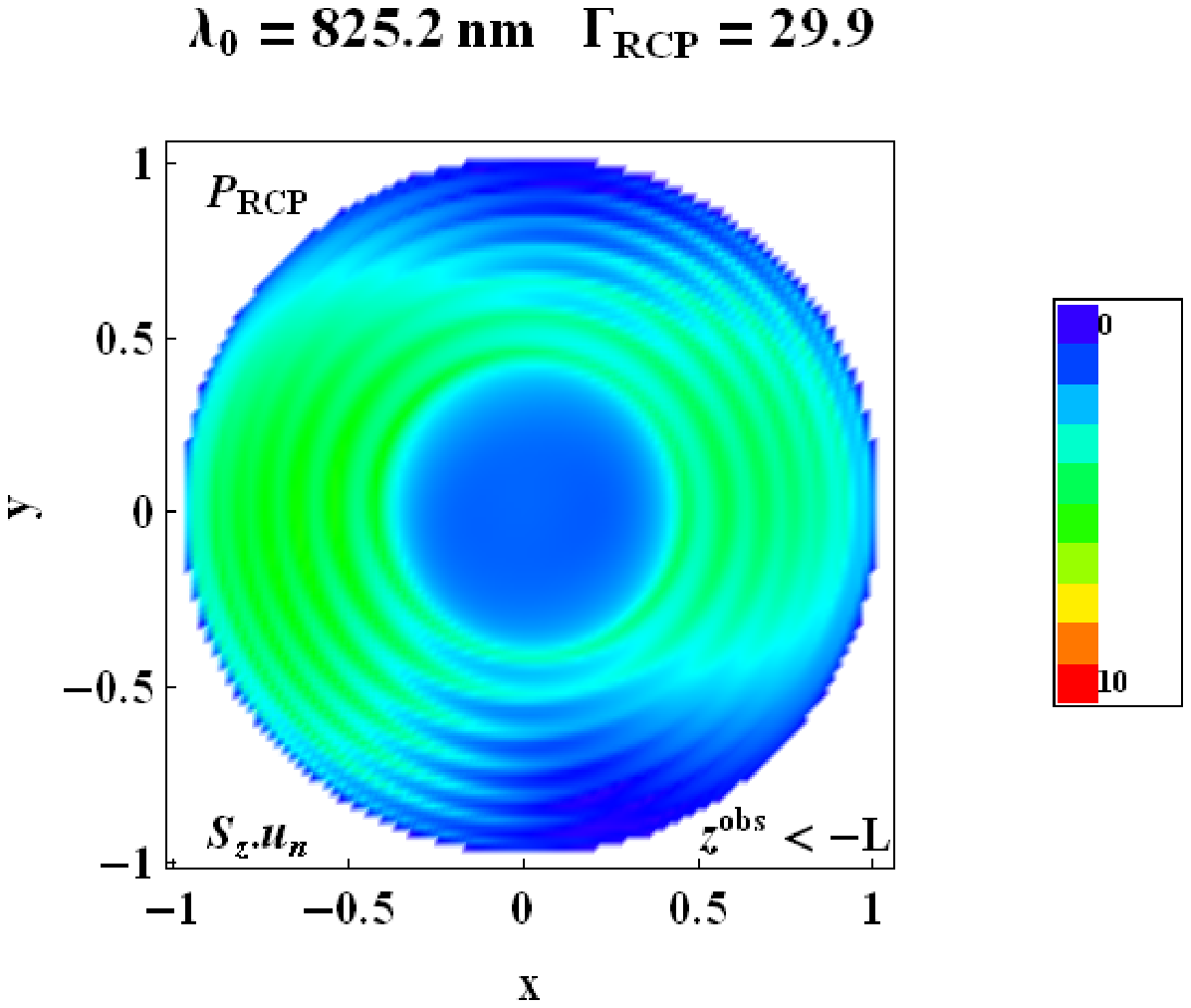}}
\subfigure[]{\includegraphics[width=2in]{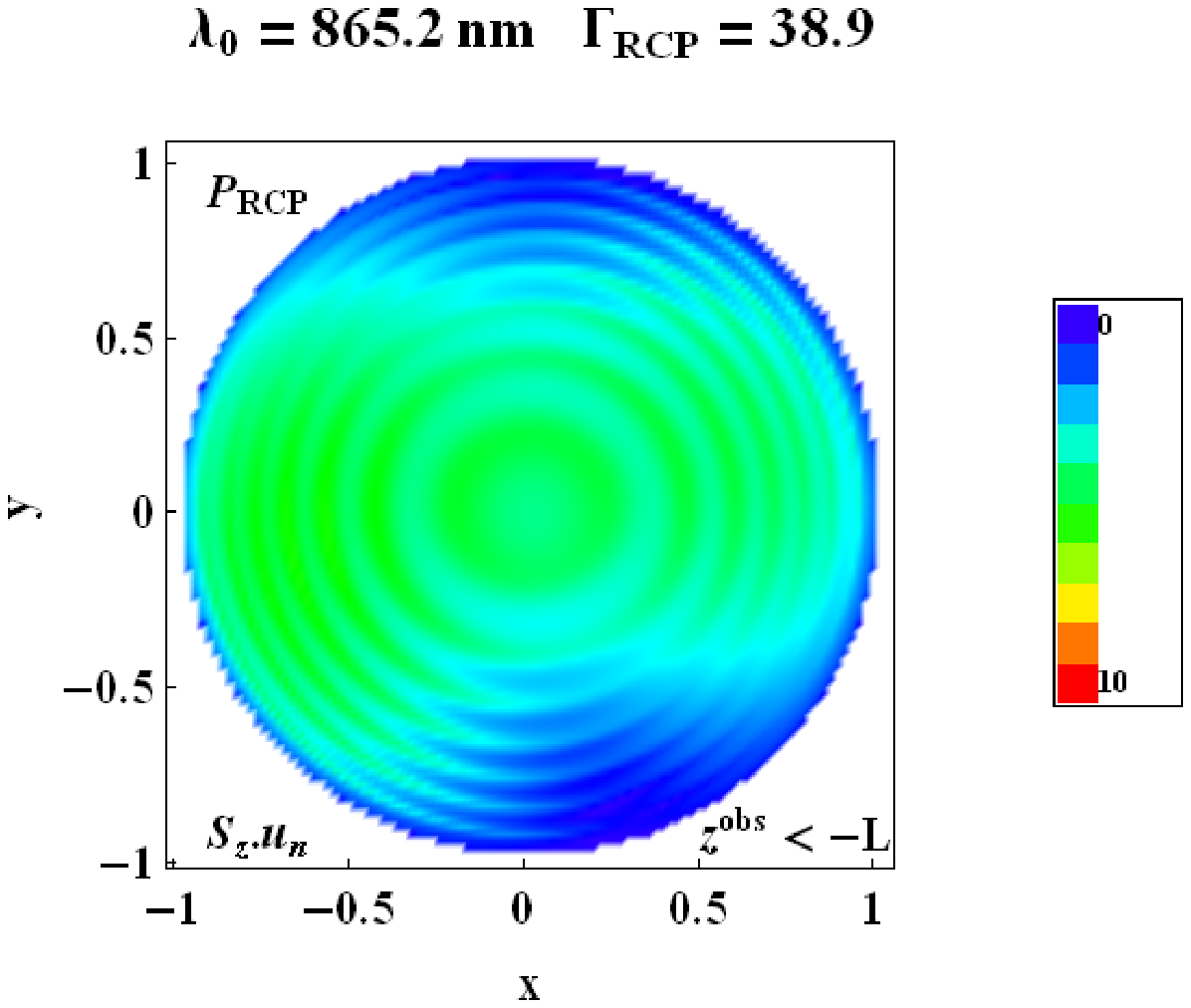}}
\caption{As Fig.~\ref{fn1.0} but with
 $\lambdao \in \lec 745.2 \mbox{nm}, 825.2 \mbox{nm},
865.2 \mbox{nm}\ric$ and $\textit{n}_{l} = 1.25$. The relative
permittivity parameters $\lec \eps_{a2}, \eps_{b2}, \eps_{c2} \ric$
for the  infiltrated CSTF were computed using the non--extended
version of the Bruggeman homogenization formalism.} \label{fn1.25}
\end{figure}

\begin{figure}[!ht]
\centering
\subfigure[]{\includegraphics[width=2in]{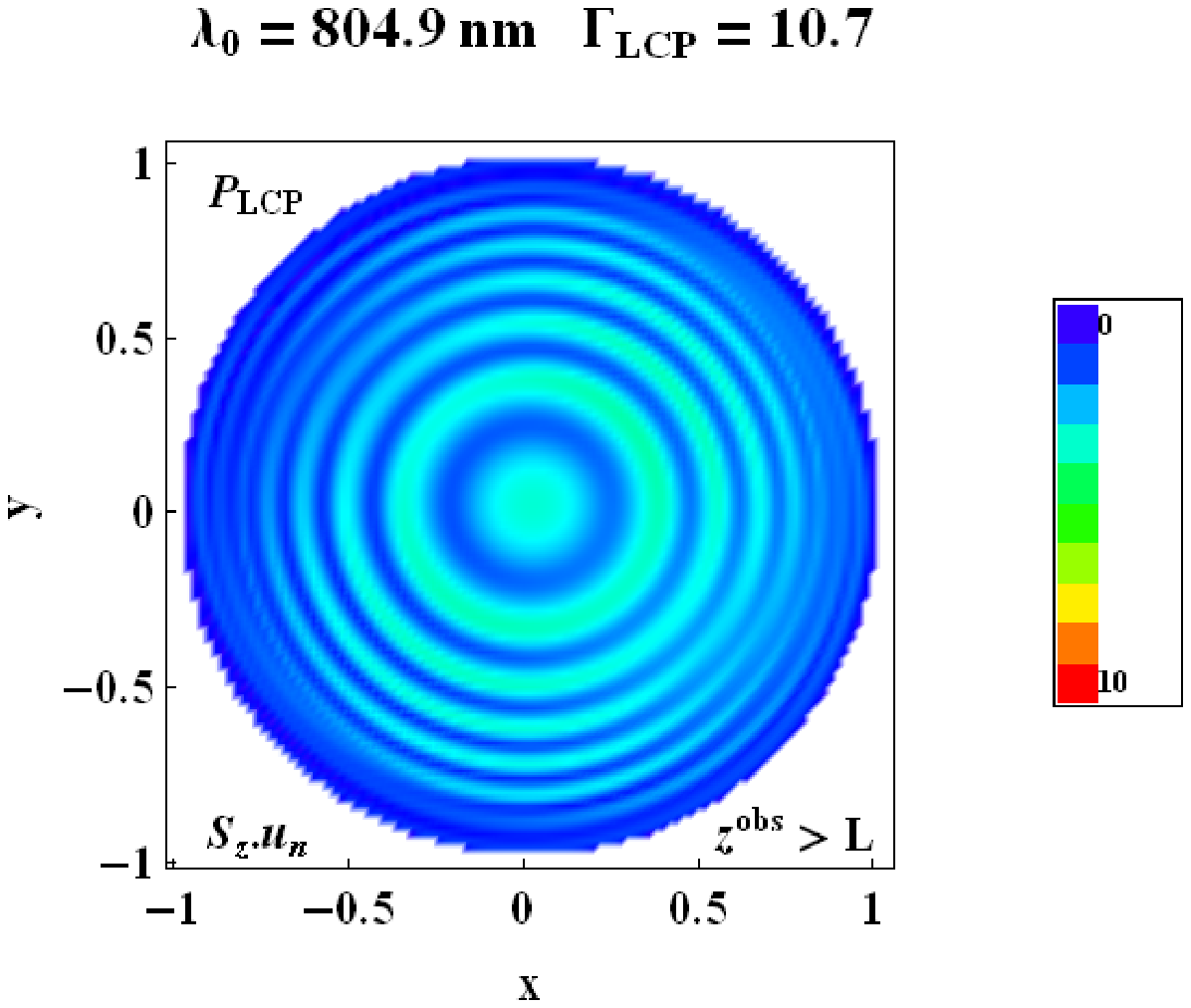}}
\subfigure[]{\includegraphics[width=2in]{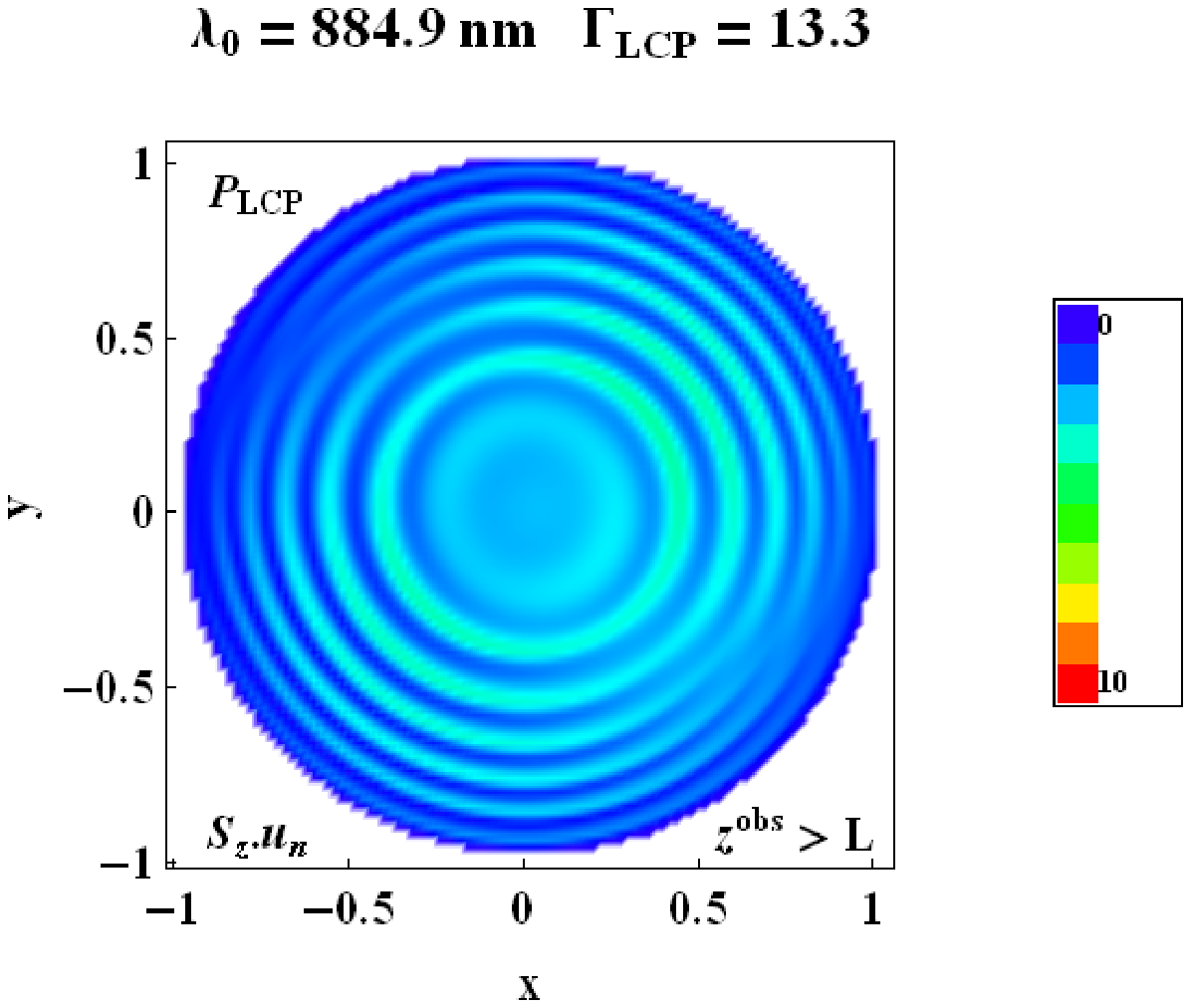}}
\subfigure[]{\includegraphics[width=2in]{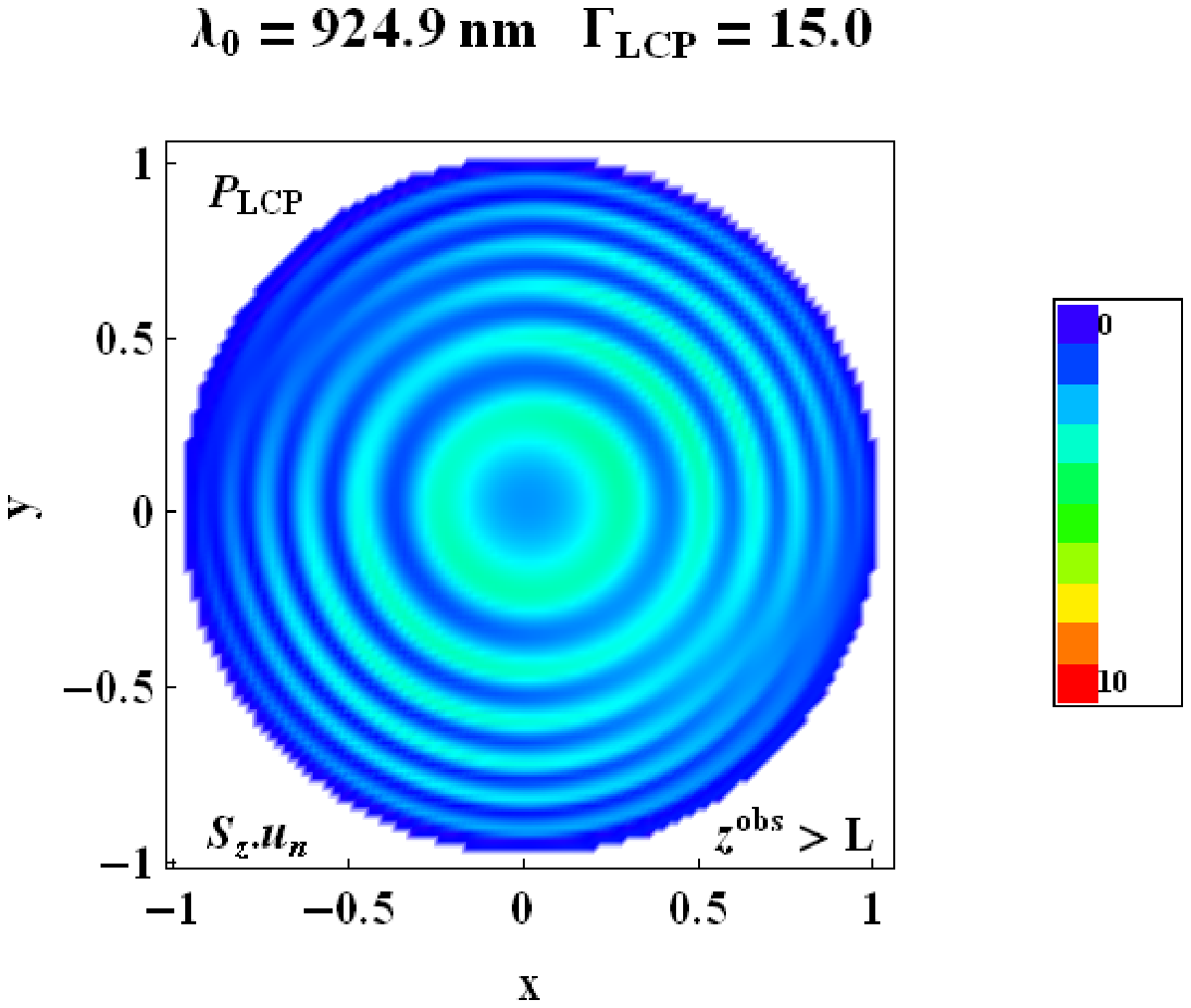}}\\
\subfigure[]{\includegraphics[width=2in]{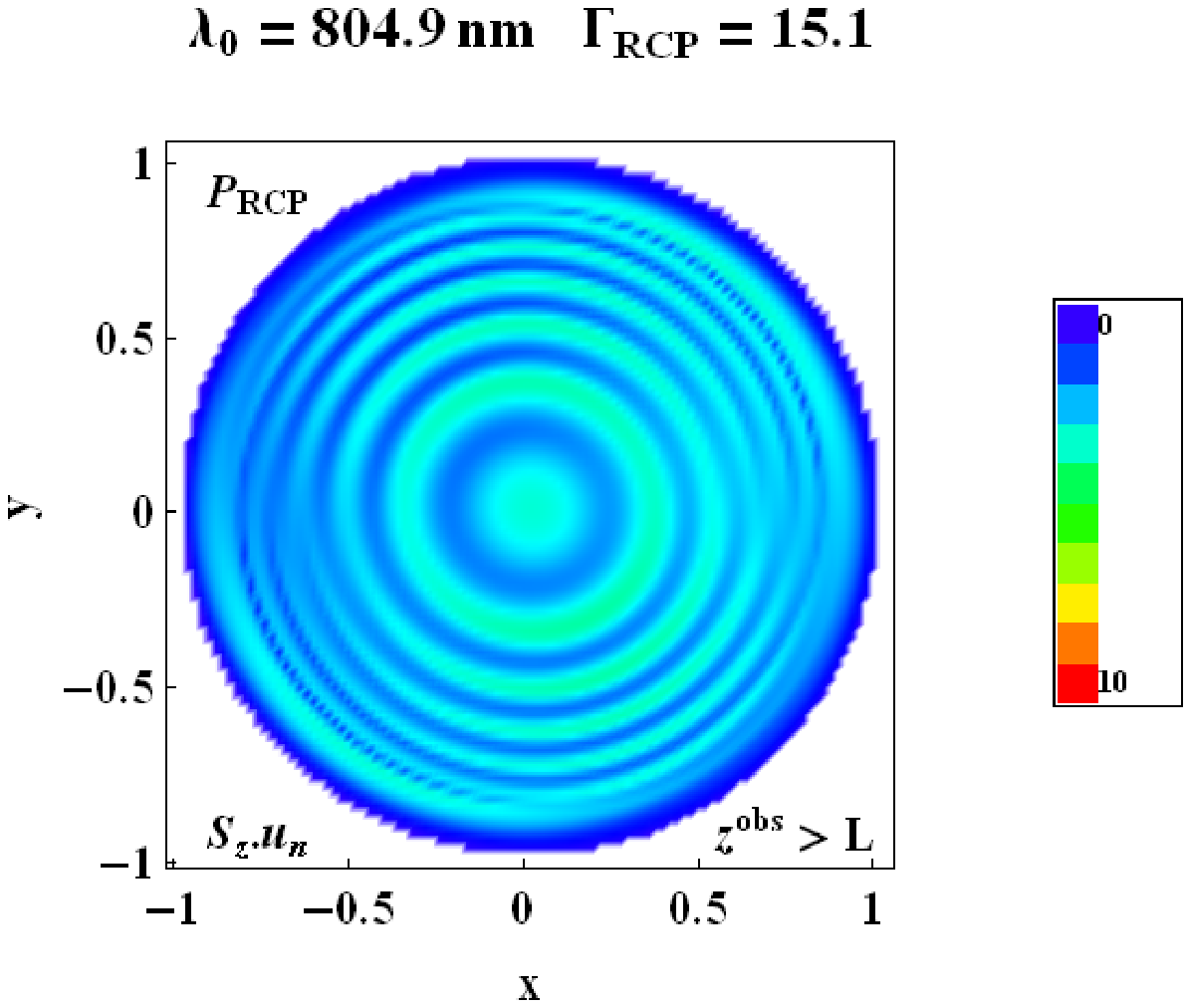}}
\subfigure[]{\includegraphics[width=2in]{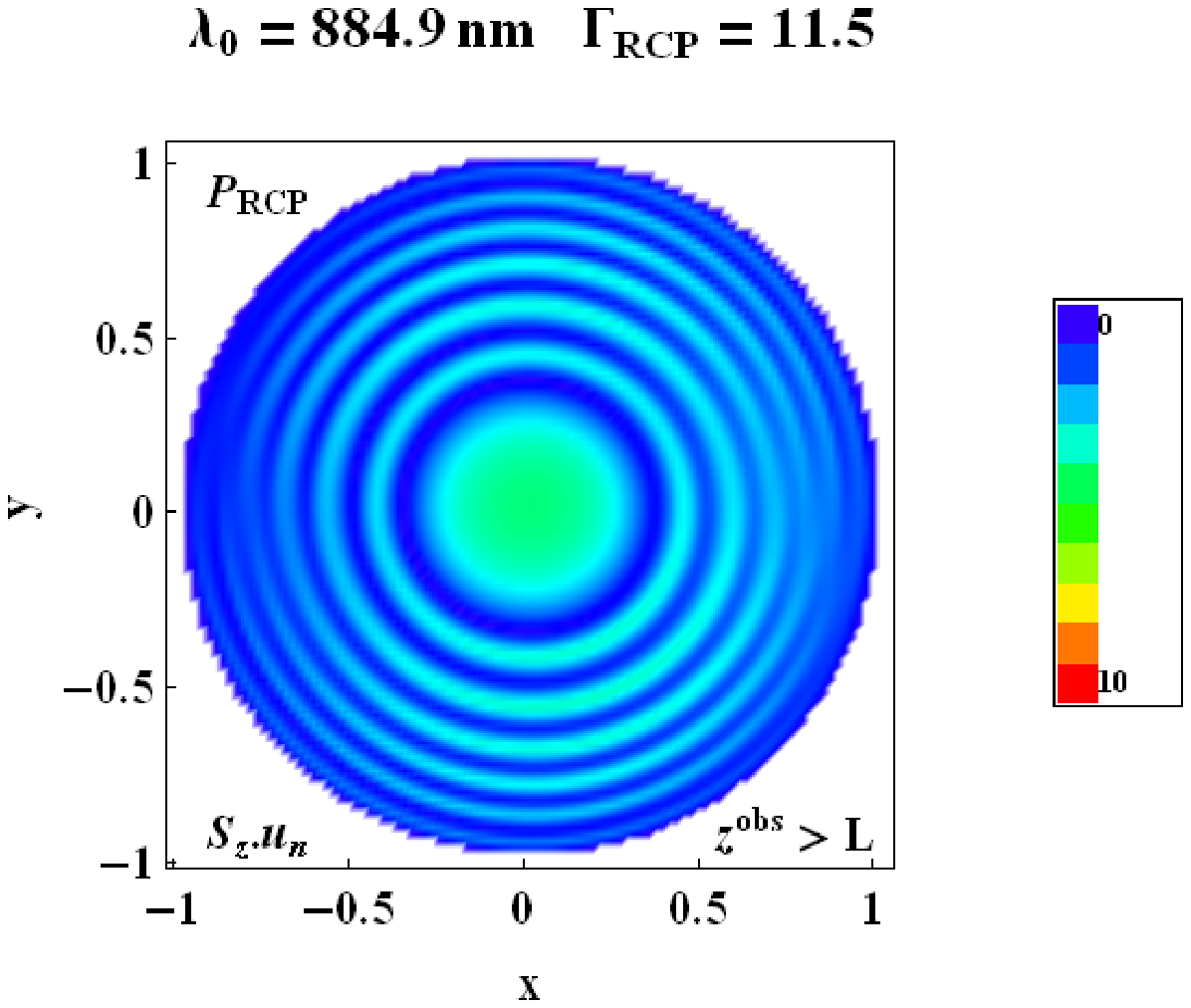}}
\subfigure[]{\includegraphics[width=2in]{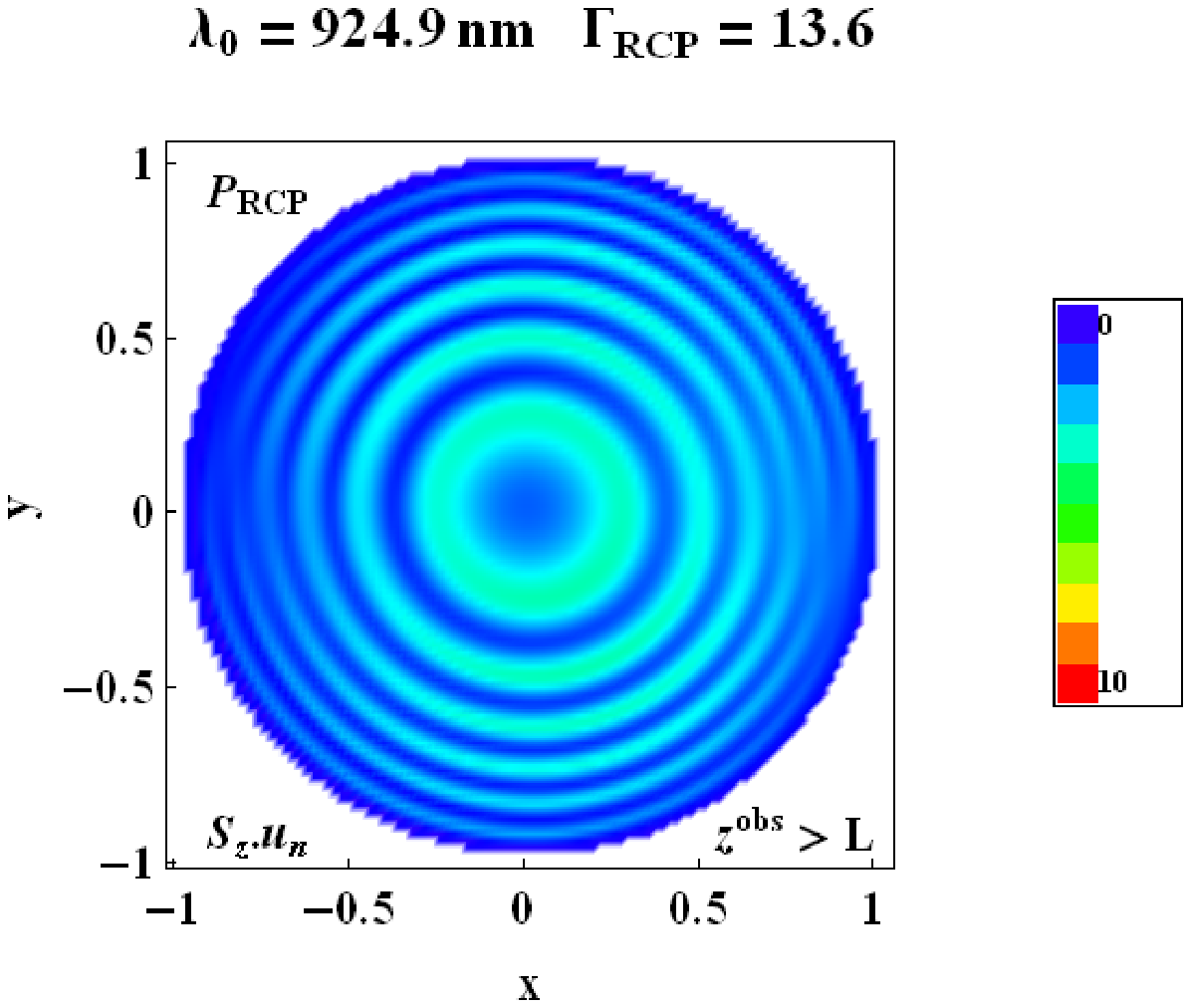}}\\
\subfigure[]{\includegraphics[width=2in]{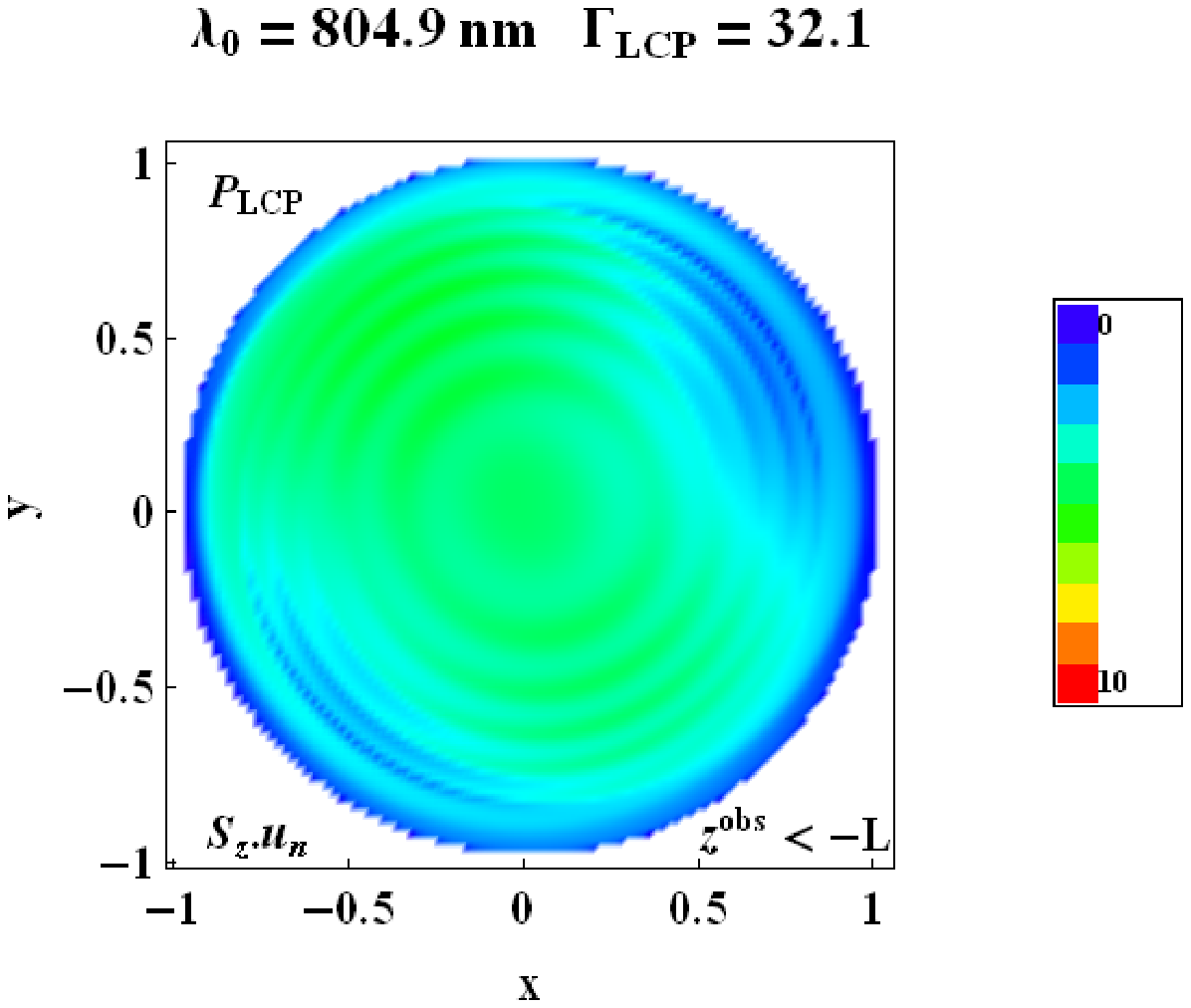}}
\subfigure[]{\includegraphics[width=2in]{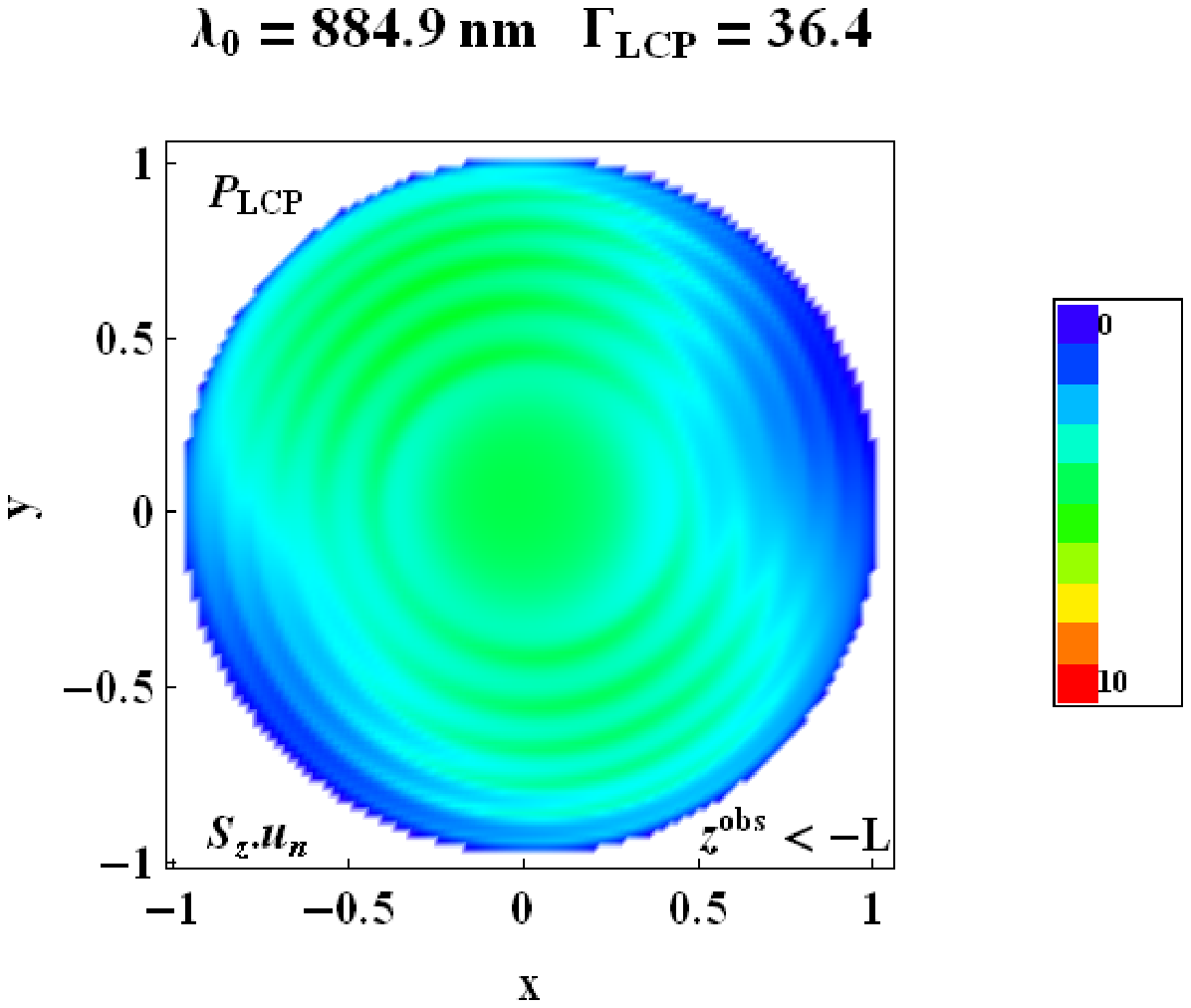}}
\subfigure[]{\includegraphics[width=2in]{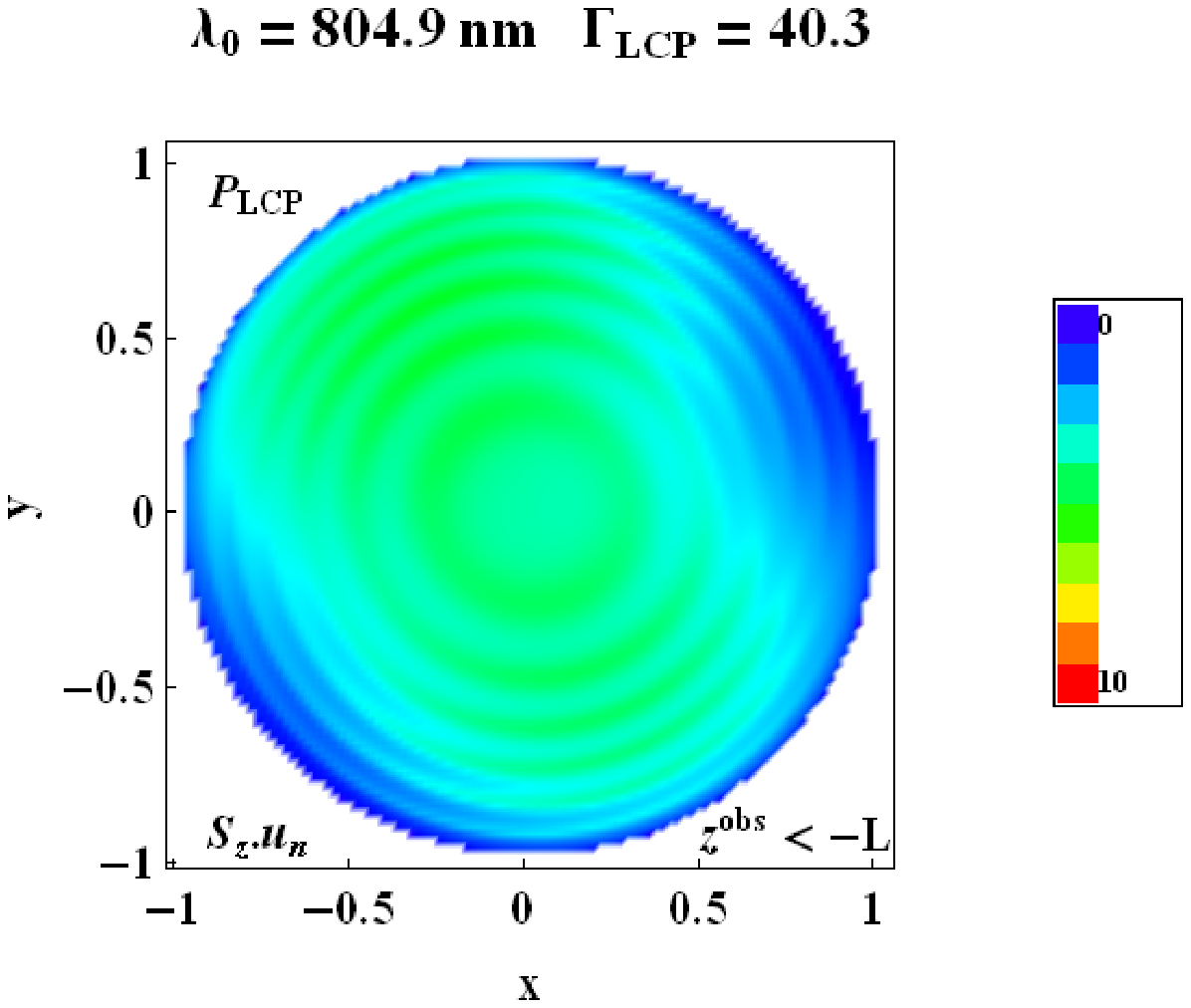}}\\
\subfigure[]{\includegraphics[width=2in]{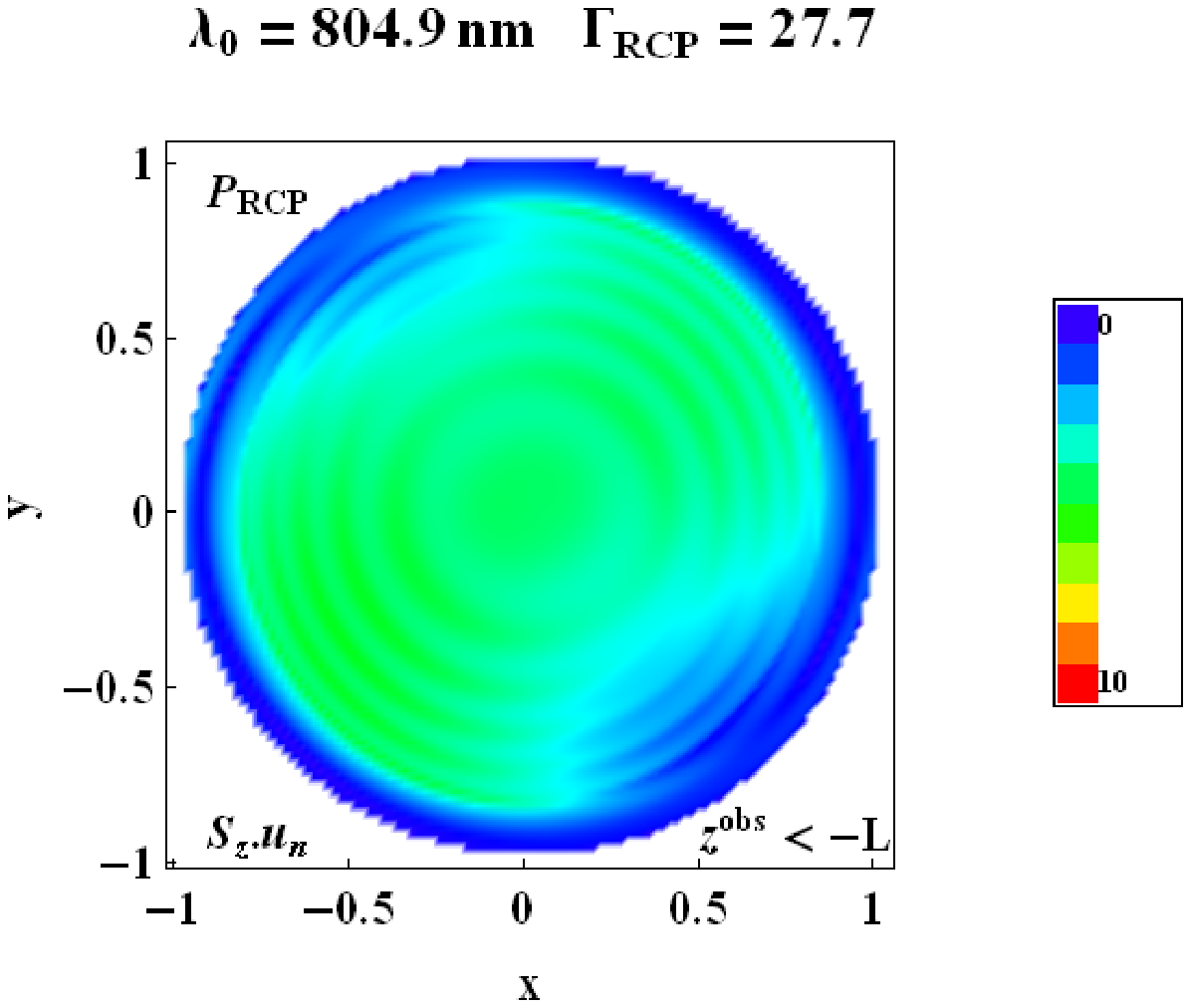}}
\subfigure[]{\includegraphics[width=2in]{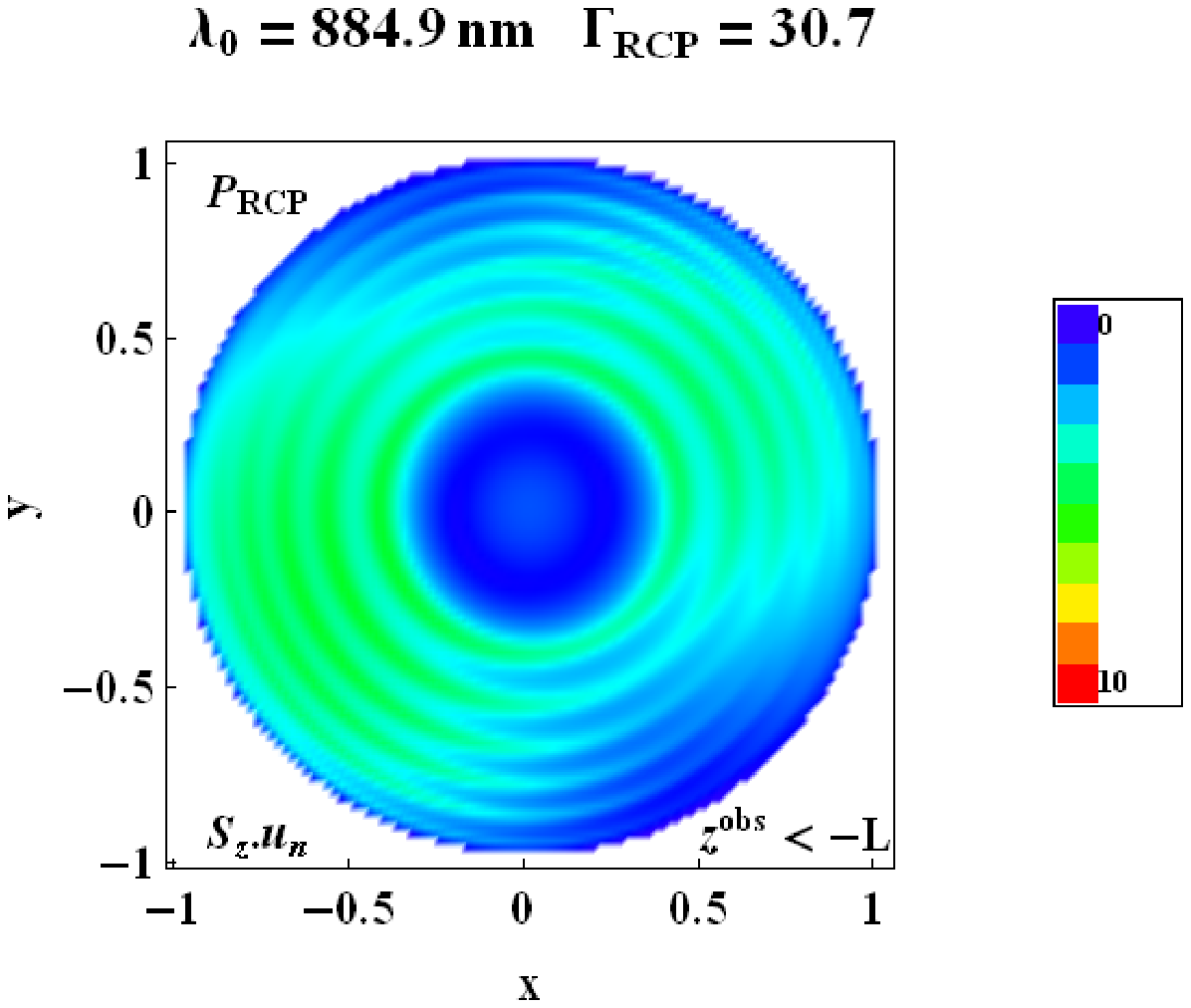}}
\subfigure[]{\includegraphics[width=2in]{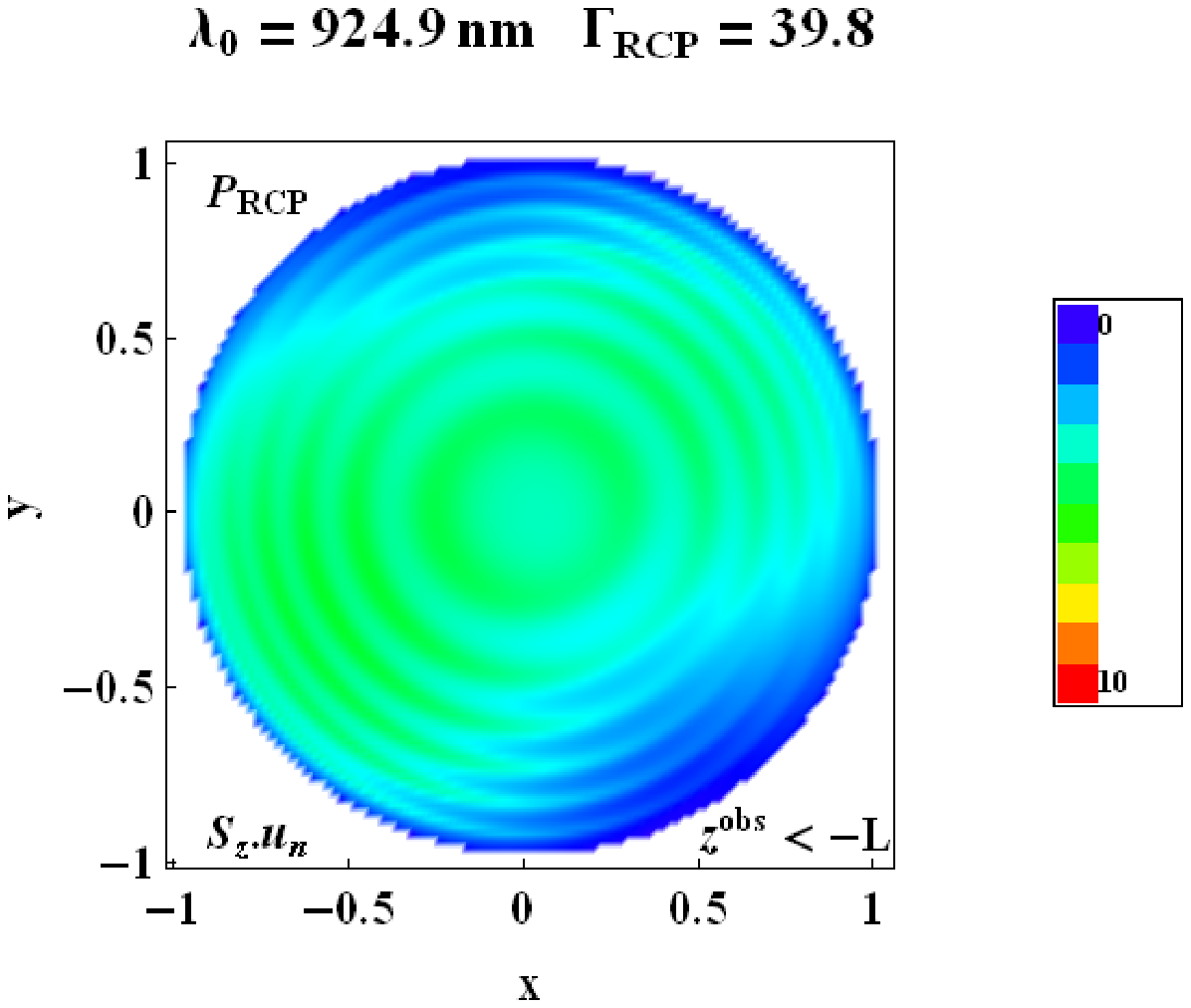}}
\caption{As Fig.~\ref{fn1.25} but with
 $\lambdao \in \lec 804.9 \mbox{nm}, 884.9 \mbox{nm},
924.9 \mbox{nm}\ric$ and $\textit{n}_{l} = 1.5$.} \label{fn1.5}
\end{figure}

The results of infiltration can be observed in Figs.~\ref{fn1.25}
and \ref{fn1.5}, which show the projections of $\left| \#P_{LCP} (
\#r\obs, \omega) \right|$ and $\left| \#P_{RCP} (\#r\obs, \omega )
\right|$ (as before, scaled by a factor of $10^{13} \omega^{-2}
\left| p \right|^{-2} $) onto the $z=0$ plane  for $\nl = 1.25$ and
$\nl = 1.5$, respectively. The relative permittivity parameters
$\lec \eps_{a2}, \eps_{b2}, \eps_{c2} \ric$ for the  infiltrated
CSTF were computed using the non--extended version of the Bruggeman
homogenization formalism (or, equivalently, the extended version
with $\eta = 0$). In keeping with Fig.~\ref{fn1.0}, results are
presented for $\lambdao = \lambdao^{Br} (   \eps_{a2}, \eps_{b2},
\eps_{c2}, 0^\circ )$, as well as for $\lambdao =  \lambdao^{Br} (
\eps_{a2}, \eps_{b2}, \eps_{c2}, 0^\circ ) - 80 $ nm and $\lambdao =
\lambdao^{Br} ( \eps_{a2}, \eps_{b2}, \eps_{c2}, 0^\circ ) + 40 $
nm. As for the uninfiltrated case,  we see that RCP radiation is
   preferentially emitted through the pupil at $z = L$ at
   whereas
LCP radiation is
   preferentially emitted through the pupil at $z = -L$, at the
   centre Bragg wavelength $\lambdao^{Br} (   \eps_{a2}, \eps_{b2}, \eps_{c2},
\theta\obs )$ for  $\vert \cos\theta\obs\vert \simeq 1$. Two effects
in particular  of infiltration  are apparent from Figs.~\ref{fn1.25}
and \ref{fn1.5} as the refractive index of the infiltrating fluid
increases: (i) the distinction between LCP/RCP emission for $
\vert\cos\theta^{obs}\vert \simeq 1$ becomes increasingly red
shifted; and (ii) the blue shift in the distinction between LCP/RCP
emission for oblique angles of incidence becomes less pronounced.

\begin{figure}[!ht]
\centering
\subfigure[]{\includegraphics[width=3.5in]{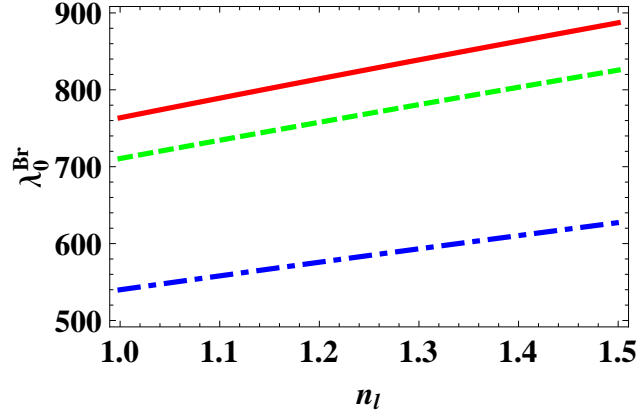}}\\
\subfigure[]{\includegraphics[width=3.5in]{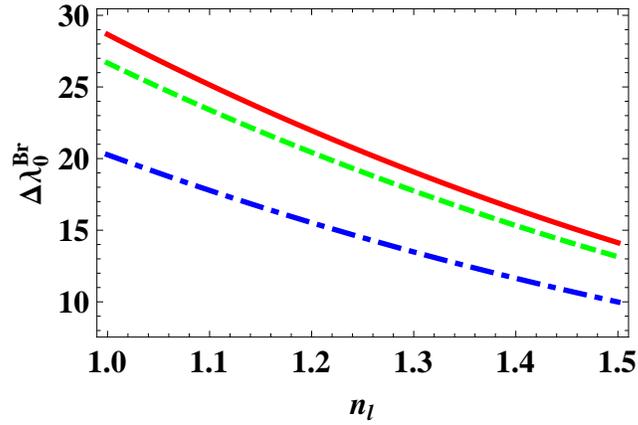}}
\caption{$\lambda^{Br}_{0}$ and $\Delta \lambda^{Br}_{0}$  plotted
against $\textit{n}_{l}$ for $\theta\obs = 0^\circ$ (solid, red
curve), $\theta\obs = 30^\circ$ (dashed, green curve) and
$\theta\obs = 60^\circ$ (broken dashed, blue curve). The relative
permittivity parameters $\lec \eps_{a2}, \eps_{b2}, \eps_{c2} \ric$
for the infiltrated CSTF were computed using the non--extended
version of the Bruggeman homogenization formalism.} \label{fBragg}
\end{figure}

In fact, the main effects of infiltration are quite well predicted
via the empirical relations \r{Bragg} and \r{Bragg_width}. To see
this, the centre Bragg wavelength and the
full--width--at--half--maximum bandwidth are plotted versus $\nl$ in
Fig.~\ref{fBragg}. As in Figs.~\ref{fn1.25} and \ref{fn1.5}, the
non--extended version of the Bruggeman homogenization formalism was
used to estimate the relative permittivity parameters $\lec
\eps_{a2}, \eps_{b2}, \eps_{c2} \ric$. The red shift in the centre
wavelength, and the reduction in the full--width--at--half--maximum
bandwidth, resulting from the refractive index of the infiltrating
fluid being increased are obvious from Fig.~\ref{fBragg}~---~and
these effects are in complete agreement with the emission patterns
observed in Figs.~\ref{fn1.0}--\ref{fn1.5}.

\begin{figure}[!ht]
\centering
\subfigure[]{\includegraphics[width=2in]{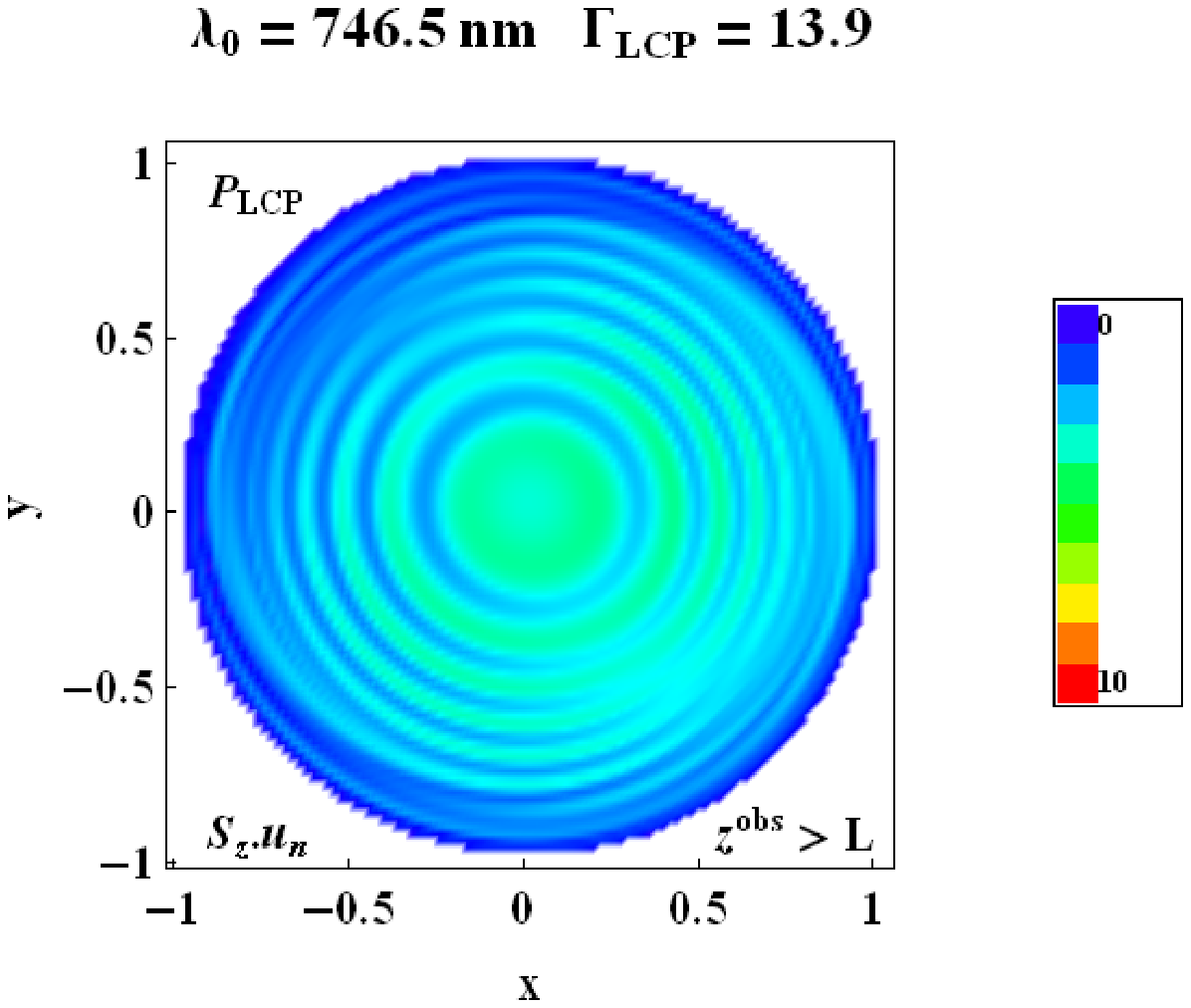}}
\subfigure[]{\includegraphics[width=2in]{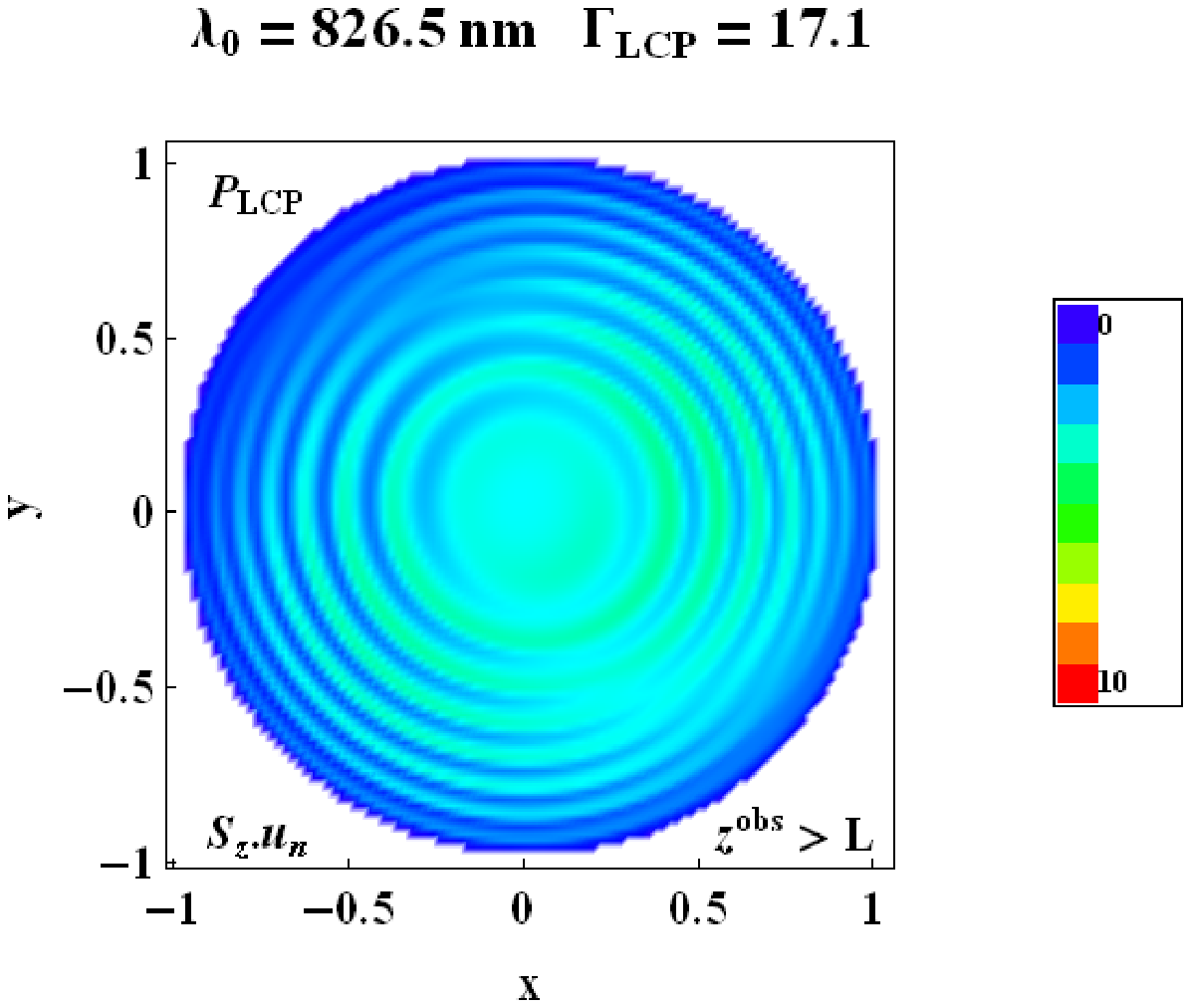}}
\subfigure[]{\includegraphics[width=2in]{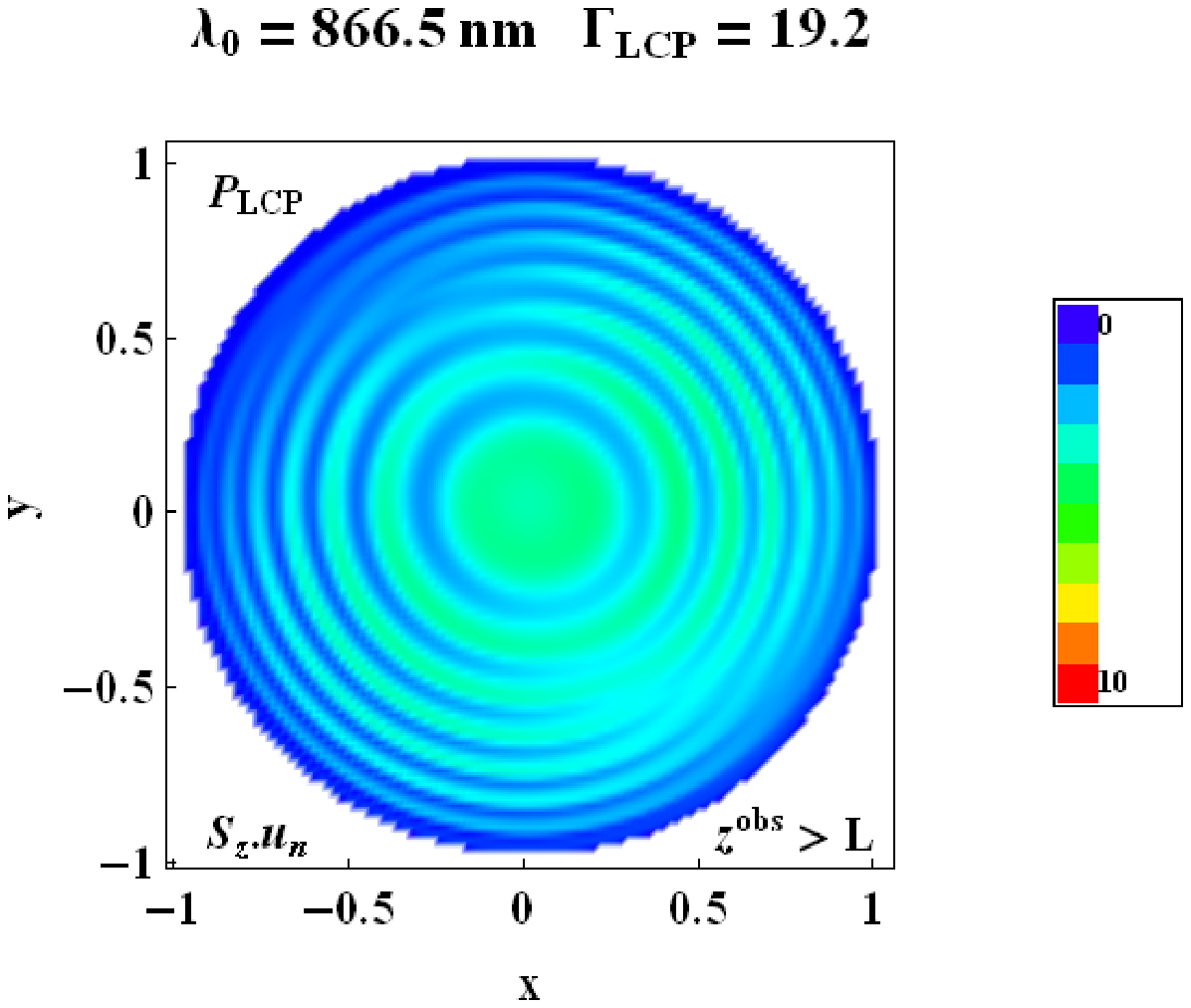}}\\
\subfigure[]{\includegraphics[width=2in]{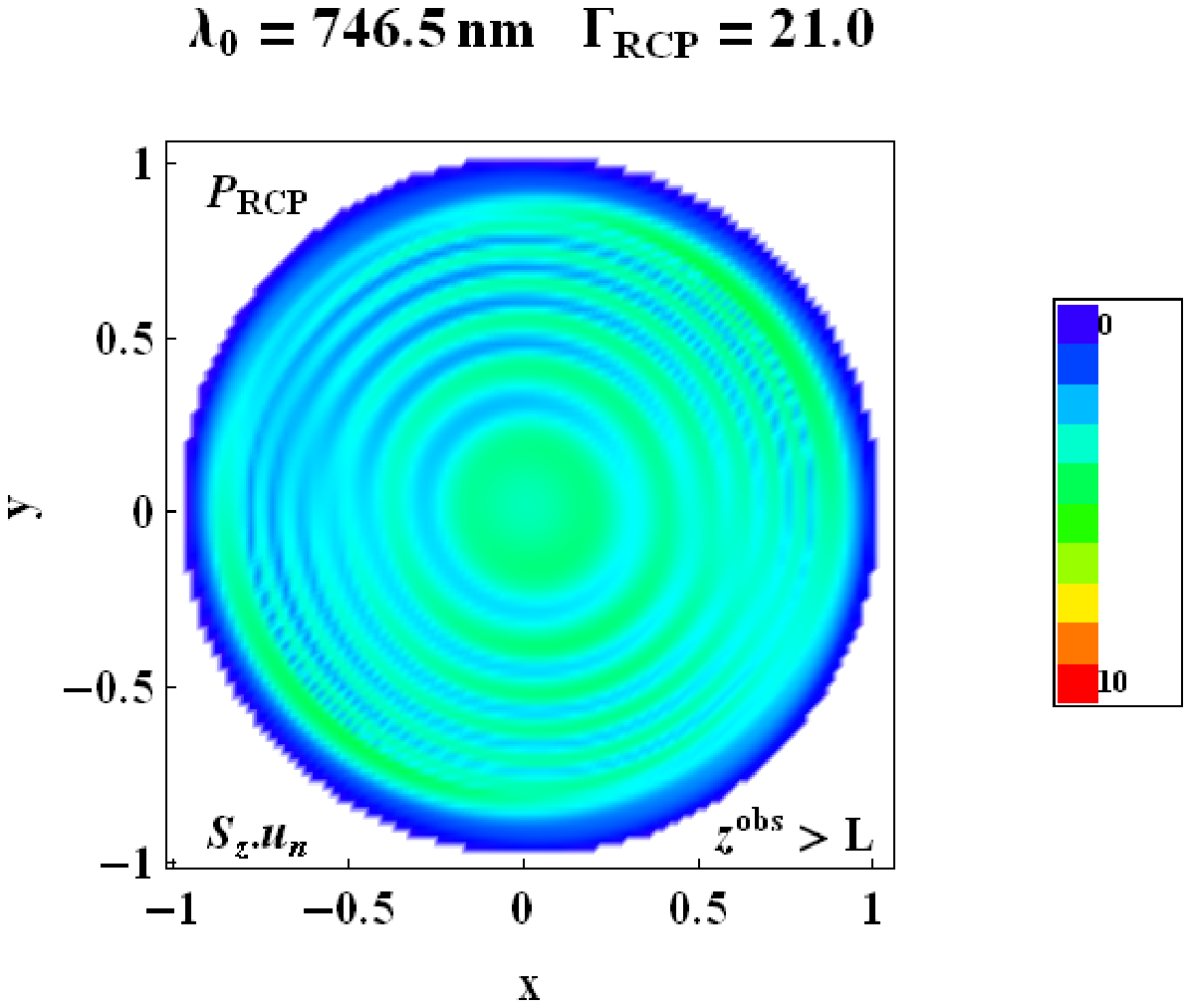}}
\subfigure[]{\includegraphics[width=2in]{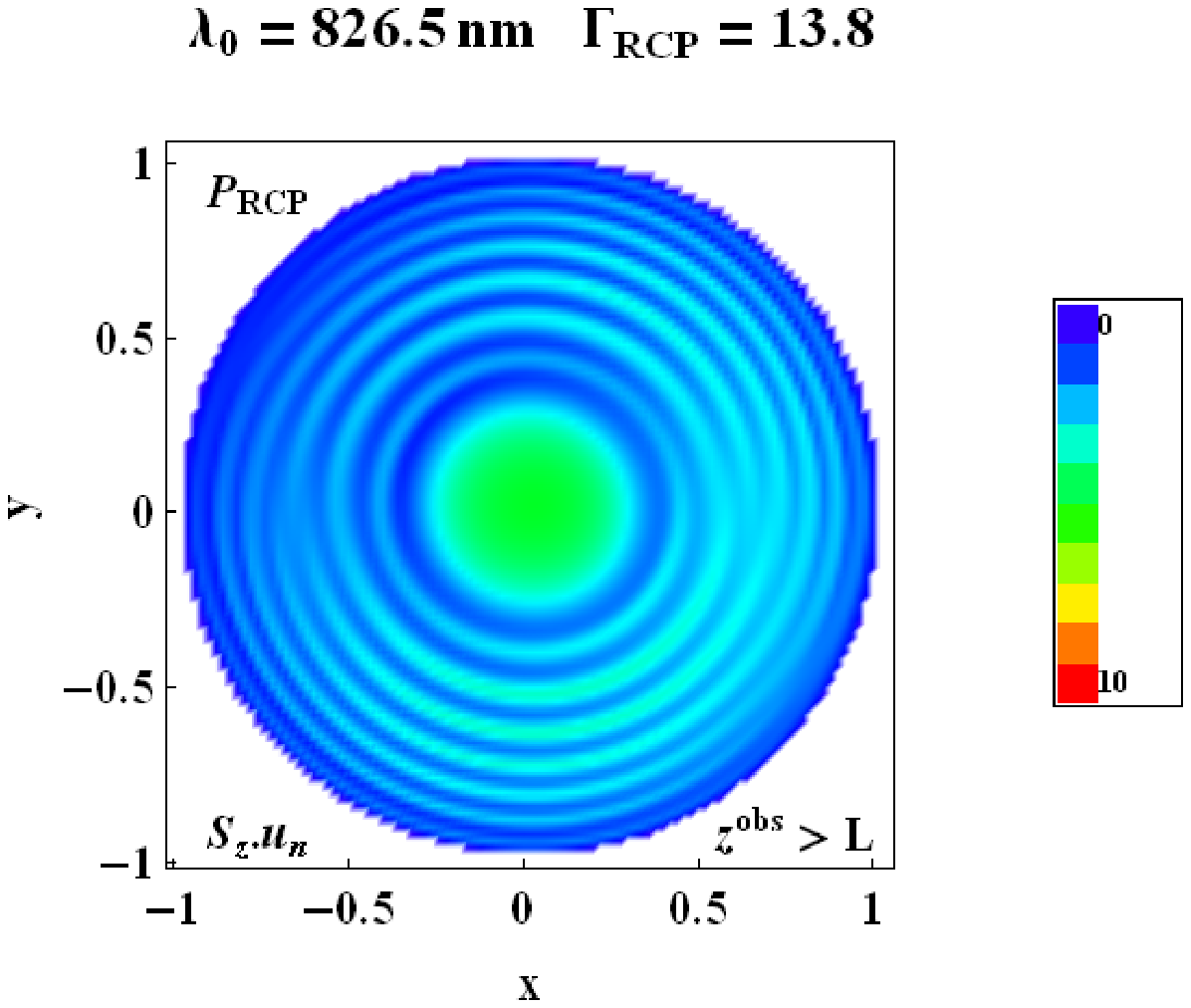}}
\subfigure[]{\includegraphics[width=2in]{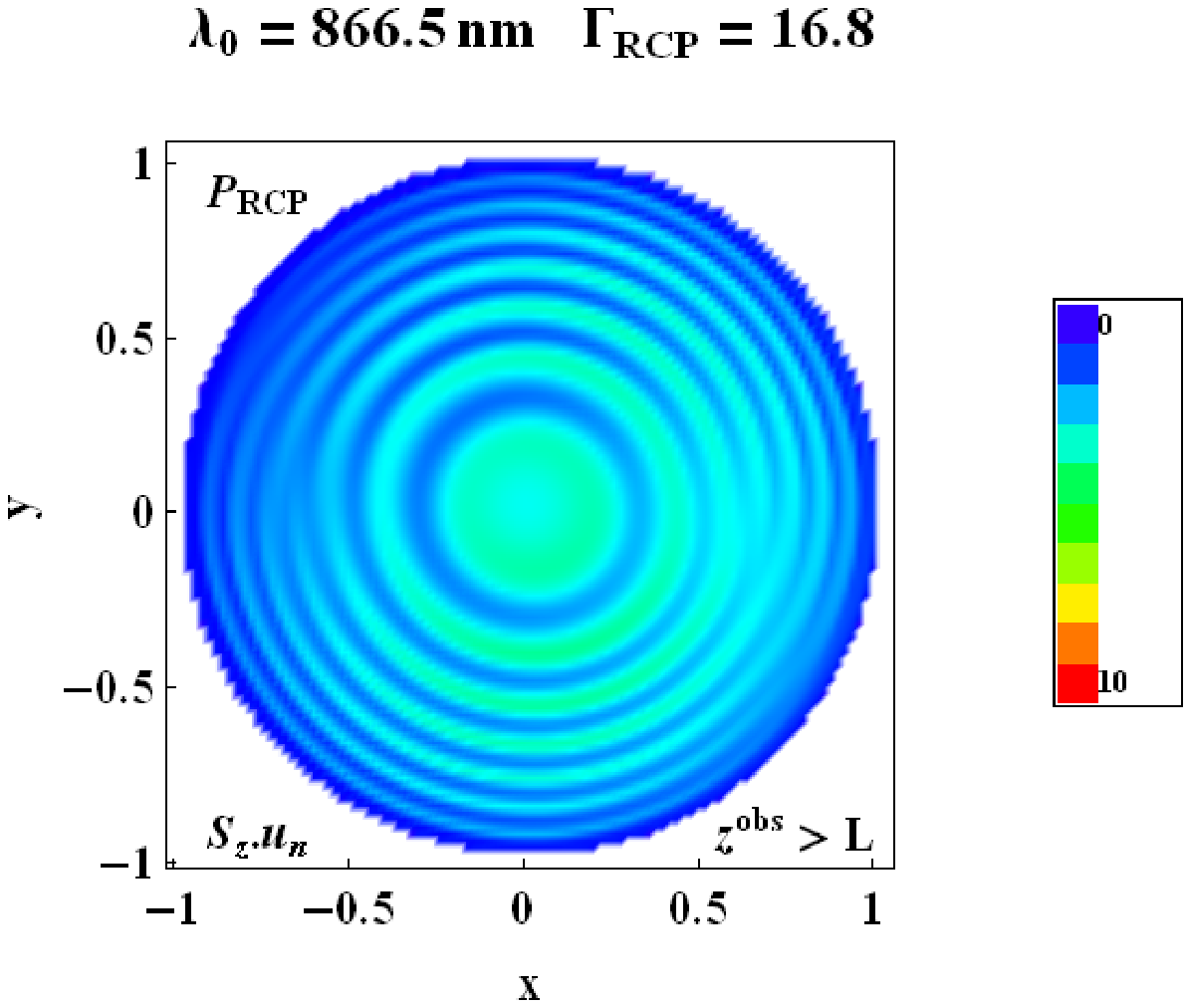}}\\
\subfigure[]{\includegraphics[width=2in]{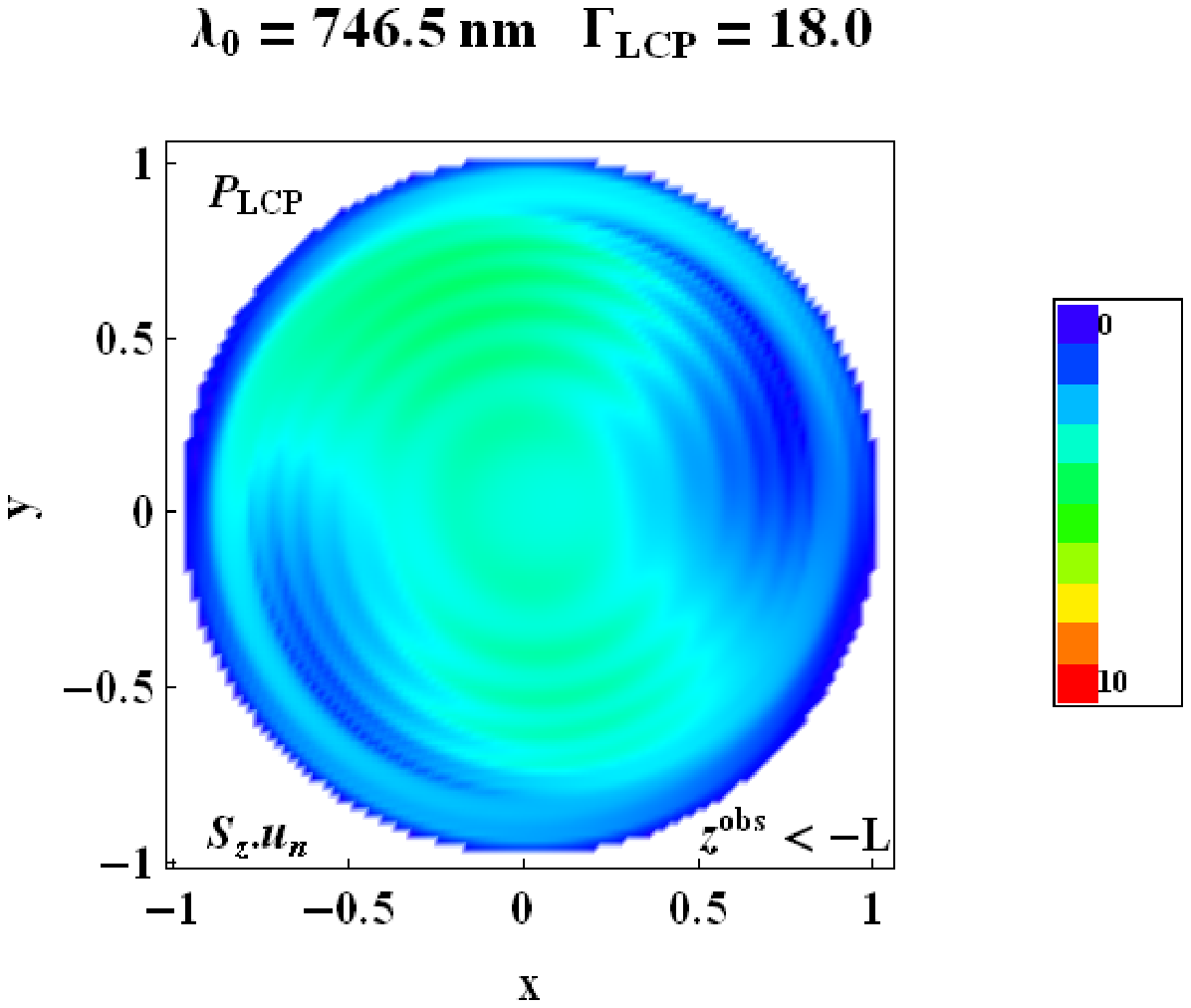}}
\subfigure[]{\includegraphics[width=2in]{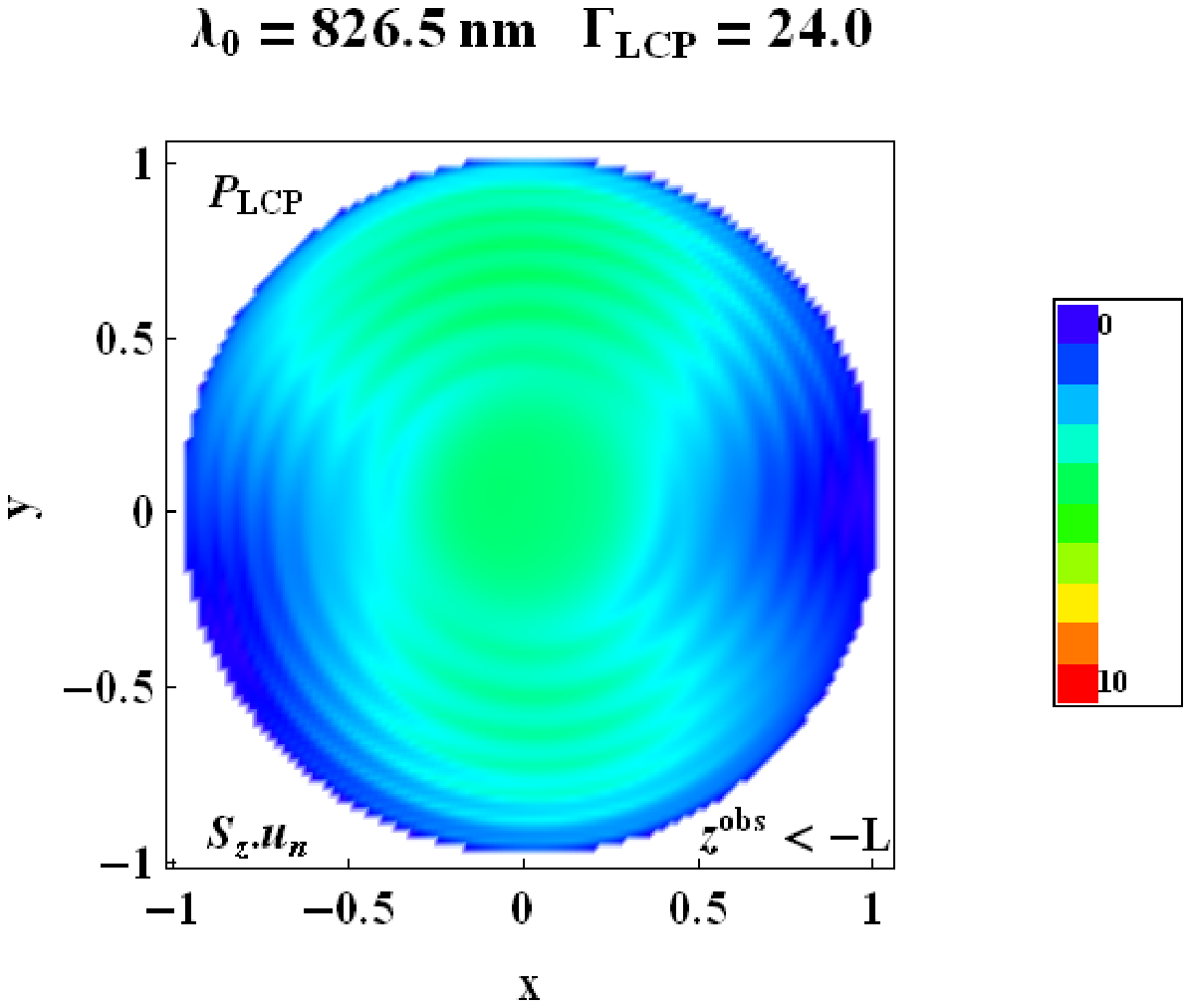}}
\subfigure[]{\includegraphics[width=2in]{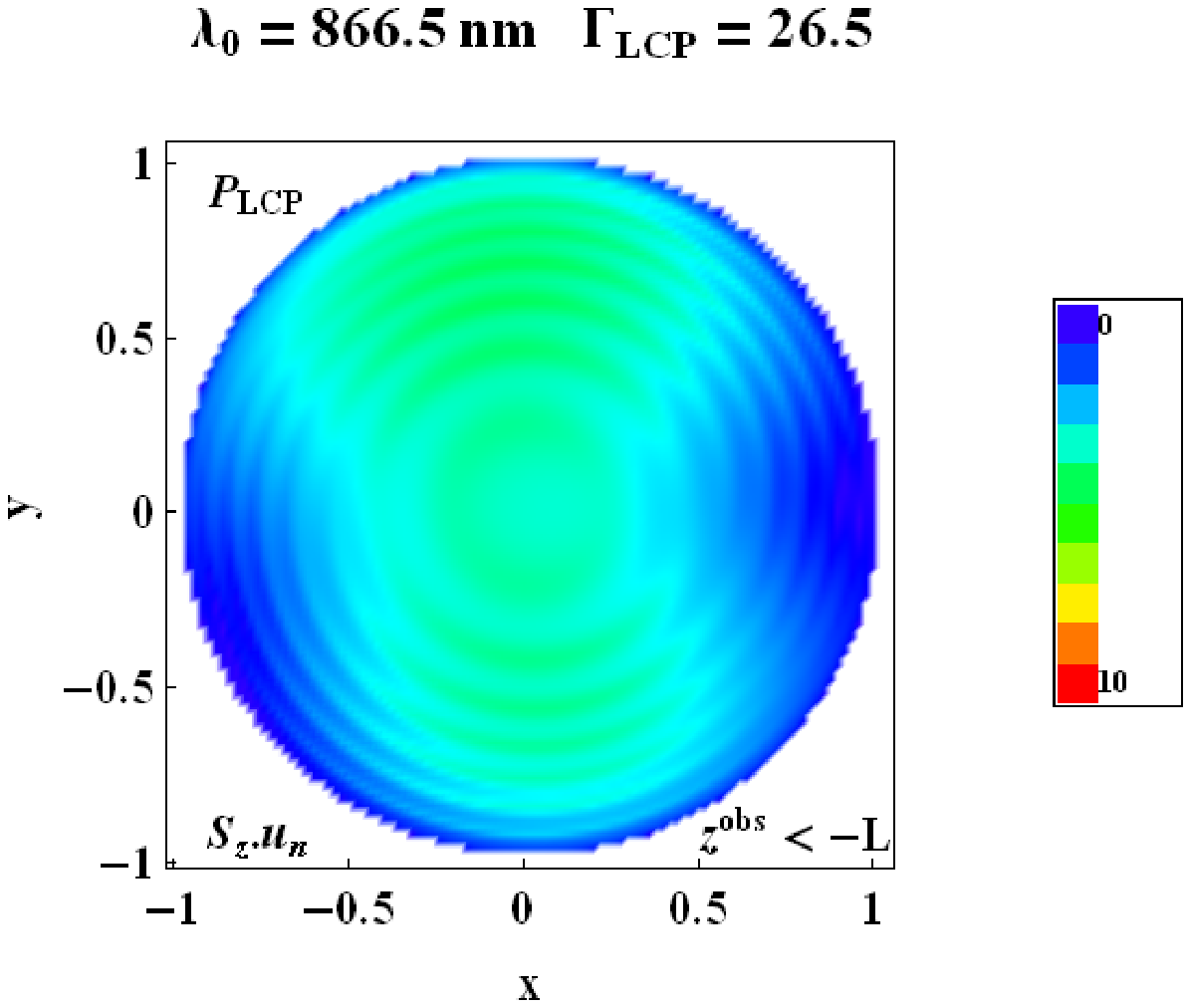}}\\
\subfigure[]{\includegraphics[width=2in]{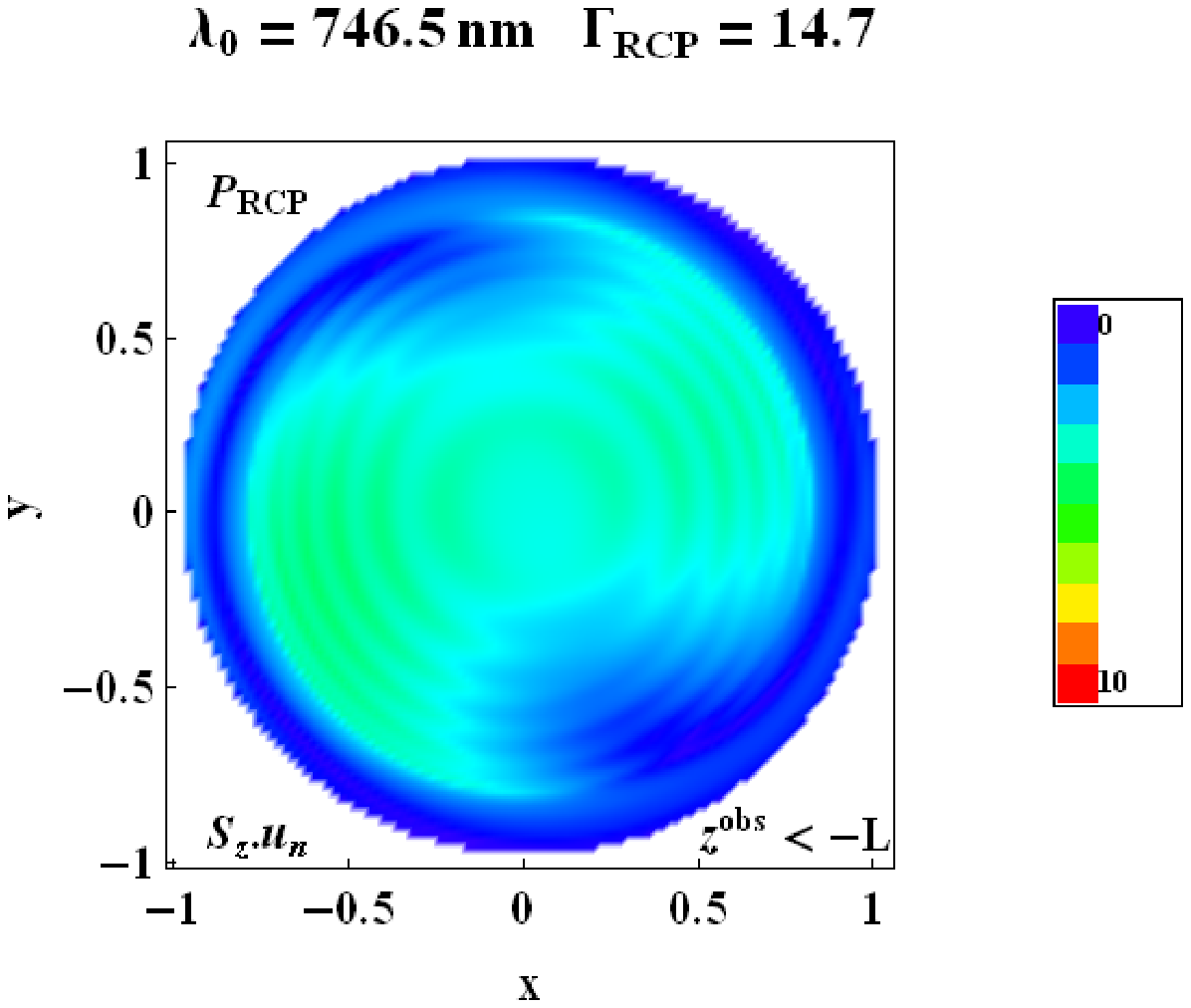}}
\subfigure[]{\includegraphics[width=2in]{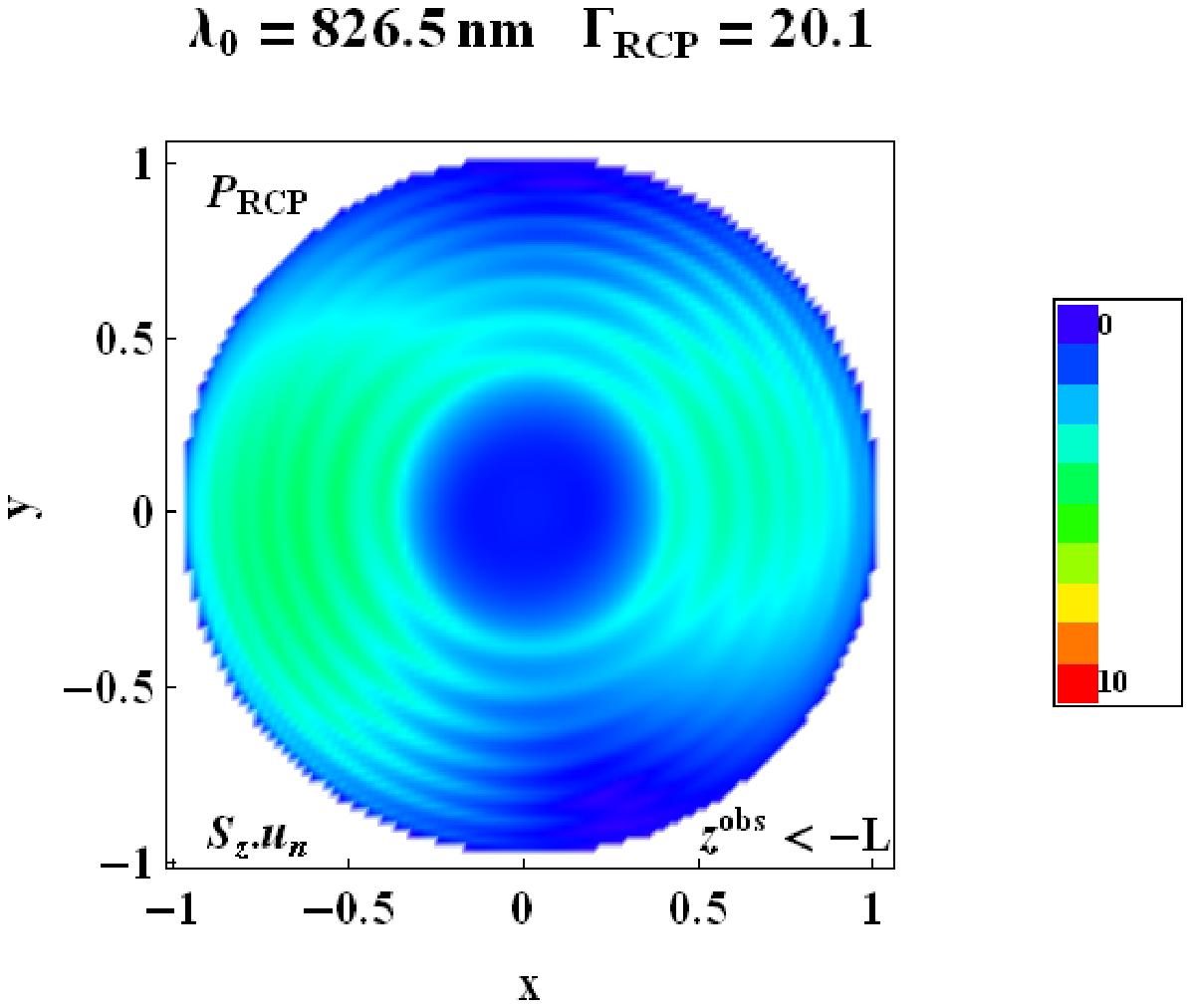}}
\subfigure[]{\includegraphics[width=2in]{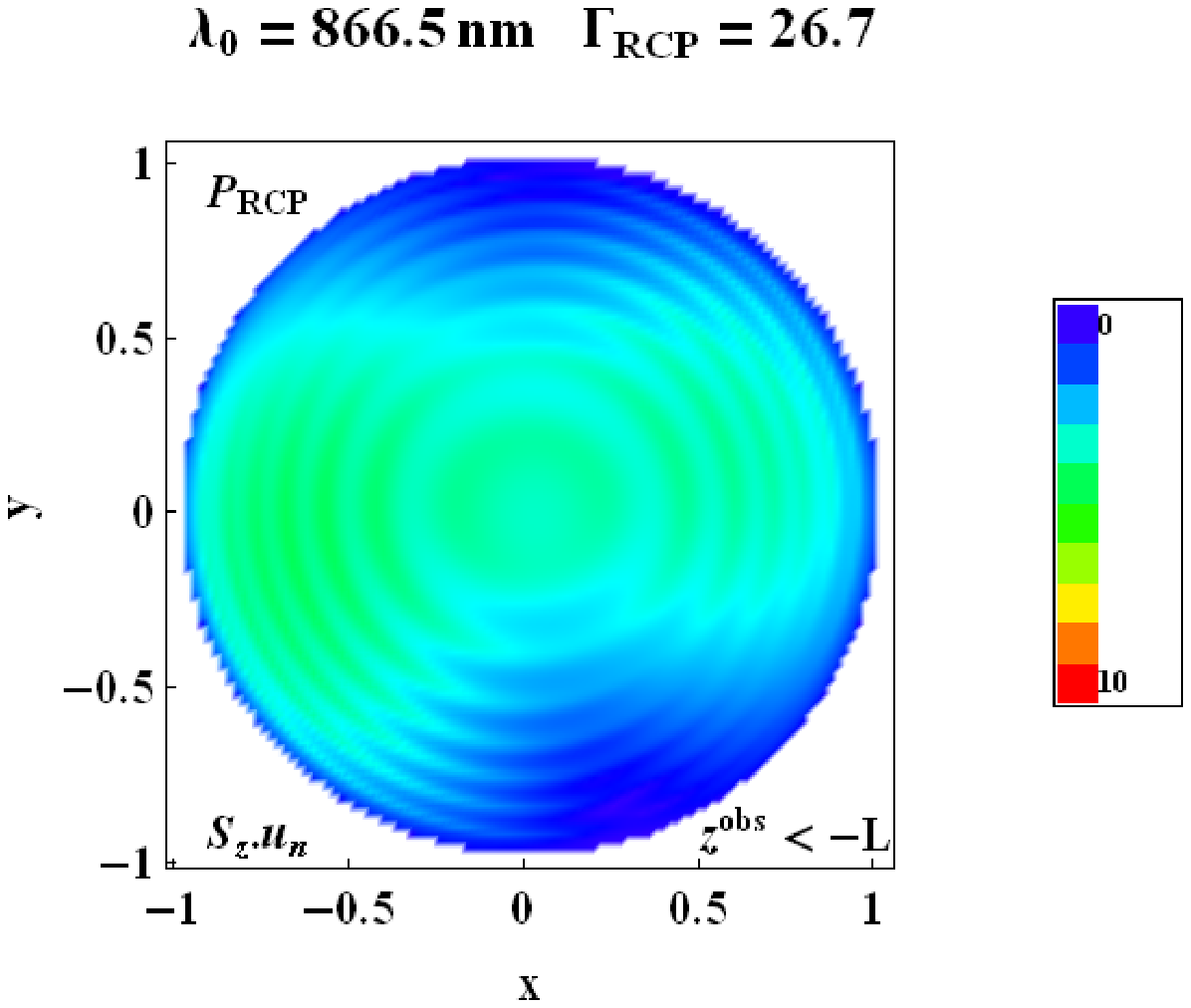}}
\caption{As Fig.~\ref{fn1.25} but the relative permittivity
parameters $\lec \eps_{a2}, \eps_{b2}, \eps_{c2} \ric$ for the
infiltrated CSTF were computed using the extended version of the
Bruggeman homogenization formalism with $\eta = 0.1 /\ko$, and
 $\lambdao \in \lec 746.5 \mbox{nm}, 826.5 \mbox{nm},
866.5 \mbox{nm}\ric$. } \label{fn1.25e}
\end{figure}

We consider now the influence of the linear size of the ellipsoidal
particles which represent the CSTF's helical nanowires, per the
homogenization model described in \S\ref{inv_homog}.
 As a representative example, we repeat the calculations of
 Fig.~\ref{fn1.25} but here using the extended Bruggeman homogenization
 formalism to estimate the relative permittivity parameters $\lec \eps_{a2}, \eps_{b2},
\eps_{c2} \ric$ of the  infiltrated CSTF. We set the size parameter
to be $\eta = 0.1 / \ko$. The corresponding  projections of $\left|
\#P_{LCP} ( \#r\obs, \omega) \right|$ and $\left| \#P_{RCP}
(\#r\obs, \omega ) \right|$ (as before, scaled by a factor of
$10^{13} \omega^{-2} \left| p \right|^{-2} $) onto the $z=0$ plane
are provided in Fig.~\ref{fn1.25e} for $\nl = 1.25$. By comparing
Figs.~\ref{fn1.25} and \ref{fn1.25e}, we see that the size parameter
has a relatively minor but significant influence on the emission
patterns. The centre wavelength for the distinction between LCP/RCP
emission
 is
slightly higher in Fig.~\ref{fn1.25e} as compared to
Fig.~\ref{fn1.25}. Furthermore, the total energy flux emitted from
the CSTF~---~as estimated by the scalar parameter $\Gamma_{LCP,
RCP}$~---~is substantially smaller when we consider $\eta = 0.1
/\ko$ as opposed to $\eta = 0$. This is most noticeable for
radiation emitted through the CSTF pupil at $z= -L$, which is a
consequence of the dipole source being closer to the $z=L$ pupil.

\begin{figure}[!ht]
\centering
\subfigure[]{\includegraphics[width=3.5in]{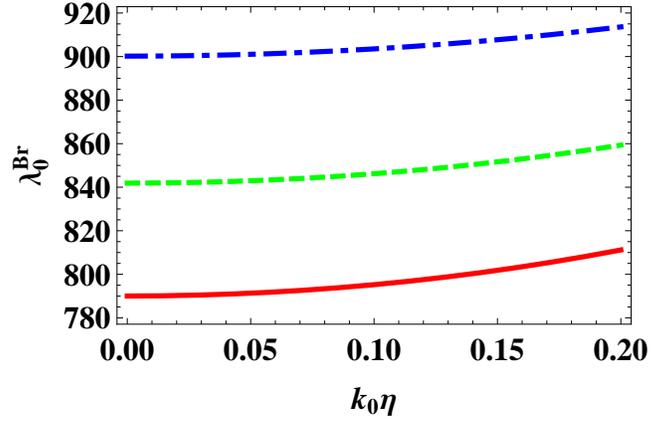}}
\subfigure[]{\includegraphics[width=3.5in]{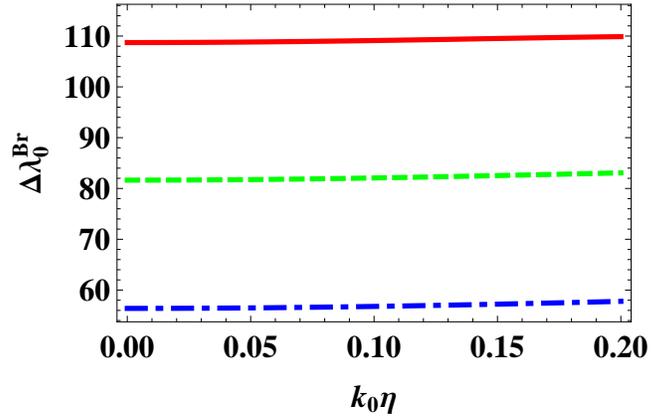}}
\caption{$\lambda^{Br}_{0}$ and $\Delta \lambda^{Br}_{0}$  plotted
against $\ko \eta$ for $\nl = 1.0$ (solid, red curve), $\nl = 1.25$
(dashed, green curve) and $\nl = 1.5$ (broken dashed, blue curve),
as computed using the extended  Bruggeman homogenization formalism.
The angle $\theta\obs = 0^\circ$.} \label{fBragge}
\end{figure}

The main effects of the size parameter $\eta$ may be estimated quite
well using the empirical relations \r{Bragg} and \r{Bragg_width}. In
Fig.~\ref{fBragge}, $\lambdao^{Br} (\eps_{a2}, \eps_{b2}, \eps_{c2},
0^\circ)$ and $\Delta \lambdao^{Br} (\eps_{a2}, \eps_{b2},
\eps_{c2}, 0^\circ)$ are plotted versus $\ko \eta$ for $\nl \in \lec
1.0, 1.25, 1.5 \ric$. The modest increase in the centre Bragg
wavelength is clear; and we note that the rate of increase is
greatest when the refractive index of the infiltrating fluid is
smallest. The full--width--at--half--maximum bandwidth is relatively
insensitive to the size parameter $\eta$, regardless of the value of
$\nl$. Qualitatively similar results are found when angles
$\theta\obs > 0^\circ$ are considered.

\section{Closing remarks}

Using a spectral Green function formalism, in conjunction with an
inverse homogenization formalism, the effect of infiltration on the
emission from a dipole source embedded within a CSTF has been
characterized. Based on  numerical studies, our conclusions may be
summarized as:
\begin{itemize}
\item The centre wavelength for the preferential emission of LCP/RCP
radiation is red shifted as the refractive index of the infiltrating
fluid increases from unity. Furthermore, the  red shift is
accentuated when the size of the ellipsoidal particles which
represent the helical nanowires of the CSTF is increased.
\item The bandwidth of the preferential LCP/RCP emission regime
decreases as the refractive index of the infiltrating fluid
increases from unity.
\item The main effects of infiltration may be reasonably predicted using
the simple empirical formulas \r{Bragg} and \r{Bragg_width} which
provide estimates of the centre Bragg wavelength and the
corresponding full--width--at--half--maximum bandwidth.
\end{itemize}
Through the elucidation of the effects of infiltration, a further
step towards the practical realization of CSTF--based  optical
sensors, including biosensors which harness chemiluminescent
emission from inside a CSTF, has been taken.

\section*{Acknowledgements}
SSJ is supported by Universiti Tun Hussein Onn Malaysia.  During
part of this study TGM was supported by a  Royal Academy of
Engineering/Leverhulme Trust Senior Research Fellowship.

\section*{Appendix}

In order to  estimate the reference relative permittivity dyadic of
an infiltrated CSTF, a local homogenization procedure is carried out
within a plane parallel to $z=0$. Two components are to be
homogenized: (i) a planar array of similarly--aligned ellipsoidal
particles~---~of refractive index $n_s$ and specified by the linear
size parameter $\eta$ and shape dyadic $\=U$ per eq. \r{r_U}~---~
which are the building blocks of the helical nanowires; and (ii)
void regions~---~represented as a collection of spherical particles,
of radius specified  the linear size parameter $\eta$~---~which are
infiltrated with a fluid of refractive index $n_\ell$. The linear
size parameter $\eta$ is taken to be much smaller than the
wavelength(s) but not vanishingly small.

From  the nonlinear
 Bruggeman equation \c{M_depol}
\begin{equation}
\le 1- f \ri  \=\alpha_{\,\ell/(2)} +  f \=\alpha_{\,s/(2)}  =
\=0\,,
\end{equation}
 $\=\eps^{(2)}_{\,cstf}$
 can be extracted
 by standard numerical methods, such as the Jacobi technique \c{Ch6_MLW98}.
Herein, the polarizability density dyadics
\begin{equation}
\left. \begin{array}{l}
 \=\alpha_{\,\ell/(2)} = \le \sqrt{n_\ell} \, \=I
- \=\eps^{(2)}_{\,cstf} \ri \cdot \les \=I +  i \omega  \=D_{\,(2)}
(\=A_{\,\ell}) \cdot  \le \sqrt{n_\ell} \,\=I -
\=\eps^{(2)}_{\,cstf}
\ri \ris^{-1} \vspace{6pt}\\
\=\alpha_{\,s/(2)} = \le \sqrt{n_s} \, \=I - \=\eps^{(2)}_{\,cstf}
\ri \cdot \les \=I +  i \omega  \=D_{\,(2)} (\=A_{\,s}) \cdot \le
\sqrt{n_s} \,\=I - \=\eps^{(2)}_{\,cstf} \ri \ris^{-1}
\end{array}\right\},
\end{equation}
where
 the depolarization dyadics
\begin{equation} \l{depol}
\=D_{\,(2)} (\=A) = \frac{1}{4 \pi i \omega} \int^{2 \pi}_0 d
\varphi \int^\pi_0 d \vartheta \sin \theta \le \=W^0 + \eta^2 \=W^+
\ri
\end{equation}
with
\begin{equation}
\left.
\begin{array}{l}
\=A_{\,s} = \displaystyle{\sin^2 \vartheta \cos^2 \varphi \, \#u_n
\, \#u_n + \frac{\cos^2 \vartheta}{\gamma^2_\tau} \, \#u_\tau \,
\#u_\tau + \frac{\sin^2 \vartheta \sin^2
\varphi}{\gamma^2_b} \,\#u_b \,\#u_b} \vspace{6pt}\\
\=A_{\,\ell} = \sin^2 \vartheta \cos^2 \varphi \,  \#u_n  \, \#u_n +
\cos^2 \vartheta \, \#u_\tau \,  \#u_\tau  + \sin^2 \vartheta \sin^2
\varphi \,\#u_b \,\#u_b
\end{array}
\right\}.
\end{equation}
The dyadic integrands in eq. \r{depol} are given as
\begin{equation}
\=W^0 =  \frac{1}{\mbox{tr} \le \,\=\eps^{(2)}_{\,cstf} \cdot \=A
\,\ri} \, \=A \,
\end{equation}
and
\begin{eqnarray}
 && \=W^+ =
\frac{1}{3 \, \tau} \lec \,  \les \,
 \frac{3 \le \kappa_+ -
\kappa_-  \ri}{2 } + i \eta \le \kappa^{\frac{3}{2}}_+  -
\kappa^{\frac{3}{2}}_-  \ri \ris
  \=a  +
i  \eta \ko^2  \,\le \kappa^{\frac{1}{2}}_+ - \kappa^{\frac{1}{2}}_-
\ri  \=b \, \ric,  \l{W_def}
\end{eqnarray}
with the 3$\times$3 dyadics
\begin{eqnarray}
  \=a &=&
  \les 2 \,\=\eps^{(2)}_{\,cstf} - \mbox{tr} \le \, \=\eps^{(2)}_{\,cstf} \, \ri
\, \=I \, \ris \cdot \=A - \mbox{tr} \le \,\=\eps^{(2)}_{\,cstf}
\cdot \=A\,\ri \, \=I\,
 \nonumber \\ && -  \, \frac{  \mbox{tr} \les \, \le \=\eps^{(2)}_{\,cstf} \ri^{adj} \cdot\=A\,\ris -
\lec \, \mbox{tr} \les \, \le\=\eps^{(2)}_{\,cstf}\ri^{adj} \, \ris
\, \mbox{tr} \le \, \=A \, \ri \, \ric }{
 \mbox{tr} \le \, \=\eps^{(2)}_{\,cstf} \cdot \=A \, \ri } \,  \=A \,,
\\
 \=b &=&
 \le \=\eps^{(2)}_{\,cstf}\ri^{adj} -  \frac{  \det \le \, \=\eps^{(2)}_{\,cstf} \, \ri }
{ \mbox{tr} \le \, \=\eps^{(2)}_{\,cstf} \cdot \=A \, \ri} \, \=A \,
\end{eqnarray}
and scalar quantities
\begin{eqnarray}
  \tau &=&   \le \lec \mbox{tr} \les \, \le
\=\eps^{(2)}_{\,cstf}\ri^{adj}\cdot\=A\,\ris -
 \mbox{tr} \les \, \le \=\eps^{(2)}_{\,cstf}\ri^{adj} \, \ris \, \mbox{tr} \le
\, \=A \, \ri \, \ric^2 - 4 \det \le \, \=\eps^{(2)}_{\,cstf} \, \ri
\mbox{tr} \le \, \=A \, \ri \,
 \mbox{tr} \le \, \=\eps^{(2)}_{\,cstf} \cdot \=A \, \ri \ri^{\frac{1}{2}}, \nonumber \\ && \\
 \kappa_{\pm}  &=& \ko^2 \frac{ \lec \, \mbox{tr} \les \, \le
\=\eps^{(2)}_{\,cstf}\ri^{adj} \, \ris \, \mbox{tr} \le \, \=A \,
\ri \, \ric - \mbox{tr} \les \, \le
\=\eps^{(2)}_{\,cstf}\ri^{adj}\cdot\=A\,\ris \pm \tau}{2 \,
\mbox{tr} \le \, \=A \, \ri \,
 \mbox{tr} \le \, \=\eps^{(2)}_{\,cstf} \cdot \=A \, \ri}.
\end{eqnarray}
Numerical methods are generally needed to evaluate the surface
integral on the right side of eq. \r{depol} \c{M_JNP_2008}.

\newpage

\begin{figure}[!ht]
\centering
\includegraphics[width=4.9in]{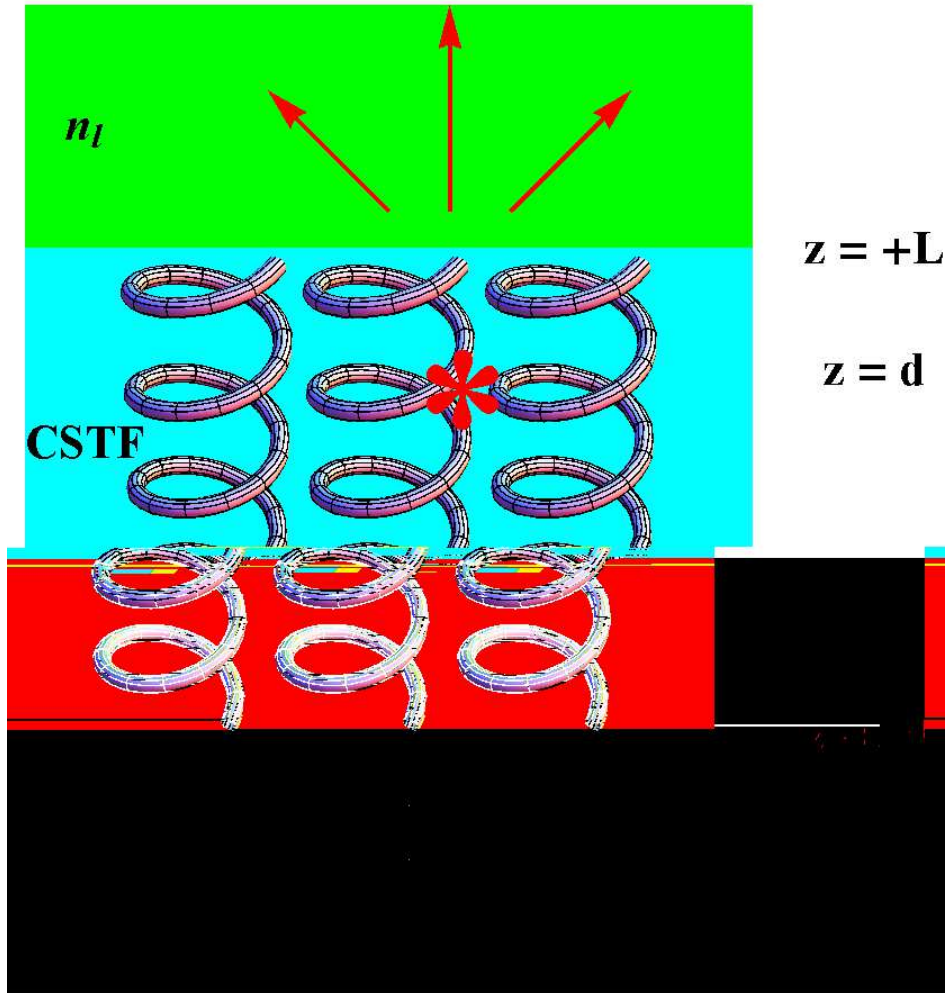}
 \caption{\l{schematic}
 A schematic diagram of the scenario under investigation: radiation emitted from a dipole
 source embedded within an infiltrated CSTF occupying $-L < z < L$.
 }
\end{figure}

\newpage

\begin{figure}[!ht]
\centering
\subfigure[]{\includegraphics[width=2.9in]{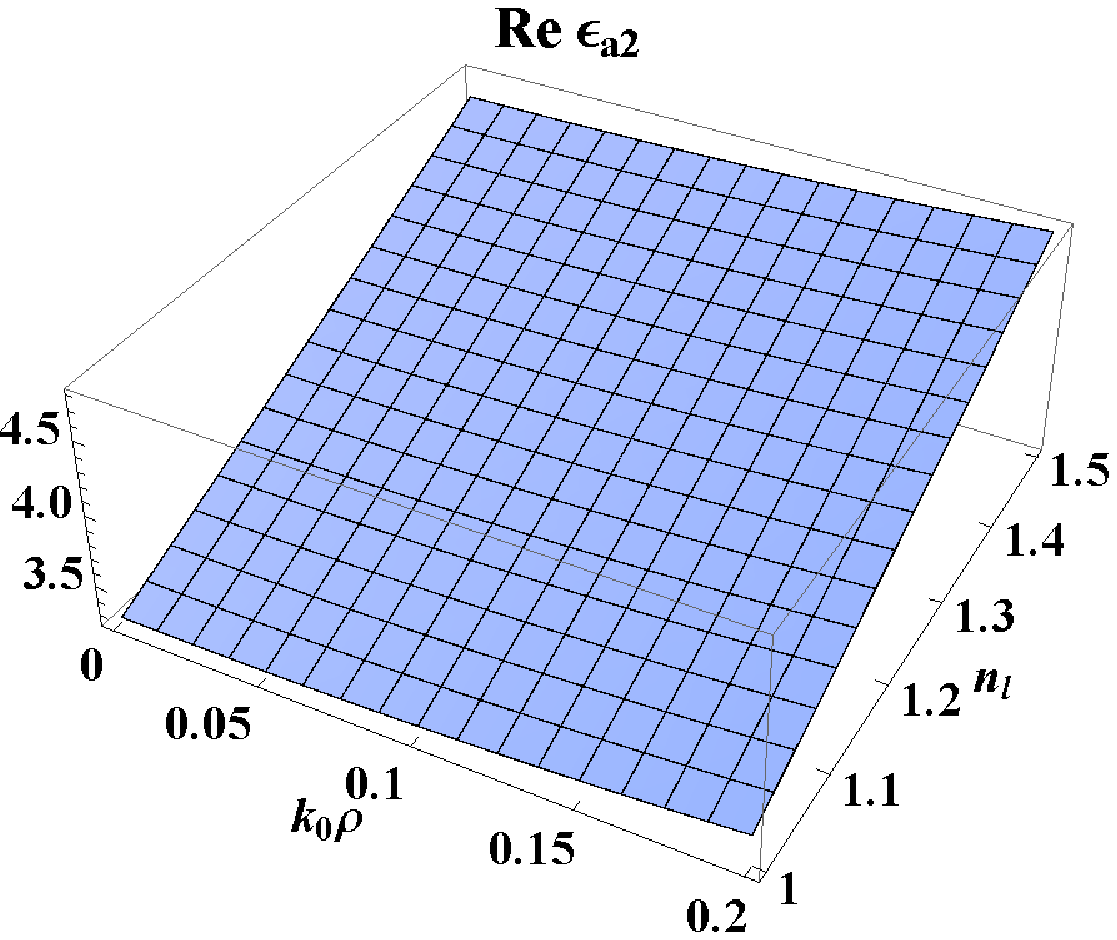}}
\subfigure[]{\includegraphics[width=2.9in]{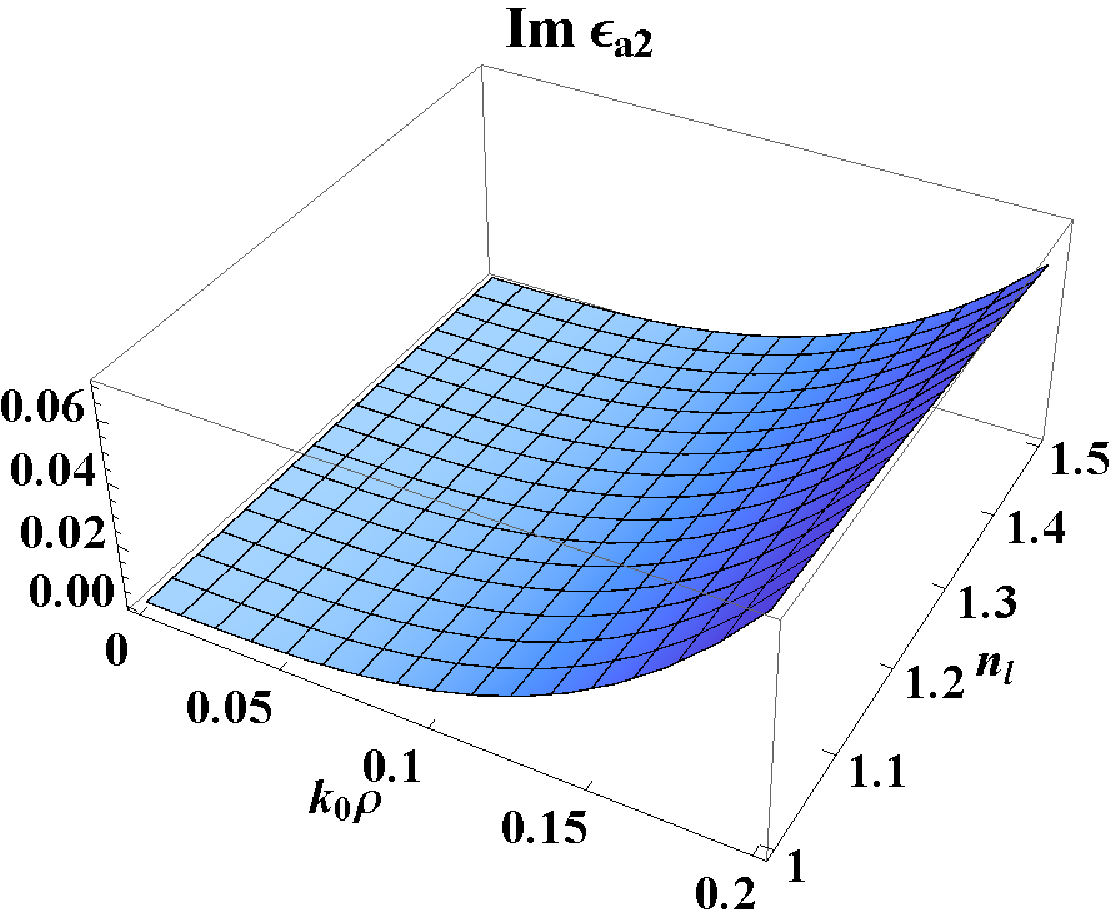}}
\subfigure[]{\includegraphics[width=2.9in]{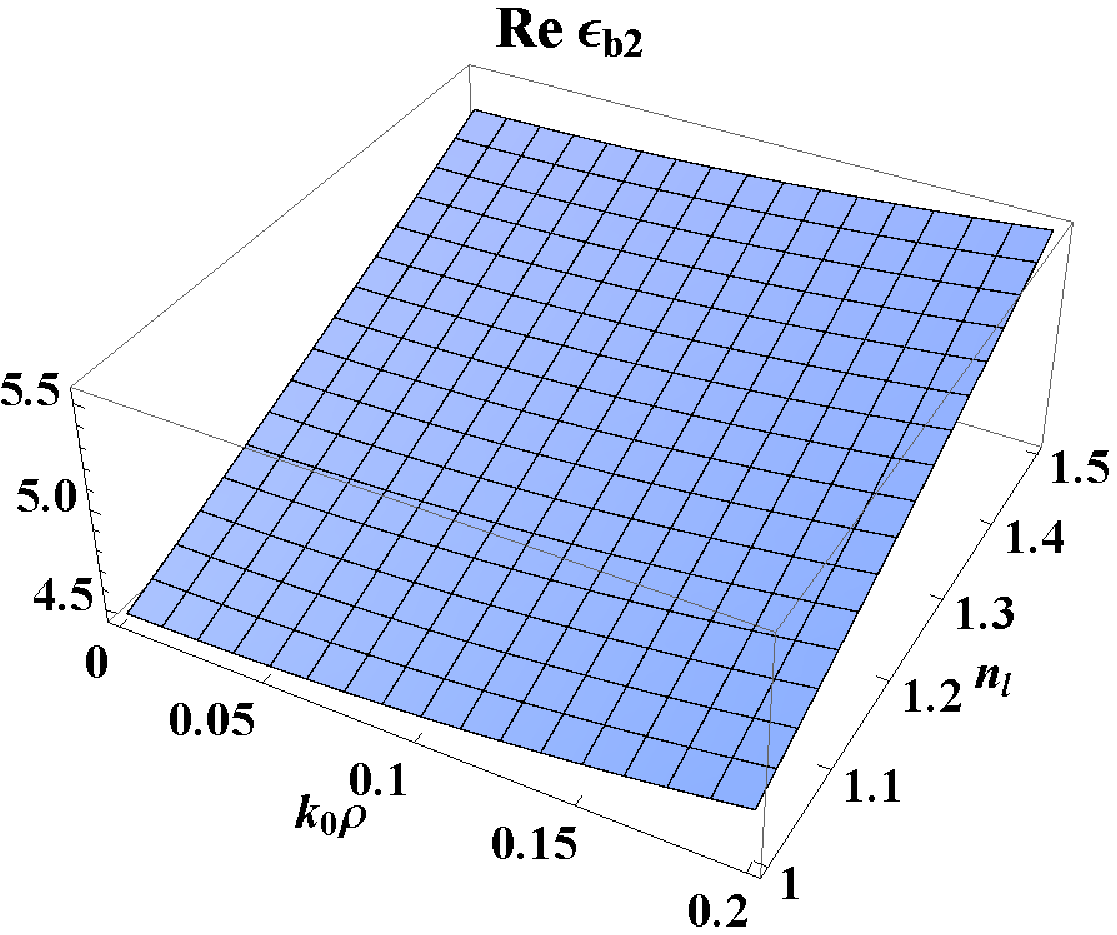}}
\subfigure[]{\includegraphics[width=2.9in]{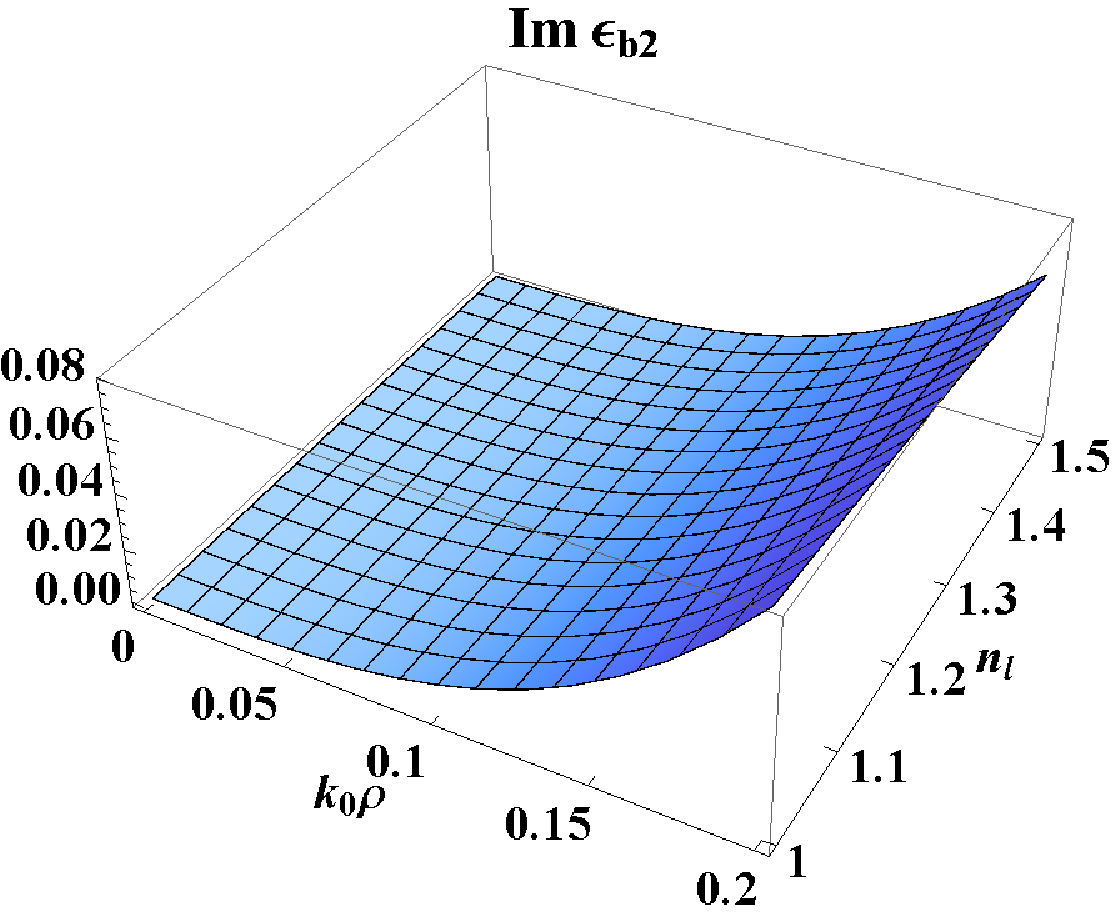}}
\subfigure[]{\includegraphics[width=2.9in]{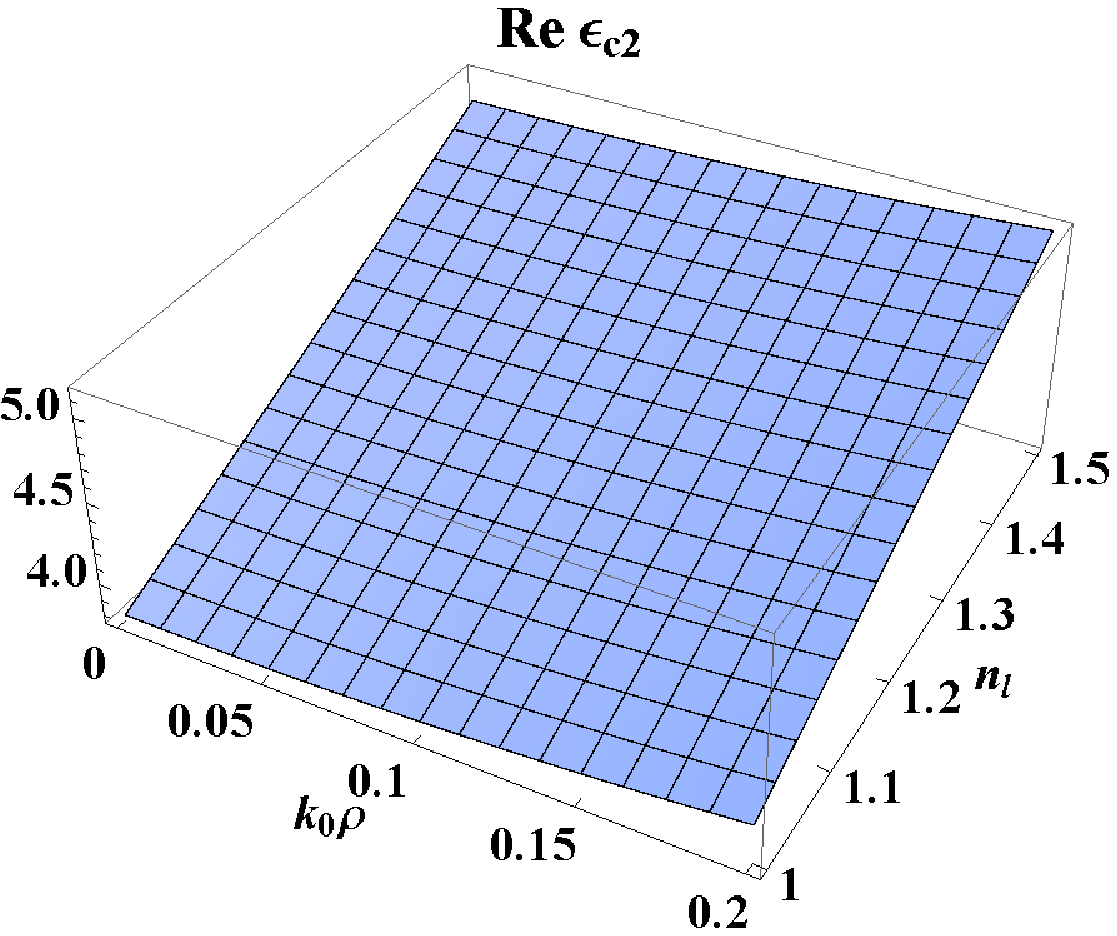}}
\subfigure[]{\includegraphics[width=2.9in]{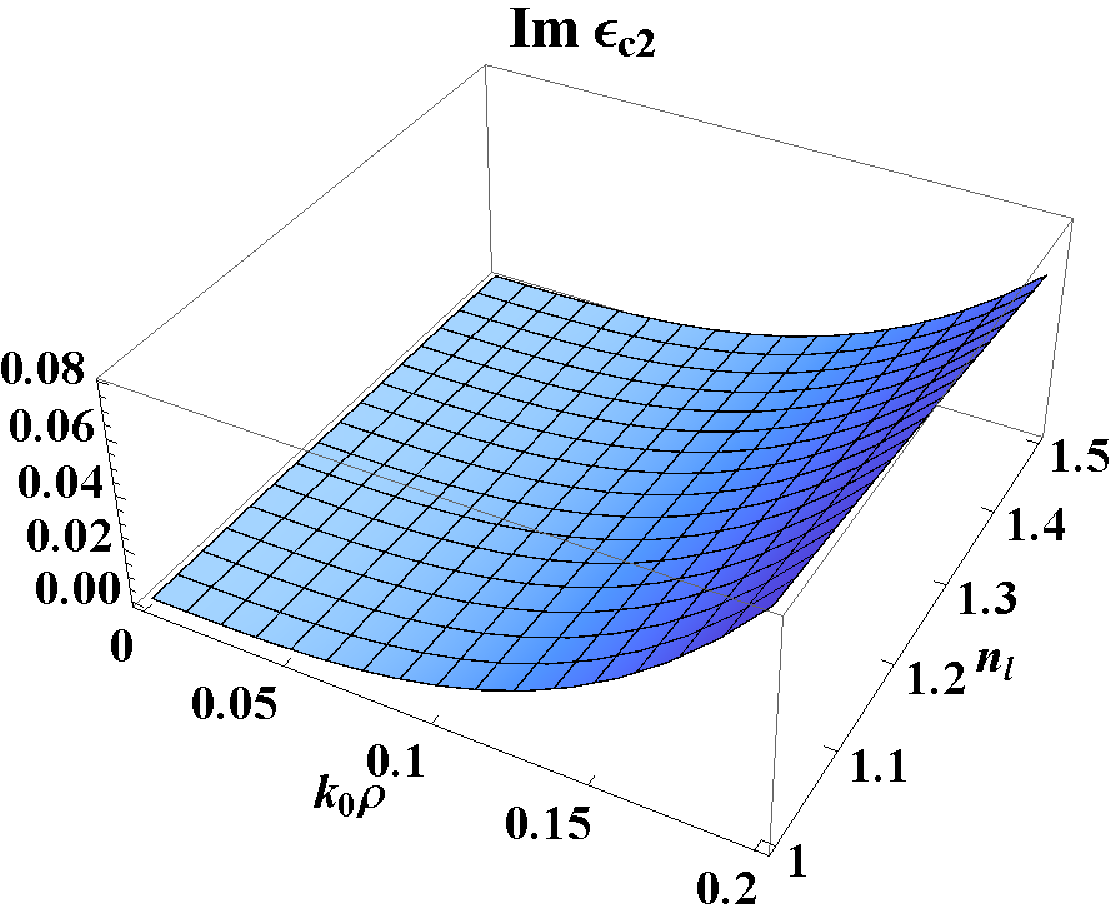}}
 \caption{\l{constitutive_params}
 Real and imaginary parts of the relative permittivity
 parameters
 $\eps_{a2}$,  $\eps_{b2}$ and  $\eps_{c2}$
  versus refractive index  $n_\ell \in (1, 1.5 )$ and relative size parameter $\ko \eta \in ( 0, 0.2 )$. }
\end{figure}

\newpage

\newpage

\newpage

\newpage

\newpage

\newpage

\end{document}